\documentclass[9.5pt,double,journal,compsoc]{IEEEtran} 

\usepackage{textcomp}
\usepackage{xcolor}
\usepackage{tikz}

\usepackage{amsfonts}
\usepackage{graphicx}
\usepackage{amsmath}
\usepackage{bm}
\usepackage[version=4]{mhchem}
\usepackage{siunitx}
\usepackage[ruled,vlined]{algorithm2e}
\usepackage{longtable,tabularx}
\usepackage{amssymb}
\usepackage{setspace}
\usepackage{subcaption}
\usepackage{array}
\usepackage{pdfpages}
\usepackage{lipsum}
\usepackage{wrapfig}
\usepackage{hyperref}
\usepackage{xr}
\usepackage{stmaryrd}
\usepackage{float}
\usepackage{pifont}% http://ctan.org/pkg/pifont
\usepackage[justification=centering]{caption}
\usepackage{enumitem}

\newcommand{\cmark}{\ding{51}}%
\newcommand{\xmark}{\ding{55}}%

\newcommand{\mathbm}[1]{\bm{\mathcal{#1}}}

\ifCLASSOPTIONcompsoc
  \usepackage[nocompress]{cite}
\else
  \usepackage{cite}
\fi

\begin{document}
\title{To Fold or not to Fold: Graph Regularized Tensor Train for Visual Data Completion}
\author{Le~Xu,
        Lei~Cheng,
        Ngai~Wong,
        and Yik-Chung~Wu% <-this % stops a space
\IEEEcompsocitemizethanks{
    \IEEEcompsocthanksitem Le Xu, Ngai Wong and Yik-Chung Wu are with the Department of Electrical and Electronic Engineering, The University of Hong Kong, Hong Kong, (Email: xule@eee.hku.hk, nwong@eee.hku.hk, ycwu@eee.hku.hk).
    \IEEEcompsocthanksitem Lei Cheng is with the College of Information Science and Electronic, Zhejiang University, P. R. China, (Email: lei\_cheng@zju.edu.cn).}% <-this % stops a space
}% \thanks{Manuscript received August 25.}}

\markboth{Journal of \LaTeX\ Class Files,~Vol.~14, No.~8, May~2025}%
{Shell \MakeLowercase{\textit{et al.}}: Bare Advanced Demo of IEEEtran.cls for IEEE Computer Society Journals}

\IEEEtitleabstractindextext{
\begin{abstract}
Tensor train (TT) representation has achieved tremendous success in visual data completion tasks, especially when it is combined with tensor folding. However, folding an image or video tensor breaks the original data structure, leading to local information loss {as nearby pixels may be assigned into different dimensions and become far away from each other}. In this paper, to fully preserve the local information of the original visual data, we explore not folding the data tensor, and at the same time adopt graph information to regularize local similarity between nearby entries. To overcome the high computational complexity introduced by the graph-based regularization in the TT completion problem, we propose to break the original problem into multiple sub-problems with respect to each TT core fiber, instead of each TT core as in traditional methods. Furthermore, to avoid heavy parameter tuning, a sparsity-promoting probabilistic model is built based on the generalized inverse Gaussian (GIG) prior, and an inference algorithm is derived under the mean-field approximation. Experiments on both synthetic data and real-world visual data show the superiority of the proposed methods.
\end{abstract}

}

\maketitle

\IEEEdisplaynontitleabstractindextext
\IEEEpeerreviewmaketitle

\IEEEraisesectionheading{\section{Introduction}}
\label{sec:intro}
\IEEEPARstart{A}{s} a high dimensional generalization of matrices, tensors have shown their superiority on representing multi-dimensional data \cite{zhou2022optimal,liu2020low,sofuoglu2021multi}. In particular, since they can recover the latent {low-rank} structure of color images or videos which naturally appear in high dimensions, tensors have been widely adopted in image processing problems and achieved superior performance over matrix-based methods \cite{liu2022efficient,baust2016combined,jiang2020framelet}. There are many different ways to decompose a tensor, among which the tensor train (TT) decomposition \cite{oseledets2012solution} and its variant tensor ring (TR) decomposition \cite{zhao2016tensor}, have conspicuously shown their advantages in image completion recently.

Basically, TT/TR completion methods target to recover the missing values of a partially observed tensor by assuming the tensor obeys a TT/TR format. With known TT/TR ranks, one can either directly minimize the square error between the observed tensor and the recovered tensor (e.g., sparse tensor train optimization (STTO) \cite{yuan2018high} and tensor ring completion by alternative least squares (TR-ALS) \cite{wang2017efficient}), or by adopting multiple matrix factorizations to approximate the tensor unfoldings along various dimensions (e.g., tensor completion by parallel matrix factorization (TMAC-TT) \cite{bengua2017efficient} and parallel matrix factorization for low TR-rank completion (PTRC) \cite{yu2020low}.

However, the TT/TR ranks are generally unknown in practice, and the choice of the TT/TR ranks significantly affects the performance of the algorithm. Instead of determining the TT/TR ranks by trial and error, methods like simple low-rank tensor completion via TT (SiLRTC-TT) \cite{bengua2017efficient} and robust tensor ring completion (RTRC) \cite{huang2020robust}, try to minimize the TT/TR rank by applying the nuclear-norm regularization on different modes of the unfolded tensor. While this strategy seems to lift the burden of determining TT/TR ranks, it actually shifts the burden to tuning the regularization parameters for balancing the relative weights among the recovery error and the regularization terms. To avoid heavy parameter tuning, probabilistic tensor train completion (PTTC) \cite{xu2021overfitting} and tensor ring completion based on the variational Bayesian framework (TR-VBI) \cite{long2021bayesian} were proposed. They are based on probabilistic models, which has the ability to learn the TT/TR ranks and regularization parameters automatically.

\begin{figure*}[tb!]
    \centering
    \begin{subfigure}{0.13\linewidth}
        \centering
        \captionsetup{justification=centering}
        \includegraphics[width=0.95\linewidth]{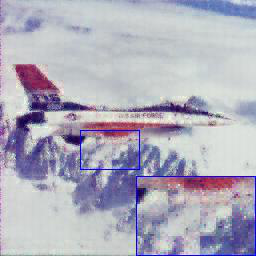}
        \caption{Ket-folding\\(24.75dB)\\\text{ }}
        \label{subfig:blocksilrtc}
    \end{subfigure}
    \begin{subfigure}{0.13\linewidth}
        \centering
        \captionsetup{justification=centering}
        \includegraphics[width=0.95\linewidth]{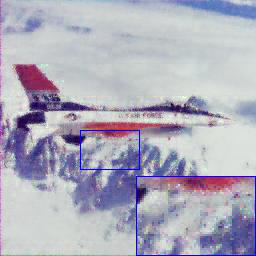}
        \caption{Improved Ket-folding\\(25.51dB)}
        \label{subfig:blockimprovedsilrtc}
    \end{subfigure}
    \begin{subfigure}{0.13\linewidth}
        \centering
        \captionsetup{justification=centering}
        \includegraphics[width=0.95\linewidth]{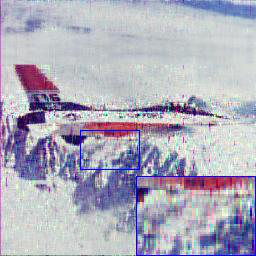}
        \caption{No folding\\(22.69dB)\\\text{ }}
        \label{subfig:nofoldingSilrtc}
    \end{subfigure}
    \begin{subfigure}{0.13\linewidth}
        \centering
        \captionsetup{justification=centering}
        \includegraphics[width=0.95\linewidth]{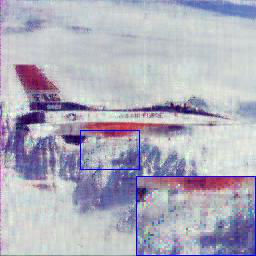}
        \caption{TTC-TV\\(23.87dB)\\\text{ }}
        \label{subfig:blockttctv}
    \end{subfigure}
    \begin{subfigure}{0.13\linewidth}
        \centering
        \captionsetup{justification=centering}
        \includegraphics[width=0.95\linewidth]{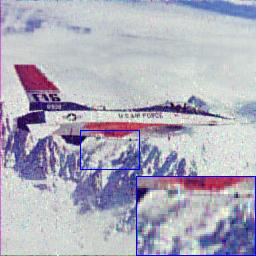}
        \caption{Proposed GraphTT-opt\\(26.60dB)}
        \label{subfig:plane60graphTTopt}
    \end{subfigure}
    \begin{subfigure}{0.13\linewidth}
        \centering
        \captionsetup{justification=centering}
        \includegraphics[width=0.95\linewidth]{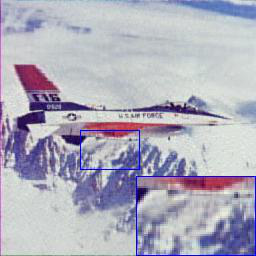}
        \caption{Proposed GraphTT-VI\\(29.81dB)}
        \label{subfig:airplane80graphTTvi}
    \end{subfigure}
    \caption{'airplane' with $60\%$ missing entries recovered by: (a)-(c). TMAC-TT under different folding strategies, (d) TTC-TV with reshape-folding, (e)-(f) proposed methods without folding.}
    \label{fig:blockeffect}
\end{figure*}

\begin{table*}[tb!]
    \centering
    \caption{Comparison between problem sizes for a video tensor with and without folding (max rank set as $16$).}
    \begin{tabular}{c|c|c}
    \hline
        Size of the folded tensor & TT cores' total size & Size of matrix to be inverted in each iteration\\
    \hline
        $[256,256,3,32]$ & 70912 & [65536,65536]\\
        $[4,4,4,4,4,4,4,4,3,8,4]$ & 3200 & [1024,1024]\\
    \hline
    \end{tabular}
    \label{tab:foldsize}
\end{table*}

Different from other tensor decompositions, most existing TT/TR methods for image completion are conducted after tensor folding, which folds the 3-dimensional images or 4-dimensional videos to a higher order tensor, e.g., a 9-dimensional tensor. 
There are two commonly adopted folding strategies. One is ket-folding, or ket augmentation (KA), which was firstly applied in TT format for compressing images \cite{latorre2005image}, and later found to be effective in image completion \cite{bengua2017efficient}. This strategy spatially breaks an image or a video into many small blocks, and then uses them to fill up a higher-order tensor. The other one is reshape-folding, which simply assigns the elements of an image/video tensor sequentially into a higher-order tensor. Together with folding, TT/TR-based methods achieve the state-of-the-art performance in image completion tasks \cite{bengua2017efficient,ko2020fast,xu2021overfitting,long2021bayesian}.

While the folding techniques improve the traditional evaluating metrics like PSNR, visual inspection of the recovered images shows that they are plagued with heavy 'block effects'. An example is shown in Fig. \ref{subfig:blocksilrtc}, where the recovered 'airplane' image by TMAC-TT from folded tensor data looks like composing of many small blocks and the edges of the blocks show obvious incoherence. The reason is that tensor folding breaks adjacent pixels into different dimensions. Pixels originally close to each other are assigned to new dimensions and become far away, leading to local information loss. Though an improved folding strategy \cite{xu2021overfitting} might help to alleviate the block effect, it is still visible in Fig. \ref{subfig:blockimprovedsilrtc}. Furthermore, since the improved folding strategy duplicates the edges of folding blocks, the image size is effectively increased, so is the computational complexity of the algorithm. In comparison, Fig. \ref{subfig:nofoldingSilrtc} shows the recovered image using the same algorithm as in Fig. \ref{subfig:blocksilrtc} but without folding. Although some parts of the image are not as clear as that in Fig. \ref{subfig:blocksilrtc}, there is no block effect.

Besides the block effect, folding also makes it less efficient to incorporate prior knowledge of image or video like local similarity, which has been widely used to aid visual data recovery, especially in matrix-based methods \cite{jiang2013graph,strahl2020scalable,CHEN2023108826}. After folding, adjacent elements are cast into different dimensions, and the local similarity is only retained within each small block. An example is illustrated in Fig. \ref{subfig:blockttctv}, where the recovered 'airplane' image by the TT completion with total variation regularization (TTC-TV) \cite{ko2020fast} is shown. Since the TV regularization of TTC-TV only enforces local similarity on each mode of the folded tensor, the block effect is still visible in the image. Surprisingly, the resulting PSNR is even lower than that from TMAC-TT with folding but no local similarity regularization (Fig. \ref{subfig:blocksilrtc} and Fig. \ref{subfig:blockimprovedsilrtc}).

Given that tensor folding and local similarity are not compatible to each other, one may suggest imposing local similarity but no folding. However, this brings another challenge, which is the large TT/TR core sizes in the model. To understand this clearly, let us focus on the TT format, as the case of TR would be similar. For a tensor of dimensions $[J_1,\ldots,J_D]$ with TT ranks $\{R_d\}_{d=1}^{D+1}$, the TT format applied to the original tensor contains a total of $\sum_{d=1}^D J_d R_d^2$ elements. The problem size is much larger than a folded tensor with number of elements $\sum_{d=1}^D \sum_{m=1}^{M} J_{d_m} R_{d_m}^2$ where $J_d = \prod_{m=1}^M J_{d_m}$ if all $R_d$ and $R_{d_m}$ take similar values, which is often the case. Furthermore, TT completion problems are usually solved in an alternative least squares (ALS) manner, which updates the TT cores iteratively by solving a quadratic sub-problem for each TT core \cite{holtz2012alternating,cichocki2017tensor,grasedyck2015alternating}. Due to the correlation induced by the local similarity among slices of the TT cores, an inverse of a $J_dR_dR_{d+1} \times J_dR_dR_{d+1}$ matrix is commonly induced in each iteration, which is both space and time consuming if the tensor is not folded. This is in contrast to TT completion without the local similarity regularization, in which only $J_d$ matrices each with size $R_dR_{d+1}\times R_dR_{d+1}$ are needed to be inverted, since the frontal TT core slices are found to have no correlation with each other \cite{yu2021robust}.

To illustrate these complications, an example is taken from a video tensor with size $256 \times 256 \times 3 \times 32$, where the first $3$ dimensions describe the size of each frame and the last is the number of frames. It can be seen from Table \ref{tab:foldsize} that without folding, the model size is more than 20 times larger than that with folding, and the matrix to be inverted in each iteration is too large to perform in a personal computer. Even under this modest setting, algorithms like TTC-TV cannot be executed in a computer with less than $32$GB RAM. In fact, graph regularization on TR decomposition (GNTR) without folding has been attempted in \cite{yu2022graph}. However, due to the above-mentioned high complexity issue, all simulations have been done with very small TR ranks like $3$ or $5$. This might be another reason why most existing TT/TR completion methods involve folding.

\begin{table*}[tb!]
    \centering
    \caption{Comparison between existing TT/TR methods and the proposed ones.}
    \scalebox{0.8}{
    \begin{tabular}{c|c|c|c|c|c|c|c|c}
    \hline
        \quad & TMAC-TT & SiLRTC-TT & STTO & TTC-TV & TR-VBI & GNTR & GraphTT-opt & GraphTT-VI \\
        \hline
        No need to fold & \xmark & \xmark & \xmark & \xmark & \xmark & not applicable  & \cmark & \cmark\\
        Graph & \xmark & \xmark & \xmark & \xmark & \xmark & \cmark & \cmark & \cmark\\
        Completion & \cmark & \cmark & \cmark & \cmark & \cmark & \xmark & \cmark & \cmark \\
        Tuning free & \xmark & \xmark & \xmark & \xmark & \cmark & \xmark & \xmark & \cmark\\
        \hline
    \end{tabular}}
    \label{tab:TTcompare}
\end{table*}

In this paper, we focus on the visual data completion problem. As visual data can hardly be accurately represented with very small ranks (e.g., TT/TR with ranks smaller than 10) \cite{long2021bayesian,ko2020fast}, this brings us to the dilemma: to fold or not to fold. Folding an image or a video tensor would reduce computational complexity \textit{but leads to block effect} due to local information loss. Not folding a tensor would not lead to block effect and allow us to induce local similarity in the formulation \textit{but incur high computational complexity}. We propose not to fold the image/video tensor, but use local similarity to boost the completion performance. {The graph regularization is adopted to incorporate such similarity due to its proven effectiveness \cite{strahl2020scalable,CHEN2023108826}.
To overcome the problem of computational complexity, we propose updating each TT core fiber as a unit rather than updating the entire TT core.}

{In addition, to eliminate the need for parameter tuning in the proposed optimization-based algorithm, we further reformulate the problem within a fully Bayesian framework. Specifically, we construct a probabilistic model for all TT cores and derive the corresponding inference algorithm. The resulting approach can be seamlessly applied to a wide range of tensor completion tasks without requiring tuning parameters such as TT ranks or regularization weights.}

To see the differences between the proposed methods and existing methods, Table \ref{tab:TTcompare} lists various TT/TR completion methods and their modeling characteristics. It can be seen that the proposed methods (GraphTT-opt and GraphTT-VI) are the first ones to embed the graph information into TT completion. As our key ideas are applicable to both TT and TR completion, in this paper we only focus on the TT completion, and the extension to the TR completion is trivial.

Notice that while a recent work GNTR \cite{yu2022graph} impose graph regularization to TR without folding, it cannot be directly adopted for the tensor completion tasks as it cannot handle missing data. In addition, as discussed before, it is not applicable for high-rank tensor since it does not handle the issue of high computational complexity. Furthermore, it heavily relies on parameter tuning.

\begin{table}[!tb]
\centering
\caption{Summary of notations.}
\scalebox{0.76}{
\begin{tabular}{| c | c  |}
 \hline
 Notation & Terminology  \\
\hline
 $\bm{y}$ & boldface lowercase letters to denote \textbf{vectors} \\
\hline
 $\bm{Y}$ &  boldface uppercase letters to denote \textbf{matrics} \\
 \hline
 $\bm{\mathcal{Y}}$ & boldface capital calligraphic letters to denote \textbf{tensors} \\
 \hline
 $\bm{\mathcal{Y}}_{i,j,k}$ & the $(i,j,k)$-th element of $\bm{\mathcal{Y}}$ \\
 \hline
 $\bm{\mathcal{Y}}_{:,:,k}$ & the $k$-th frontal slice of $\bm{\mathcal{Y}}$ \\
 \hline 
 $\bm{\mathcal{Y}}_{(d)}$ & mode-$d$ metricization of $\bm{\mathcal{Y}}$ \\
 \hline
 $\bm{\mathcal{G}}^{(d)}$ & the $d$-th TT core from $\ll \bm{\mathcal{G}}^{(1)},\bm{\mathcal{G}}^{(2)},\ldots,\bm{\mathcal{G}}^{(D)} \gg$ \\
 \hline
 $\bm{\mathcal{G}}^{(<d)}$ & $d$-th order tensor composed from $\bm{\mathcal{G}}^{(1)},\ldots,\bm{\mathcal{G}}^{(d)}$\\
 \hline
 $\bm{\mathcal{G}}^{(>d)}$ & $(D-d)$-th order tensor composed from $\bm{\mathcal{G}}^{(d+1)}, \ldots,\bm{\mathcal{G}}^{(D)}$\\
 \hline
 $\bm{I}_n$ & identity matrix with size $n\times n$ \\
 \hline
  $\mathbb{E}\llbracket.\rrbracket$ & expectation of the variables \\
  \hline
  $\mathcal{N}(\bm{\mu},\bm{\Sigma})$ & Gaussian distribution with mean $\bm{\mu}$ and covariance $\bm{\Sigma}$ \\
  \hline
  $\text{Gamma}(\alpha,\beta)$ & Gamma distribution with shape $\alpha$ and rate $\beta$ \\
  \hline
  $\otimes$ & Kronecker product\\
  \hline
  $\ast$ &  entry-wise product \\
  \hline
\end{tabular}}
\label{tab:notation}
\end{table}

The contributions of this paper are summarized below:

{1. Graph regularization is incorporated into the TT model to eliminate the need for tensor folding, thereby avoiding block artifacts. To address the resulting computational burden, we propose updating TT core fibers independently in each iteration, which improves efficiency significantly.}

2. Bayesian modeling and inference algorithm of the above problem are derived. This gets rid of the tedious tuning of TT ranks and regularization parameters.

3. Experiments show the proposed graph-regularized TT completion methods without folding perform better than existing methods while not causing block effect in the recovered data. A sneak preview of the performance of the proposed methods is presented in Fig. \ref{fig:blockeffect}.

The notations adopted in this paper are summarized in Table. \ref{tab:notation}.

\section{Preliminaries}
\label{sec:preliminaries}

\subsection{Tensor train completion and the ALS solution}
\label{subsec:pre_tt}
In this subsection, we first introduce the tensor train completion problem. Through a sketch of the widely used ALS algorithm, some properties of TT are also introduced, which are important in the proposed algorithms in later sections.

\begin{figure}[tb!]
    \centering
    \includegraphics[width=0.95\linewidth]{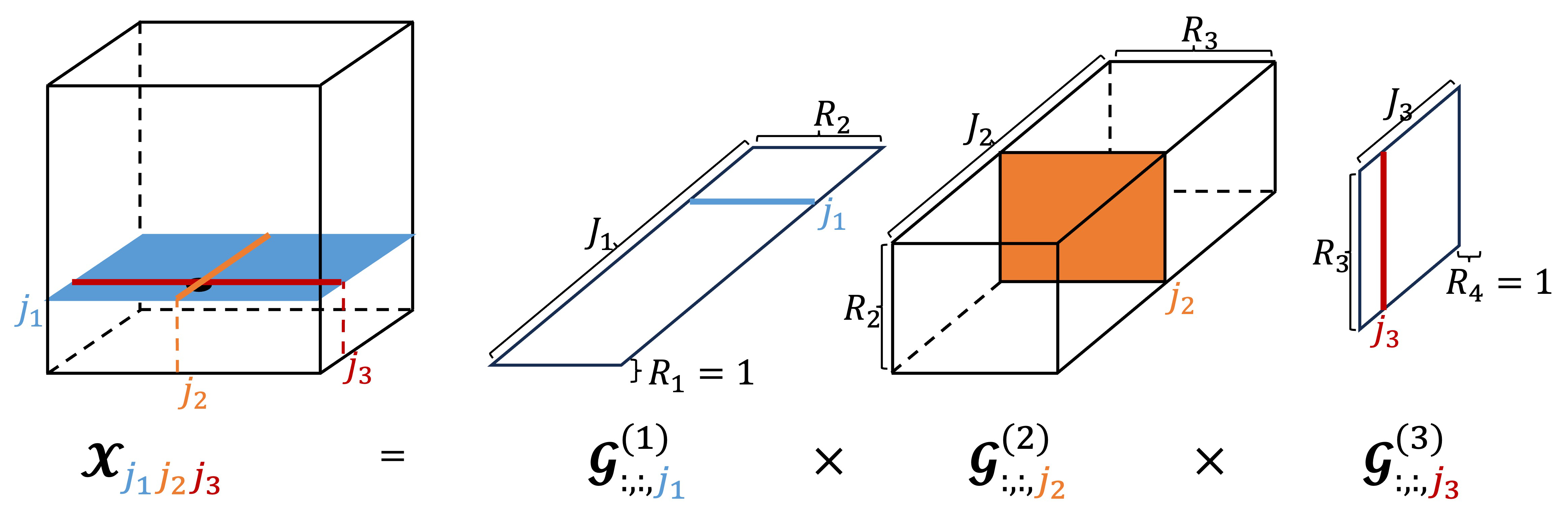}
    \caption{{An illustration of the TT decomposition.}}
    \label{fig:ttdemo}
\end{figure}

\textbf{\textit{Definition 1}} \cite{sixteenoseledets2011tensor}\textbf{\textit{.}} A $D$-th order tensor $\mathbm{X}\in \mathbb{R}^{J_1\times J_2 \times \ldots \times J_D}$ is in the TT format if its elements can be expressed as
\begin{align}
    \bm{\mathcal{X}}_{j_1j_2\ldots j_D} & = \bm{\mathcal{G}}_{:,:,j_1}^{(1)}\bm{\mathcal{G}}_{:,:,j_2}^{(2)},\ldots\bm{\mathcal{G}}_{:,:,j_D}^{(D)},\nonumber \\
    & \triangleq \ll \bm{\mathcal{G}}^{(1)},\bm{\mathcal{G}}^{(2)},\ldots,\bm{\mathcal{G}}^{(D)} \gg_{j_1j_2\ldots j_D}
    \label{eqn:TTdefinition}
\end{align}
in which $\{\mathbm{G}^{(d)} \in \mathbb{R}^{R_d \times R_{d+1} \times J_d}\}_{d=1}^{D}$ are the TT cores, and $\{R_d\}_{d=1}^{D+1}$ are the TT ranks. As can be seen from (\ref{eqn:TTdefinition}), to express $\bm{\mathcal{X}}_{j_1j_2\ldots j_D}$, the $\{j_d\}_{d=1}^{D}$-th frontal slices are selected from the TT cores respectively and then multiplied consecutively. As the product of these slices must be a scalar, $R_1$ and $R_{D+1}$ must be $1$. For the other TT ranks $\{R_d\}_{d=2}^D$, they control the model complexity and are generally unknown. {Fig. \ref{fig:ttdemo} demonstrates the TT decomposition of a 3rd-order tensor.}

Suppose $\bm{\mathcal{X}}=\ll \bm{\mathcal{G}}^{(1)},\bm{\mathcal{G}}^{(2)},\ldots  ,\bm{\mathcal{G}}^{(D)} \gg$ is the tensor to be recovered, and the observed tensor is
\begin{align}
    \mathbm{Y} = \mathbm{O} \ast (\mathbm{X} + \mathbm{W}),
    \label{eqn:observed}
\end{align}
where $\mathbm{W}$ is a noise tensor which is composed of element-wise independent zero-mean Gaussian noise, and $\mathbm{O}$ is an indicator tensor with the same size as $\mathbm{X}$ with its element being $1$ if the corresponding element in $\mathbm{X}$ is observed, and $0$ otherwise. The basic TT completion problem is formulated as
\begin{align}
    \min_{\bm{\mathcal{G}}^{(1)},\bm{\mathcal{G}}^{(2)},\ldots  ,\bm{\mathcal{G}}^{(D)}} & \left\| \bm{\mathcal{O}} \ast 
    \left( \bm{\mathcal{Y}} - \ll \bm{\mathcal{G}}^{(1)},\bm{\mathcal{G}}^{(2)},\ldots  ,\bm{\mathcal{G}}^{(D)} \gg \right) \right\|_F^2, \nonumber \\
    \textit{s.t. } & \text{TT ranks} = [1,R_2,\ldots,R_D,1].
    \label{eqn:basicttcompletion}
\end{align}

To solve problem (\ref{eqn:basicttcompletion}), ALS is commonly adopted \cite{grasedyck2015alternating,yu2021robust}, which updates each TT core iteratively until convergence. However, due to the complicatedly coupled tensor cores in the TT format, the following definition and a related property are needed to obtain the solution for each update step.

\textbf{\textit{Definition 2.}} The mode-$d$ matricization of tensor $\mathbm{X}$ is denoted as $\bm{X}_{(d)} \in \mathbb{R}^{J_d \times (J_1 \ldots J_{d-1} J_{d+1} \ldots J_D)}$, which is obtained by stacking the mode-$d$ fibers $\mathbm{X}_{j_1,\ldots,j_{d-1},:,j_{d+1},\ldots,j_D}$ as columns of a matrix. Specifically, the mapping from an element of $\mathbm{X}$ to $\bm{X}_{(d)}$ is as follows
\begin{align}
    \mathbm{X}_{j_1\ldots j_D} \rightarrow {\bm{X}_{(d)}}_{j_d,i}, \text{with }{i = j_1 + \prod_{\substack{k=2 \\ k\neq d}}^D \bigg( (j_k -1) \prod_{\substack{\ell=1 \\ \ell \neq d}}^{k-1} J_{\ell} \bigg)}.
\end{align}

\textbf{\textit{Property 1.}} For tensor $\mathbm{X}$ obeying the TT format in (\ref{eqn:TTdefinition}), its mode-$d$ matricization can be expressed as
\begin{align}
    \bm{X}_{(d)} = \bm{G}_{(3)}^{(d)} \times (  {\bm{G}_{(1)}^{(>d)}}\otimes {\bm{G}_{(d)}^{(<d)}} ),
    \label{eqn:modedG}
\end{align}
where $\bm{G}_{(3)}^{(d)}$ is the mode-$3$ matricization of $\mathbm{G}^{(d)}$, $\bm{G}_{(d)}^{(<d)}$ is the mode-$d$ matricization of $\mathbm{G}^{(<d)}$, with $\mathbm{G}^{(<d)} \in \mathbb{R}^{J_1 \times\ldots\times J_{d-1}\times R_d}$ and its element composed by $\mathbm{G}_{j_1,\ldots,j_{d-1},:}^{(<d)} = (\mathbm{G}_{:,:,j_1}^{(1)}\mathbm{G}_{:,:,j_2}^{(2)}\ldots \mathbm{G}_{:,:,j_{d-1}}^{(d-1)})^T$, and $\bm{G}_{(1)}^{(>d)}$ stands for the mode-$1$ unfolding of $\mathbm{G}^{(>d)} \in \mathbb{R}^{R_{d+1} \times J_{d+1}\times\ldots\times J_{D}}$, with $\mathbm{G}_{:,j_{d+1},\ldots,j_{D}}^{(>d)} = \mathbm{G}_{:,:,j_{d+1}}^{(d+1)}\mathbm{G}_{:,:,j_{d+2}}^{(d+2)}\ldots \mathbm{G}_{:,:,j_{D}}^{(D)}$.

Using {\textit{Property 1}}, the subproblem of updating the TT core $\mathbm{G}^{(d)}$ can be reformulated as
\begin{align}
    \min_{\mathbm{G}^{(d)}} \quad &\left\| \bm{O}_{(d)} \ast \bigg(\bm{Y}_{(d)} -  \bm{G}_{(3)}^{(d)}\times (\bm{G}_{(1)}^{(>d)}\otimes \bm{G}_{(d)}^{(<d)})\bigg) \right\|_F^2 \nonumber\\
    = \min_{\mathbm{G}^{(d)}} \quad & \sum_{j_d=1}^{J_d} \Bigg\|  {\bm{O}_{(d)}}_{j_d,:} \ast {\bm{Y}_{(d)}}_{j_d,:} - {\bm{G}_{(3)}^{(d)}}_{j_d,:} \nonumber \\
    & \quad \quad \times \bigg( {\bm{O}_{(d)}}_{j_d,:} \odot (\bm{G}_{(1)}^{(>d)}\otimes \bm{G}_{(d)}^{(<d)})  \bigg)  \Bigg\|_F^2.
\label{eqn:basicopt_wrtTTcore}
\end{align}
From the first line of (\ref{eqn:basicopt_wrtTTcore}), it is clear that the sub-problem is quadratic with respect to each TT core. Moreover, from the second line of (\ref{eqn:basicopt_wrtTTcore}) it is worth noticing that different frontal slices from the same TT core $\{ {\bm{G}_{(3)}^{(d)}}_{j_d,:} = \text{vec}(\mathbm{G}_{:,:,j_d}^{(d)})^T\}_{j_d=1}^{J_d}$ are independent of each other. Thus in each iteration, the solution is obtained by updating its frontal slices in parallel, with
\begin{align}
    {\bm{G}_{(3)}^{(d)}}_{j_d,:}^T = &{\bigg( {\bm{O}_{(d)}}_{j_d,:} \odot (\bm{G}_{(1)}^{(>d)}\otimes \bm{G}_{(d)}^{(<d)})  \bigg)^T}^\dagger  \nonumber \\
    &\quad \quad \quad \times \bigg({\bm{O}_{(d)}}_{j_d,:} \ast {\bm{Y}_{(d)}}_{j_d,:}\bigg)^T,
    \label{eqn:lssolution_wrtTTcoreslice}
\end{align}
in which the superscript $\dagger$ denotes the Moore-Penrose pseudo inverse.

Since each TT core slice takes the size $R_d\times R_{d+1}$, the computational complexity for updating one TT core according to (\ref{eqn:lssolution_wrtTTcoreslice}) is $\mathcal{O}(R_d^3 R_{d+1}^3 J_d)$, and the storage required for the matrix inverse is of $\mathcal{O}(R_d^2 R_{d+1}^2)$. With high TT ranks, the ALS algorithm would be computationally complicated, which unfortunately is often the case for visual data. This problem would be much more severe when the independence among frontal slices is lost under graph regularization, as will be shown in Section \ref{subsec:graphLap}. Even though matrix inverse can be avoided by gradient methods (e.g., STTO \cite{yuan2018high}), it converges slowly and cannot reduce the storage complexity.

\subsection{Graph Laplacian}
\label{subsec:graphLap}
To introduce smoothness among the entries in a matrix or tensor, graph Laplacian-based regularization has recently been widely adopted \cite{chen2013simultaneous,strahl2020scalable,CHEN2023108826}. For an undirected weighted graph $(\mathcal{V},\mathcal{E})$ with vertices $\mathcal{V}=\{v_1,\ldots,v_N\}$ and the set of edges $\mathcal{E}$, its graph Laplacian is
\begin{align}
    \bm{L} = \bm{D} - \bm{A},
    \label{eqn:graphdef}
\end{align}
in which $\bm{A}$ is the weight matrix with $\bm{A}_{ij}$ being the weight between $v_i$ and $v_j$, and $\bm{D}$ is a diagonal matrix with $\bm{D}_{ii} = \sum_{j=1}^{N} A_{ij}$. For a vector $\bm{x}\in \mathbb{R}^{N}$, $\text{tr}(\bm{x}^T\bm{L}\bm{x})$ would introduce smoothness among elements of $\bm{x}$ according to $\bm{L}$ since
\begin{align}
    \text{tr}(\bm{x}^T \bm{L} \bm{x}) = \frac{1}{2} \sum_{i=1}^{N} \sum_{j=1}^N \bm{A}_{ij} ||\bm{x}_i - \bm{x}_j||_2^2.
    \label{eqn:graphregu_intro}
\end{align}

\begin{figure*}[tb!]
    \centering
    \includegraphics[width=0.8\linewidth]{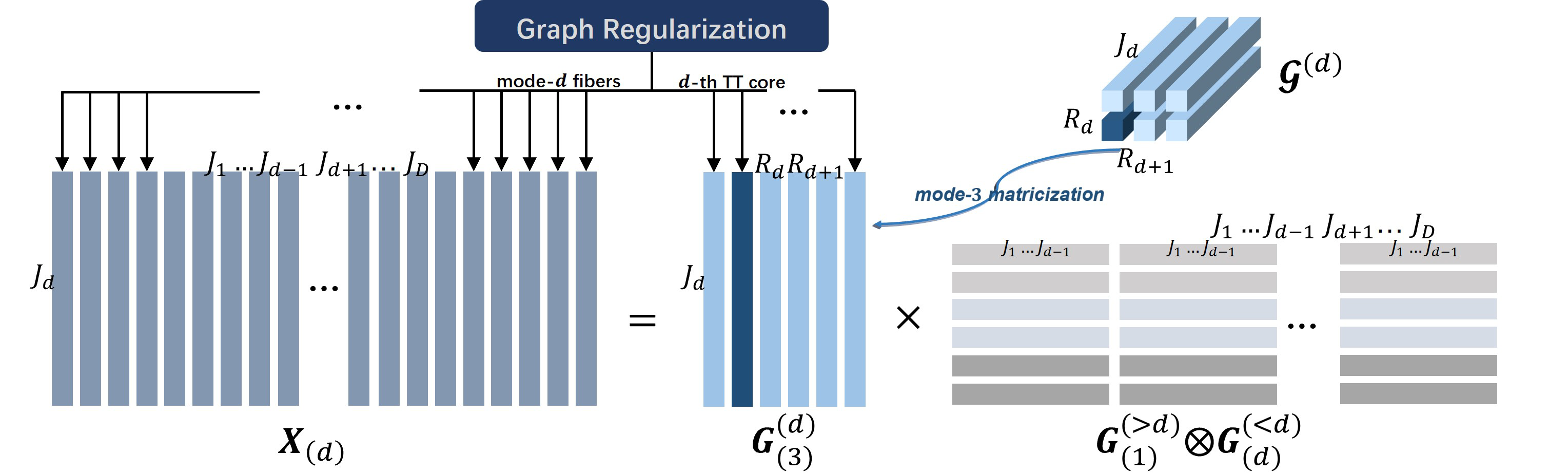}
    \caption{Tensor unfolding and Graph regularization.}
    \label{fig:graphonXd}
\end{figure*}

To apply the graph regularization (\ref{eqn:graphregu_intro}) in visual data, a direct but naive implementation is to specify the connections between every two entries in the data tensor. However, such method requires a very large graph Laplacian, e.g., a graph Laplacian with size $196008\times 196008$ for a $256\times 256 \times 3$ image, which would bring a heavy computational burden. 

Fortunately, from the experience of graph regularized matrix factorization \cite{shahid2016fast,CHEN2023108826}, graph information can be added to rows and columns respectively, rather than the vectorization of the matrix. Furthermore, for matrix decomposition $\bm{A} = \bm{V}^T \bm{W}$, the graph regularization on the rows and columns can be formulated using the latent matrices, as $\text{tr}(\bm{V} \bm{L}_{\bm{1}} \bm{V}^T )$ and $\text{tr}(\bm{W}^T \bm{L}_{\bm{2}} \bm{W} )$, respectively \cite{paradkar2017graph}.

From (\ref{eqn:modedG}), it is noticed that $\bm{G}_{(3)}^{(d)}$ and $\bm{G}_{(1)}^{(>d)} \otimes \bm{G}_{(d)}^{(<d)}$ can be seen as the left and right factor matrices of $\bm{X}_{(d)}$, respectively. Then inspired by the graph regularized matrix decomposition, if we would like to regularize the mode-$d$ fibers of $\bm{\mathcal{X}}$, we can formulate it using $\bm{G}_{(3)}^{(d)}$ as follows,
\begin{align}
    &\quad \text{tr}({\bm{G}_{(3)}^{(d)}}^T \bm{L}^{(d)} {\bm{G}_{(3)}^{(d)}})  =  \sum_{j_d=1}^{J_d} \sum_{k_d=1}^{J_d} \bm{L}_{j_d,k_d}^{(d)} {\bm{G}_{(3)}^{(d)}}_{j_d,:}^T {\bm{G}_{(3)}^{(d)}}_{k_d,:}\nonumber \\
    &= \sum_{r_d=1}^{R_d} \sum_{r_{d+1}=1}^{R_{d+1}} {\mathbm{G}_{r_d,r_{d+1},:}^{{(d)}^T}} \bm{L}^{(d)} {\mathbm{G}_{r_d,r_{d+1},:}^{(d)}},      \text{for } d = 1 \text{ to } D,
    \label{eqn:graphTTcore}
\end{align}
where the second line is due to \textit{Definition 2} and it states that graph Laplacian is applied to each fiber of the TT cores. The regularization (\ref{eqn:graphTTcore}) is depicted in the right-hand side of Fig. \ref{fig:graphonXd}. Fig. \ref{fig:graphonXd} provides an illustration for the proposed regularization, which indicates mode-$d$ fibers of $\bm{\mathcal{X}}$ (i.e., columns of $\bm{X}_{(d)}$) are linear combinations of columns from $\bm{G}_{(3)}^{(d)}$. Therefore the graph regularization (\ref{eqn:graphTTcore}) extends to all mode-$d$ fibers of $\bm{\mathcal{X}}$.

Based on the above analysis, if we want to introduce smoothness on rows and columns of a $3$rd order tensor $\bm{\mathcal{X}}$ (i.e., columns of $\bm{X}_{(1)}$ and $\bm{X}_{(2)}$), graph regularization can be applied on $\bm{G}_{(3)}^{(1)}$ and $\bm{G}_{(3)}^{(2)}$, respectively. On the other hand, if we want to model similarity among a $D$-th order dataset $\bm{\mathcal{X}}$, in which $\bm{\mathcal{X}}_{:,\ldots,:,i}$ is the $i$-th data sample, the graph regularization can be applied on the $\bm{G}_{(3)}^{(D)}$.
Since we try to solve the image completion problem under noise corruption, we will mainly focus on the first case. To find an appropriate weight matrix $\bm{A}$, it is usually assumed that pixels that are spatially close tend to be similar to each other. This leads to a commonly adopted weighting matrix $\bm{A}^{(d)}\in \mathbb{R}^{J_d \times J_d}$ with element $\bm{A}_{j_d,k_d}^{(d)} = \text{exp}(\alpha |j_d-k_d|^2)$, which models the correlation between the $j_d$-th row and $k_d$-th row of $\bm{X}_{(d)}$, and $\alpha$ is a manually chosen parameter.

On the other hand, it can be observed from Fig. \ref{fig:graphonXd} that graph regularization using (\ref{eqn:graphTTcore}) on the columns of $\bm{G}_{(3)}^{(d)}$ also introduces correlations among frontal slices of the TT core $\mathbm{G}^{(d)}$. Thus the update cannot be done in a slice-by-slice manner as in equation (\ref{eqn:lssolution_wrtTTcoreslice}). If we insist on updating one TT core as a unit in each iteration, the least squares (LS) solution will lead to inversion of a matrix of size $R_dR_{d+1}J_d\times R_dR_{d+1}J_d$, which takes time complexity $\mathcal{O}(J_d^3 R_d^3 R_{d+1}^3)$. Then both the storage and time complexity will be much larger than those required in (\ref{eqn:lssolution_wrtTTcoreslice}). This makes the algorithm extremely implementation-unfriendly. Therefore a new update strategy is required.

\section{Graph regularized TT completion with core fiber update}
\label{sec:graphTTopt}
As mentioned in the last section, we use (\ref{eqn:graphTTcore}) as the graph information to regularize the TT completion problem. This results in the following formulation
\begin{align}
     & \min_{\bm{\mathcal{G}}^{(1)},\bm{\mathcal{G}}^{(2)},\ldots  ,\bm{\mathcal{G}}^{(D)}} \left\| \bm{\mathcal{O}} \ast 
    \left( \bm{\mathcal{Y}} - \ll \bm{\mathcal{G}}^{(1)},\bm{\mathcal{G}}^{(2)},\ldots  ,\bm{\mathcal{G}}^{(D)} \gg - {\mathbm E} \right) \right\|_F^2 \nonumber \\
     & \quad \quad \quad + \sum_{d=1}^{D} \beta_d {\text{tr}(\bm{G}_{(3)}^{(d)}}^{T} \bm{L}^{(d)} {\bm{G}_{(3)}^{(d)}}) {+ \beta_{\mathbm E} \| \mathbm E\|_1} , \nonumber \\
     & \quad s.t.\text{ TT rank} = [1,R_2,...,R_D,1],
    \label{eqn:knownrank}
\end{align}
in which $\bm{L}^{(d)}$ and $\beta_d$ are the graph Laplacian and the regularization parameter for the $d$-th mode of the TT model, respectively. {In addition, to improve robustness towards potential outliers, we explicitly incorporate an $\ell_1$-penalty term---$\beta_{\mathbm E} \| \mathbm E\|_1$, with its effectiveness shown in prior works\cite{tibshirani1996regression,candes2011robust,goldfarb2014robust}.} If there is no graph information on a particular mode, then the corresponding $\bm{L}^{(d)}$ is set as $\bm{I}_{J_d}$, which regularizes the power of $\mathbm{G}^{(d)}$. Such a regularization is vital since the TT format is invariant by inserting a non-singular matrix among successive TT cores, i.e., $\ll \bm{\mathcal{G}}^{(1)},\ldots,\bm{\mathcal{G}}^{(d)},\bm{\mathcal{G}}^{(d+1)},\ldots,\bm{\mathcal{G}}^{(D)} \gg$ will be the same as 
$\ll \bm{\mathcal{G}}^{(1)},\ldots  ,\bar{\bm{\mathcal{G}}}^{(d)},\bar{\bm{\mathcal{G}}}^{(d+1)},\ldots,\bm{\mathcal{G}}^{(D)} \gg$ with 
$\bar{\bm{G}_{(2)}^{(d)}}^T = {\bm{G}_{(2)}^{(d)}}^T\bm{M}$ and $\bar{\bm{G}_{(1)}^{(d+1)}} = \bm{M}^{-1}{\bm{G}_{(1)}^{(d)}}$. If there is no regularization on a particular TT core, then the other TT cores can transfer their power to this TT core through the aforementioned process and thus all the regularization terms will be close to zero and (\ref{eqn:knownrank}) reduces to the traditional TT completion problem.

{Problem \eqref{eqn:knownrank} can be solved with a block coordinate descent (BCD) framework, which alternates between updating the unknown variables---the TT cores $\{\mathbm G^{(d)}\}_{d=1}^D$ and the outliers $\mathbm E$---until convergence.} For the TT cores, it is observed that the problem becomes quadratic if we focus on one particular TT core while fixing the others,
\begin{align}
    & \min_{\mathbm{G}^{(d)}} \left\| \bm{O}_{(d)} \ast \bigg(\bm{Y}_{(d)} {- \bm{E}_{(d)}} - \bm{G}_{(3)}^{(d)} \times (\bm{G}_{(1)}^{(>d)}\otimes \bm{G}_{(d)}^{(<d)}) \bigg) \right\|_F^2 \nonumber \\
    & \quad \quad \quad + \beta_d \text{tr}({\bm{G}_{(3)}^{(d)}}^{T} \bm{L}^{(d)} {\bm{G}_{(3)}^{(d)}})
\label{eqn:opt_wrtTTcore}
\end{align}
where various notations were introduced in Section \ref{subsec:pre_tt}. Different from the second line of (\ref{eqn:basicopt_wrtTTcore}) where the frontal slices $\{{\bm{G}_{(3)}^{(d)}}_{j_d,:}\}_{j_d=1}^{J_d}$ are independent of each other, the regularization introduces correlation between these slices as reflected in the first line of (\ref{eqn:graphTTcore}). As discussed in Section \ref{subsec:graphLap}, this would lead to a computationally expensive matrix inverse if we insist on the closed-form update of $\bm{G}_{(3)}^{(d)}$ based on (\ref{eqn:opt_wrtTTcore}).

To bypass this problem, we notice from the second line of (\ref{eqn:graphTTcore}) that the regularization on $\bm{\mathcal{G}}^{(d)}$ can be separated into $R_dR_{d+1}$ independent regularization terms, each of which regularizes a TT core fiber $\mathbm{G}_{r_d,r_{d+1},:}^{(d)}$ (equivalently ${\bm{G}_{(3)}^{(d)}}_{:,p}$ with $p = (r_{d+1}-1)R_d + r_d$) as a block of variables instead of a TT core. 
Putting (\ref{eqn:graphTTcore}) into (\ref{eqn:opt_wrtTTcore}), the problem becomes
\begin{align}
    & \min_{\mathbm{G}^{(d)}} \bigg\| \bm{O}_{(d)} \ast \bigg(\bm{Y}_{(d)} - \sum_{p=1}^{R_dR_{d+1}} {\bm{G}_{(3)}^{(d)}}_{:,p} \Big[ \bm{G}_{(1)}^{(>d)}\otimes \bm{G}_{(d)}^{(<d)} \Big]_{p,:}  \nonumber \\
    & \quad \quad \quad {- \bm{E}_{(d)}}\bigg) \bigg\|_F^2 + \beta_d \sum_{p=1}^{R_dR_{d+1}} {\bm{G}_{(3)}^{(d)}}_{:,p}^T \bm{L}^{(d)} {\bm{G}_{(3)}^{(d)}}_{:,p},
    \label{eqn:opt_wrtTTcore_fiberform}
\end{align}
Focusing on the terms that are only related to ${\bm{G}_{(3)}^{(d)}}_{:,p}$, (\ref{eqn:opt_wrtTTcore_fiberform}) simplifies to
\begin{align}
    & \min_{{\bm{G}_{(3)}^{(d)}}_{:,p}} \Bigg\| \bm{O}_{(d)} \ast \bigg(\bm{Y}_{(d)} - \sum_{q=1,q\neq p }^{R_dR_{d+1}} {\bm{G}_{(3)}^{(d)}}_{:,q} \Big[ \bm{G}_{(1)}^{(>d)}\otimes \bm{G}_{(d)}^{(<d)} \Big]_{q,:} \nonumber \\
    & \quad \quad \quad  {- \bm{E}_{(d)}} \bigg) - \bm{O}_{(d)}\ast \big({\bm{G}_{(3)}^{(d)}}_{:,p} \Big[ \bm{G}_{(1)}^{(>d)}\otimes \bm{G}_{(d)}^{(<d)} \Big]_{p,:}\big)  \Bigg\|_F^2 \nonumber \\
    & \quad \quad \quad + \beta_d {\bm{G}_{(3)}^{(d)}}_{:,p}^T \bm{L}^{(d)} {\bm{G}_{(3)}^{(d)}}_{:,p},
    \label{eqn:opt_wrtTTcorefiber}
\end{align}
which is quadratic with respect to the TT core fiber ${\bm{G}_{(3)}^{(d)}}_{:,p}$, and the closed-form solution is shown in Appendix A to be
\begin{align}
    {\bm{G}_{(3)}^{(d)}}_{:,p} =  {\bm{\Upsilon}^{(d,p)}}^{-1}{\bm{\mu}^{(d,p)}},
    \label{eqn:opt_solfiber}
\end{align}
with
\begin{align}
    & \bm{\Upsilon}^{(d,p)} =  \text{diag}\bigg( \bm{O}_{(d)} (\Big[ \bm{G}_{(1)}^{(>d)}\otimes \bm{G}_{(d)}^{(<d)} \Big]_{p,:}^T \nonumber \\
    & \quad \quad \quad \quad \quad \ast \Big[ \bm{G}_{(1)}^{(>d)}\otimes \bm{G}_{(d)}^{(<d)} \Big]_{p,:}^T) \bigg) + \beta_d \bm{L}^{(d)},
\end{align}
\begin{align}
    & \bm{\mu}^{(d,p)} =\bigg(\bm{O}_{(d)} \ast \Big(\bm{Y}_{(d)} {- \bm{E}_{(d)}}- \sum_{q=1,q\neq p }^{R_dR_{d+1}} {\bm{G}_{(3)}^{(d)}}_{:,q} \nonumber \\
    & \quad \quad \times \Big[ \bm{G}_{(1)}^{(>d)}\otimes \bm{G}_{(d)}^{(<d)} \Big]_{q,:} \Big)\bigg) \Big[ \bm{G}_{(1)}^{(>d)}\otimes \bm{G}_{(d)}^{(<d)} \Big]_{p,:}^T.
\end{align}

{The update of $\mathbm E$ relegates to solving the following problem
\begin{align} \label{eq:outlier_problem_opt}
    \min_{\mathbm E} \| \mathbm O \ast \big(\mathbm Y - \ll \bm{\mathcal{G}}^{(1)},\bm{\mathcal{G}}^{(2)},\ldots  ,\bm{\mathcal{G}}^{(D)} \gg - \mathbm E \big)\|_F^2 + \beta_{\mathbm E} \| \mathbm E\|_1,
\end{align}
which admits a closed-form solution via the element-wise soft-thresholding operator \cite{tibshirani1996regression,candes2011robust},
\begin{align} \label{eq:outlier_update_opt}
    \mathbm E = \rm{soft}_{\beta_{\mathbm E}/2} \big(\mathbm O \ast \big(\mathbm Y - \ll \bm{\mathcal{G}}^{(1)},\bm{\mathcal{G}}^{(2)},\ldots  ,\bm{\mathcal{G}}^{(D)} \gg \big) ,
\end{align}
where $\mathrm{soft}_{\beta}(\bm{X})$ applies soft-thresholding element-wisely as:
\begin{align}
       {\rm soft}_{\beta} (x) \triangleq {\rm sign} (x) {\rm max}(|x|-\beta,0),
\end{align}
with $\mathrm{sign}(x)$ denoting the sign of $x$ and $\max(\cdot, \cdot)$ returning the larger of the two.
}

{The whole algorithm is outlined in Algorithm \ref{alg:graphTTC_opt}. It follows a proximal BCD framework to solve problem \ref{eqn:knownrank}, which alternates between updates of the TT core fibers and the outliers. For each TT core, the basic LS problem in (\ref{eqn:opt_wrtTTcorefiber}) is solved from $p=1$ to $R_d R_{d+1}$. Then various TT cores are updated from $d=1$ to $D$ at the outer iteration. For the outliers, the update is obtained using \eqref{eq:outlier_update_opt}---the proximal operator associated with problem \eqref{eq:outlier_problem_opt}.}

{Algorithm \ref{alg:graphTTC_opt} is guaranteed to converge to a stationary point of problem (\ref{eqn:knownrank}) \cite{xu2013block,wen2019nonconvex,bolte2014proximal}, as each update is feasible and achieves the global optimum of its respective subproblem\footnote{Due to the positive semidefinite Laplacian matrices \cite{deng2010graph}, \eqref{eqn:opt_solfiber} always exist.}.}
Notice that even if we choose to update a whole TT core as a block of variables, the corresponding algorithm is still under the proximal BCD framework, and the convergent point is still a stationary point of (\ref{eqn:knownrank}). In this sense, while updating each TT core and updating each TT core fiber in the BCD lead to different solutions, they achieve the same quality of convergent point. Experiments on synthetic data will be provided to compare the two updating mechanisms in Section \ref{subsec:synthetic}, which show the proposed fiber update using (\ref{eqn:opt_wrtTTcorefiber}) performs similarly to the core update based on (\ref{eqn:opt_wrtTTcore}).

For a fiber from the $d$-th TT core, it takes around $\bm{O}({J_d}^3)$ to compute the solution to (\ref{eqn:opt_wrtTTcorefiber}). Apart from that, it takes around $\bm{O}(R^4\prod_{k=1}^{D}  J_k)$ to obtain $\bm{\Upsilon}^{(d,p)}$ and $\bm{\mu}^{(d)}$. Since there are common terms for different fibers in the same TT core, in general, it takes complexity $\bm{O}(R^2 {J_d}^3 + R^4\prod_{k=1}^{D} J_k)$ for the update of each TT core.
In contrast, if we update one TT core as a whole, it would take $\bm{O}(R^6J_d^3 + R^4\prod_{k=1}^{D} J_k)$ to compute the closed-form solution, which is much more complicated. Furthermore, the storage complexity required for the matrix to be inverted is $\bm{O}(R^4 J_d^2) $ for the core update, while that of the proposed algorithm is only $\bm{O}(J_d^2)$. Suppose we use $64$-bit double type data, with $J_d = 256$ and $R_d=R_{d+1}=16$, then the amount of RAM required for the matrix in the core update is $32$GB, while the proposed fiber update only requires $512$KB.

\begin{algorithm}[!tb]
\SetAlgoLined
 \textbf{initialization:} Input the observed tensor $\bm{\mathcal{Y}}$ and indicator tensor $\mathbm{O}$. Set TT ranks $\{R_d\}_{d=1}^{D+1}$, the graph Laplacian $\{\bm{L}^{(d)}\}_{d=1}^D$, and regularization parameters $\{\bm{\beta}_{d}\}_{d=1}^{D}$ and {$\beta_{\mathbm{E}}$};
 
 \While{Not Converged}{
    \textbf{For} $d=1$ \textbf{to} $D-1$
    
    \quad \textbf{For} $p=1$ \textbf{to} $R_dR_{d+1}$
    
    \quad \quad Update ${\bm{G}_{(3)}^{(d)}}_{:,p}$ according to (\ref{eqn:opt_solfiber});
    
    \quad \textbf{end}
    
    \textbf{end}

    {Update $\mathbm E$ according to \eqref{eq:outlier_update_opt};}
 }
 \caption{TT completion with graph regularization (GraphTT-opt).}
 \label{alg:graphTTC_opt}
\end{algorithm}

\section{A Bayesian treatment to graph TTC}
\label{sec:bayesiantt}
In the last section, Algorithm \ref{alg:graphTTC_opt} is provided to solve the graph regularized TT completion problem. However, a critical drawback of Algorithm \ref{alg:graphTTC_opt} is that it heavily relies on parameter tuning, like most optimization-based methods do. 
For TT completion with graph regularization, the burden of parameter tuning is even heavier than that of traditional matrix completion or canonical polyadic (CP) completion, since the TT model has multiple TT ranks, and for each TT core there is an individual regularization parameter for the graph information. 
For example, for a $4$-th order tensor $\mathbm{Y}$, there are three TT-ranks, four graph regularization parameters and one outlier-related regularization parameter to be tuned.

To solve this problem, the probabilistic model, which has shown its ability to perform matrix/tensor completion without the need of parameter tuning \cite{twofourcheng2017probabilistic,cheng2020learning,twofivezhao2015bayesian,xu2021overfitting,babacan2014bayesian}, is adopted in this section.

\subsection{The generalized hyperbolic model for TT with graph information embedding}
Firstly, from (\ref{eqn:observed}), due to the additive white Gaussian noise, the log-likelihood of the observed tensor $\mathbm{Y}$ is
\begin{align}
    & \ln{\left(p(\bm{\mathcal{Y}}|\bm{\mathcal{O}},\{\bm{\mathcal{G}}^{(d)}\}_{d=1}^{D},{\mathbm E,}\tau)\right)}  = \frac{|\Omega|}{2}\ln{\tau}-\frac{\tau}{2} \Big\|\bm{\mathcal{O}} \ast \big(\bm{ \mathcal{Y}} \nonumber \\
    & \quad \quad - \ll \bm{\mathcal{G}}^{(1)},\bm{\mathcal{G}}^{(2)},\ldots  ,\bm{\mathcal{G}}^{(D)} \gg {- \mathbm E} \big) \Big\|_F^2 + \text{const},
\label{eqn:ttlikelihood}
\end{align}
where $\tau$ is the inverse of the noise variance, $\Omega$ denotes the set of indices of the observed entries, and $|\Omega|$ denotes the cardinality of $\Omega$, which equals the number of observed entries. For the noise precision $\tau$, it is assumed to follow a Gamma distribution as 
\begin{align}
    p( \tau | {\alpha}_{\tau},{\beta}_{\tau} ) = \text{Gamma}(\tau|{\alpha}_{\tau},{\beta}_{\tau}).
    \label{eqn:tauprior}    
\end{align}

To enable estimation of the TT-ranks during inference, a sparsity-inducing prior distribution with graph information is adopted for each TT core
\begin{align}
    & \quad p( \bm{ \mathcal{G} }^{(d)} | \bm{ z }^{(d)},\bm{ z }^{(d+1)} ) = \nonumber \\
    & \prod_{k=1}^{S_d}\prod_{\ell=1}^{S_{d+1}}\mathcal{N}\Big(\bm{\mathcal{G}}_{k,\ell,:}^{(d)} | \bm{0}, \bm{z}_{k}^{(d)}\bm{z}_{\ell}^{(d+1)} {\bm{L}^{(d)}}^{-1}\Big), \forall d \in \{1,\ldots,D\},
\label{eqn:Gcoreprior}
\end{align}
\begin{align}
    &p( \bm{ z }^{(d)} | \bm{a}^{(d)},\bm{b}^{(d)}, \bm{\lambda}^{(d)} ) = \prod_{k=1}^{S_d}\text{GIG}(\bm{z}_{k}^{(d)} | \bm{a}_{k}^{(d)},\bm{b}_{k}^{(d)},\bm{\lambda}_{k}^{(d)}), \nonumber \\
    & \quad \quad \quad \quad \forall d \{2,\ldots,D\},
    \label{eqn:zprior}
\end{align}
where $S_d$ and $S_{d+1}$ are upper bound of $R_d$ and $R_{d+1}$ respectively \cite{oseledets2012solution,holtz2012manifolds}, which are set as large numbers in practice so that learning the TT ranks in the inference algorithm is possible. $\bm{z}^{(d)}=[\bm{z}_1^{(d)},\ldots,\bm{z}_{S_{d}}^{(d)}]$ controls the variance of all mode-$3$ fibers in both $\mathcal{G}^{(d)}$ and $\mathcal{G}^{(d+1)}$. In particular, $\bm{z}^{(1)}$ and $\bm{z}^{(D+1)}$ are scalars and set as $1$ so that the expression in (\ref{eqn:Gcoreprior}) is applicable for the first and last TT cores. As in (\ref{eqn:zprior}), each element of $\bm{z}^{(d)}$ is modeled to follow a generalized inverse Gaussian (GIG) distribution, which is controlled by the hyperparameters $\bm{a}^{(d)},\bm{b}^{(d)},\bm{\lambda}^{(d)}$ and is defined as
\begin{align}
    &\text{GIG}(\bm{z}_{k}^{(d)} | \bm{a}_{k}^{(d)},\bm{b}_{k}^{(d)},\bm{\lambda}_{k}^{(d)}) = \frac{  (\frac{\bm{a}_{k}^{(d)}}{\bm{b}_{k}^{(d)}})^{\frac{\bm{\lambda}_k^{(d)}}{2}}  }{  2 K_{\bm{\lambda}_{k}^{(d)}}(\sqrt{\bm{a}_{k}^{(d)}\bm{b}_{k}^{(d)}}) } {\bm{z}_{k}^{(d)}}^{\bm{\lambda}_{k}^{(d)}-1} \nonumber \\
    &\quad \quad \quad \times  \text{exp}\Bigg(-\frac{1}{2}(\bm{a}_{k}^{(d)}\bm{z}_{k}^{(d)} + \bm{b}_{k}^{(d)} \frac{1}{\bm{z}_{k}^{(d)}})\Bigg),
    \label{eqn:gigdiscribe}
\end{align}
where $K_.(.)$ is the modified Bessel function of the second kind. For $\bm{a}^{(d)}$ which dominantly affects the distribution of $\bm{z}^{(d)}$, it is further assigned a Gamma distribution as
\begin{align}
    p( \bm{a}^{(d)} | \bm{c}^{(d)},\bm{f}^{(d)} ) = \text{Gamma}(\bm{c}^{(d)},\bm{f}^{(d)}).
\label{eqn:aprior}    
\end{align}

Checking the marginal distribution of the TT cores under (\ref{eqn:Gcoreprior})-(\ref{eqn:gigdiscribe}), the following proposition is obtained. The proof can be found in Appendix B.

\noindent \textbf{Proposition 1.} When $\bm{a}_\ell^{(d+1)}$, $\bm{b}_\ell^{(d+1)}$ and $\bm{\lambda}_\ell^{(d+1)}$ all tend to zero for all $\ell$, then the marginal distribution of $\mathbm{G}_{k,:,:}^{(d)}$ and $\mathbm{G}_{m,:,:}^{(d+1)}$ follows
\begin{align}
    p(\mathbm{G}_{k,:,:}^{(d)}) \propto \prod_{\ell=1}^{S_{d+1}} ({\mathbm{G}_{k,\ell,:}^{(d)}}^T \bm{L}^{(d)} {\mathbm{G}_{k,\ell,:}^{(d)}})^{-J_d},\nonumber \\
    p(\mathbm{G}_{:,m,:}^{(d)}) \propto \prod_{\ell=1}^{S_{d+1}} ({\mathbm{G}_{\ell,m,:}^{(d+1)}}^T \bm{L}^{(d)} {\mathbm{G}_{\ell,m,:}^{(d+1)}})^{-J_d},
    \label{eqn:apx-proposition1}
\end{align}
respectively, for all $k$ and $m$.

\begin{figure}[!tb]
    \centering
    \begin{subfigure}{0.48\linewidth}
        \centering
        \captionsetup{justification=centering}
        \includegraphics[width=0.98\linewidth]{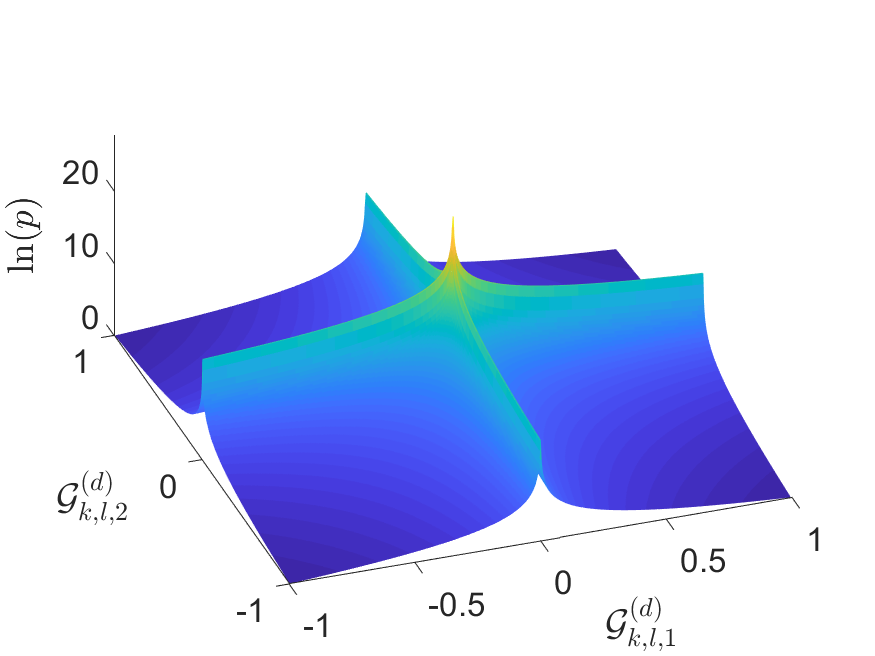}
        \subcaption{$\bm{L}^{(d)}$ is an identity matrix}
        \label{subfig:marginG_identity}
    \end{subfigure}
    \begin{subfigure}{0.48\linewidth}
        \centering
        \captionsetup{justification=centering}
        \includegraphics[width=0.98\linewidth]{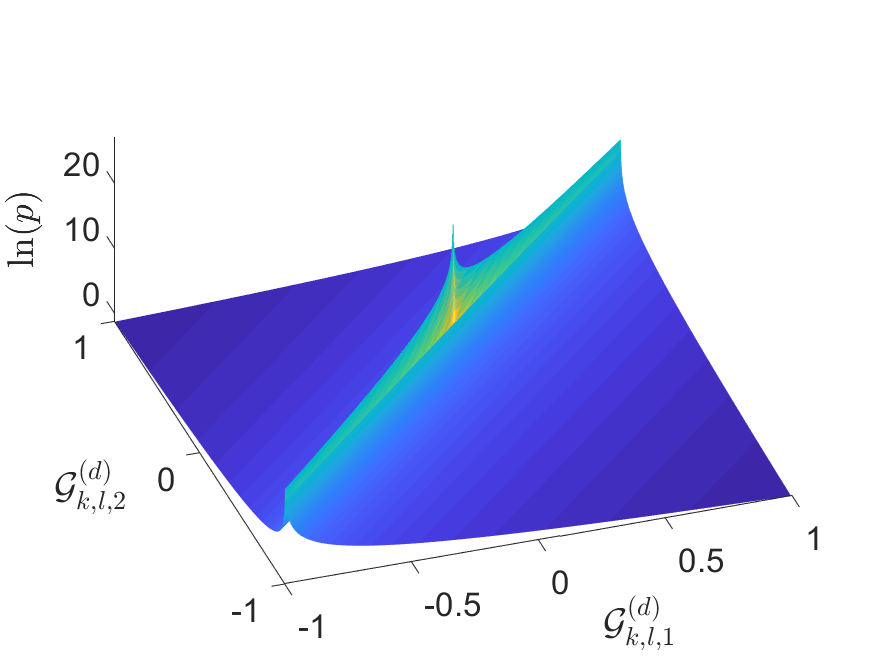}
        \subcaption{$\bm{L}^{(d)} = [1,-1;-1,1]$}
        \label{subfig:marginG_graph}
    \end{subfigure}
    \caption{Demonstration of the marginal distribution of a mode-$3$ fiber, with hyperparameters tending to $0$.}
    \label{fig:apx-jointdistribution}
\end{figure}

Proposition 1 shows the sparsity-promoting property of the proposed model. Specifically, the marginal distribution of the TT core slices will concentrate most of the probabilistic density around $0$, which indicates the initial belief that the underlying TT structure is sparse. In addition, it also has heavy tails, which allows to learn important components from the observation. An illustration of Proposition 1 is in Fig. \ref{fig:apx-jointdistribution}, using an example of a TT core fiber $\mathbm{G}_{k,\ell,:}^{(d)} \in \mathbb{R}^2$. Fig. \ref{subfig:marginG_identity} shows a traditional probabilistic TT model \cite{xu2021overfitting}, in which $\bm{L}^{(d)} = \bm{I}_2$. Different from \cite{xu2021overfitting}, graph information is incorporated in the proposed model, which makes the elements in the mode-$3$ fibers of a TT core correlated and is vividly shown in Fig. \ref{subfig:marginG_graph}. This sparsity-promoting property enables automatic identification of TT ranks, while the correlation among elements helps in missing data recovery.

{
To model the outliers $\mathbm E$, which have only a few non-zero entries, we adopt a Student's $t$-distribution---a sparsity-promoting prior that has been shown effective for outlier modeling \cite{twofourcheng2017probabilistic}. Specifically, $\mathbm E$ is modeled as
\begin{align}
    p(\mathbm E) = 
    &\prod_{j_1=1}^{J_1}\ldots\prod_{j_D=1}^{J_D} \int ~ \mathcal{N}( \mathbm E_{j_1\ldots j_D} | 0, \mathbm U_{j_1\ldots j_D}^{-1}) \nonumber \\
    \times & \text{Gamma}(\mathbm U_{j_1\ldots j_D} | \mathbm{P}_{j_1\ldots j_D},\mathbm Q_{j_1\ldots j_D}) ~ d_{\mathbm U_{j_1\ldots j_D}},
    \label{eqn:outlierprior}
\end{align}
where the Student's {$t$}-distribution is equivalently represented as a Gaussian scale mixture \cite{west1987scale}.
}

\subsection{Inference algorithm}
Given the probabilistic model (\ref{eqn:tauprior})-(\ref{eqn:aprior}), we need to infer the unknown variables $\bm{\Theta}:=\big\{\{\bm{\mathcal{G}}^{(d)}\}_{d=1}^{D},\{\bm{z}^{(d)}\}_{d=2}^{D},\{\bm{a}^{(d)}\}_{d=2}^{D},$  ${\mathbm E,\mathbm U,}\tau \big\}$ based on the observation $\mathbm{Y}$. In Bayesian inference, this is achieved through the posterior distribution $p(\bm{\Theta}|\bm{\mathcal{Y}})=p(\bm{\mathcal{Y}},\bm{\Theta})/p(\bm{\mathcal{Y}})$. However, the Bayesian graph-regularized TT model is so complicated that there is no closed-form expression for $p(\bm{\mathcal{Y}})=\int p(\bm{\mathcal{Y}},\bm{\Theta})d\bm{\Theta}$, and therefore the posterior distribution $p(\bm{\Theta}|\bm{\mathcal{Y}})$ cannot be computed explicitly. Therefore, instead of directly deriving the posterior distribution, variational inference (VI) is adopted, which tries to find a variational distribution $q(\bm{\Theta})$ that best approximates the posterior distribution by minimizing the Kullback-Leibler (KL) divergence
\begin{align}
\min_{q(\bm{\Theta})} \text{ KL}\bigg(q(\bm{\Theta})\text{ }||\text{ }p(\bm{\Theta}|\bm{\mathcal{Y}})\bigg) = \int q(\bm{\Theta}) \ln \frac{q(\bm{\Theta})}{p(\bm{\Theta}|\bm{\mathcal{Y}})} \text{d} \bm{\Theta}.
\label{eqn:KLdivergence}
\end{align}

To solve (\ref{eqn:KLdivergence}), the mean-field approximation is commonly adopted, which assumes that $q(\bm{\Theta})=\prod_{s=1}^Sq(\bm{\Theta}_s)$, where $\{\bm{\Theta}_s\}_{s=1}^{S}$ are non-overlapping partitioning of $\bm{\Theta}$ (i.e. $\bm{\Theta}_s \subset \bm{\Theta}$, $\cup_{s=1}^{S}\bm{\Theta}_s = \bm{\Theta}$, and $\bm{\Theta}_s \cap \bm{\Theta}_t = \emptyset$ for $s\neq t$). Under the mean-field approximation, the optimal solution for $\bm{\Theta}_s$ (when other variables are fixed) can be derived as \cite[pp.~737]{murphy2012probabilistic}
\begin{align}
\ln{q^*({\bm{\Theta}}_s)} = \mathbb{E}_{\bm{\Theta}\backslash{\bm{\Theta}}_s}\llbracket\ln{p(\bm{\mathcal{Y}},\bm{\Theta})}\rrbracket+\text{const},
\label{eqn:viupdate}
\end{align}
where $\mathbb{E}_{\bm{\Theta}\backslash{\bm{\Theta}}_s}\llbracket.\rrbracket$ denotes the expectation over all variables expect $\bm{\Theta}_s$. For the proposed TT model, we employ the mean-field
\begin{align}
q(\bm{\Theta}) = & \prod_{d=1}^{D}\prod_{k=1}^{S_d}\prod_{\ell=1}^{S_{d+1}} q(\mathbm{G}_{k,\ell,:}^{(d)}|\bm z^{(d),\bm z^{(d+1)}}) \prod_{d=2}^{D}q(\bm{z}^{(d)})\prod_{d=2}^{D}q(\bm{a}^{(d)}) \nonumber \\
& \times q(\mathbm E| \mathbm U) q(\mathbm U) q(\tau).
\label{eqn:meanfieldproposed}
\end{align}
Under this mean-field approximation, the optimal variational distributions of different variables are derived using (\ref{eqn:viupdate}) with $p(\mathbm Y, \bm \Theta) = p(\mathbm Y | \bm \Theta)~ p(\bm \Theta)$.
The detailed derivations are given in Appendix C and the results are presented below.

\noindent \underline{\textbf{Update $\bm{{G}}_{:,p}^{(d)}$} for $p$ from $1$ to $S_dS_{d+1}$, $d$ from $1$ to $D$}

For each fiber of $\bm{\mathcal{G}}^{(d)}$, its variational distribution follows a Gaussian distribution
\begin{align*}
q(\bm{G}_{:,p}^{(d)}) = \mathcal{N}(\bm{\nu}^{(d,p)},\bm{\Sigma}^{(d,p)}),
\end{align*}
with
\begin{align}
    & \bm{\Sigma}^{(d,p)} = \Bigg(\mathbb{E}{\big\llbracket \tau\big\rrbracket}\text{diag}\bigg( \bm{O}_{(d)} \mathbb{E}\bigg\llbracket \Big[ \bm{G}_{(1)}^{(>d)}\otimes \bm{G}_{(d)}^{(<d)} \Big]_{p,:}^T \nonumber \\
    & \ast \Big[ \bm{G}_{(1)}^{(>d)}\otimes \bm{G}_{(d)}^{(<d)} \Big]_{p,:}^T \bigg\rrbracket  \bigg) +  \mathbb{E}\bigg\llbracket\frac{1}{\bm{z}_{k_p}^{(d)}}\bigg\rrbracket \mathbb{E}\bigg\llbracket\frac{1}{\bm{z}_{\ell_p}^{(d+1)}}\bigg\rrbracket\bm{L}^{(d)} \Bigg)^{-1},
    \label{eqn:GcoreVar}
\end{align}
\begin{align}
    & \bm{\nu}^{(d,p)} =\mathbb{E} {\big\llbracket \tau\big\rrbracket} \bm{\Sigma}^{(d,p)}\bigg( \Big(\bm{O}_{(d)} \ast (\bm{Y}_{(d)} {- \bm{E}_{(d)}})\Big)  \nonumber \\
    & \times \mathbb{E}\bigg\llbracket \Big[ \bm{G}_{(1)}^{(>d)}\otimes \bm{G}_{(d)}^{(<d)} \Big]_{p,:}^T\bigg\rrbracket - \sum_{q=1,q\neq p}^{S_dS_{d+1}} \text{diag}\Big( \mathbb{E}\Big\llbracket {\bm{G}_{(3)}^{(d)}}_{:,q} \Big\rrbracket\Big) \bm{O}_{(d)} 
    \nonumber \\
    & \times \mathbb{E}\bigg \llbracket \Big[ \bm{G}_{(1)}^{(>d)}\otimes \bm{G}_{(d)}^{(<d)} \Big]_{q,:}^T \ast \Big[ \bm{G}_{(1)}^{(>d)}\otimes \bm{G}_{(d)}^{(<d)} \Big]_{p,:}^T\bigg \rrbracket\bigg).
    \label{eqn:Gcoremean}
\end{align}
Most of the expectations in (\ref{eqn:GcoreVar}) and (\ref{eqn:Gcoremean}) are trivial, except $\mathbb{E} \llbracket [ \bm{G}_{(1)}^{(>d)}\otimes \bm{G}_{(d)}^{(<d)} ]_{q,:}^T \ast [ \bm{G}_{(1)}^{(>d)}\otimes \bm{G}_{(d)}^{(<d)} ]_{p,:}^T\rrbracket$, which is discussed in detail in Appendix C.

\noindent \underline{\textbf{Update $\bm{a}^{(d)}$ from $d=2$ to $D$}}

For $\bm{a}^{(d)}$, it follows a Gamma distribution
\begin{align*}
    p( \bm{ a }^{(d)} |  \hat{\bm{c}}_k^{(d)},\hat{\bm{f}}_k^{(d)} ) = \prod_{k=1}^{S_d}\text{Gamma}(\bm{a}_{k}^{(d)} | \hat{\bm{c}}_k^{(d)},\hat{\bm{f}}_k^{(d)}), 
\end{align*}
with
\begin{align}
    \hat{\bm{c}}_k^{(d)} = \bm{c}_k^{(d)} + \frac{\hat{\bm{\lambda}}_{k}^{(d)}}{2},
    \label{eqn:cupdate}
\end{align}
\begin{align}
    \hat{\bm{f}}_k^{(d)} = \bm{f}_k^{(d)} + \frac{\mathbb{E}[\bm{z}_k^{(d)}]}{2}.
    \label{eqn:fupdate}
\end{align}

\begin{algorithm}[!tb]
\SetAlgoLined
 \textbf{initialization:} Input the observed tensor $\bm{\mathcal{Y}}$. Set initial ranks $\{S_d\}_{d=1}^{D}$ and hyperparameters $\{\bm{\lambda}^{(d)}\}_{d=2}^{D}$, $\{\bm{b}^{(d)}\}_{d=2}^{D}$, $\{\bm{c}^{(d)}\}_{d=2}^{D}$, $\{\bm{f}^{(d)}\}_{d=2}^{D}$, $\alpha_\tau$, $\beta_\tau$;
 \While{Not Converged}{
    Update $q(\bm{G}_{:,p}^{(d)})$ via (\ref{eqn:GcoreVar}) and (\ref{eqn:Gcoremean}) sequentially for $p=1,\ldots,S_dS_{d+1}$ and $d=1,\ldots,D$\;
    Update $\{q(\bm{a}^{(d)})\}_{d=2}^{D}$ via (\ref{eqn:cupdate}),
    (\ref{eqn:fupdate}) sequentially for $d=2,\dots,D$\;
    Update $\{q(\bm{z}^{(d)})\}_{d=2}^{D}$ via (\ref{eqn:aforzupdate}), (\ref{eqn:lambdaupdate}) and (\ref{eqn:bupdate}) sequentially for $d=2,\dots,D$\;
    {Update $q(\mathbm E)$ via \eqref{eq:outlier_variance_VI} and \eqref{eq:outlier_mean_VI}\;
    Update $q(\mathbm U)$ via \eqref{eq:outlier_gamma_P} and \eqref{eq:outlier_gamma_Q}\;}
    Update $q(\tau)$ via (\ref{eqn:taualpha}) and (\ref{eqn:taubeta})\;
    Rank selection\;
 }
 \caption{VI Algorithm for the probabilistic graph regularized TT model (GraphTT-VI).}
 \label{alg:graphTTC_vi}
\end{algorithm}

\noindent \underline{\textbf{Update $\bm{z}^{(d)}$ from $d=2$ to $D$}}

The variational distribution of $\bm{z}^{(d)}$ follows a GIG distribution
\begin{align*}
    p( \bm{ z }^{(d)} | \hat{\bm{a}}^{(d)},\hat{\bm{b}}^{(d)},\hat{\bm{\lambda}}^{(d)} ) = \prod_{k=1}^{S_d}\text{GIG}(\bm{z}_{k}^{(d)} | \hat{\bm{a}}^{(d)},\hat{\bm{b}}^{(d)},\hat{\bm{\lambda}}^{(d)}),
\end{align*}
with the parameters updated as
\begin{align}
    \hat{\bm{a}}_k^{(d)} = \mathbb{E}\big\llbracket{\bm{a}_k^{(d)}}\big\rrbracket,
    \label{eqn:aforzupdate}
\end{align}
\begin{align}
    \hat{\bm{\lambda}}_{k}^{(d)} = \bm{\lambda}_{k}^{(d)} - \frac{J_d S_{d+1}}{2} - \frac{J_{d-1} S_{d-1}}{2},
    \label{eqn:lambdaupdate}
\end{align}
\begin{align}
    & \hat{\bm{b}}_k^{(d)} = \bm{b}_k^{(d)} + \frac{1}{2}\sum_{\ell=1}^{S_{d-1}} \mathbb{E}\big\llbracket\frac{1}{\bm{z}_\ell^{(d-1)}}\big\rrbracket \mathbb{E}\big\llbracket{\mathbm{G}_{\ell,k,:}^{(d-1)}}^T \bm{L}^{(d-1)} {\mathbm{G}_{\ell,k,:}^{(d-1)}}\big\rrbracket \nonumber \\
    & + \sum_{\ell=1}^{S_{d+1}} \mathbb{E}\big\llbracket\frac{1}{\bm{z}_\ell^{(d+1)}}\big\rrbracket\mathbb{E}\llbracket{\mathbm{G}_{\ell,k,:}^{(d)}}^T \bm{L}^{(d)} {\mathbm{G}_{\ell,k,:}^{(d)}}\rrbracket.
    \label{eqn:bupdate}
\end{align}

\begin{table*}[!tb]
\caption{Calculation of Expectations.}
\centering
\scalebox{1}{
\begin{tabular}{| c | c  |}
 \hline
 Expectations &  Calculation  \\
\hline
 $\mathbb{E}\llbracket \mathbm{G}_{k,\ell,:}^{(d)}\rrbracket, \forall k,\ell,d$  & $\bm{\nu}^{(d,(\ell-1)S_d+k)}, \forall k,\ell,d$ \\
\hline
$\mathbb{E}\llbracket{\mathbm{G}_{:,:,j_d}^{(d)} \otimes \mathbm{G}_{:,:,j_d}^{(d)}}\rrbracket, \forall d,j_d$ &
\begin{tabular}{@{}c@{}}$\bm{\mathrm{Var}}^{(d,j_d)} + \mathbb{E}\big \llbracket \mathbm{G}_{:,:,j_d}^{(d)}\big \rrbracket \otimes \mathbb{E}\big \llbracket \mathbm{G}_{:,:,j_d}^{(d)}\big \rrbracket$, with\\ 
\tiny{\quad}\\
    \small{
    $\quad \quad \bm{\mathrm{Var}}_{i,t}^{(d,j_d)} =
    \begin{cases}
        \bm{\Sigma}_{j_d,j_d}^{(d,(\ell-1)S_d + k)},& \text{if } i \in \{(k-1)S_d+k\}_{k=1}^{S_d}\\
        \quad &  \& \quad t \in \{(\ell-1)S_{d+1}+\ell\}_{\ell=1}^{S_{d+1}}\\
        0,              & \text{otherwise}
    \end{cases}$} \\
\end{tabular}
 \\
\hline
 \normalsize$\mathbb{E}\llbracket \bm{z}_{k}^{(d)}\rrbracket, \forall k,d$ & 
 $\left(\frac{\hat{\bm{b}}_k^{(d)}}{\hat{\bm{a}}_k^{(d)}}\right)^{\frac{1}{2}}\frac{K_{\hat{\bm{\lambda}}_k^{(d)}+1}\left(\sqrt{\hat{\bm{a}}_k^{(d)}\hat{\bm{b}}_k^{(d)}}\right)}{K_{\hat{\bm{\lambda}}_k^{(d)}}\left(\hat{\bm{a}}_k^{(d)}\hat{\bm{b}}_k^{(d)}\right)}$ \\
\hline
 $\mathbb{E}\big\llbracket \frac{1}{\bm{z}_{k}^{(d)}}\big\rrbracket,\forall k,d$ & 
 $\left(\frac{\hat{\bm{b}}_k^{(d)}}{\hat{\bm{a}}_k^{(d)}}\right)^{-\frac{1}{2}}\frac{K_{\hat{\bm{\lambda}}_k^{(d)}+1}\left(\sqrt{\hat{\bm{a}}_k^{(d)}\hat{\bm{b}}_k^{(d)}}\right)}{K_{\hat{\bm{\lambda}}_k^{(d)}-1}\left(\hat{\bm{a}}_k^{(d)}\hat{\bm{b}}_k^{(d)}\right)}$\\
\hline
 $\mathbb{E}\llbracket \bm{a}_{k}^{(d)} \rrbracket, \forall k,d$ &
 ${\hat{\bm{c}}_k^{(d)}}/{\hat{\bm{f}}_k^{(d)}}$\\
\hline
$\mathbb{E}\llbracket  \mathbm E_{j_1\ldots j_D}^2 \rrbracket$ & $\mathbm V_{j_1\ldots j_D} +\mathbm M_{j_1\ldots j_D}^2$\\
\hline 
$\mathbb E \llbracket \mathbm U_{j_1\ldots j_D} \rrbracket$ & $\hat{\mathbm P}_{j_1\ldots j_D} / \hat{\mathbm Q}{j_1\ldots j_D}$\\
\hline
 $\mathbb{E}\llbracket \tau \rrbracket$ &
 ${\hat{\alpha}_\tau}/{\hat{\beta}_\tau}$\\
\hline
\end{tabular}}
\label{tab:expectation_cal}
\end{table*}

\noindent \underline{\textbf{Update $\mathbm E$} and $\mathbm U$}

The variational distribution of $\mathbm E$ follows a Gaussian distribution
\begin{align}
    q(\mathbm E) = \prod_{j_1=1}^{J_1} \ldots \prod_{j_D=1}^{J_D} \mathcal N(\mathbm E_{j_1\ldots j_D} | \mathbm M_{j_1\ldots j_D}, \mathbm V_{j_1\ldots j_D}), \nonumber
\end{align}
where the posterior variance and mean are given by
\begin{align} \label{eq:outlier_variance_VI}
    \mathbm V_{j_1\ldots j_D} = ( \mathbb E \llbracket \tau \rrbracket \mathbm O_{j_1\ldots j_D} + \mathbb E\llbracket \mathbm U_{j_1\ldots j_D} \rrbracket)^{-1},
\end{align}
\begin{align}\label{eq:outlier_mean_VI}
    &\mathbm M_{j_1 \ldots j_D} = \mathbb E\llbracket \tau \rrbracket \mathbm O_{j_1\ldots j_D}\mathbm V_{j_1\ldots j_D}\nonumber \\
    &   \times (\mathbm Y_{j_1\ldots j_D} - \mathbb E \llbracket \mathbm G_{:,:,j_1}^{(1)}\rrbracket\ldots \mathbb E \llbracket \mathbm G_{:,:,j_D}^{(D)}\rrbracket).
\end{align}

In addition, the latent Gamma variable $\mathbm U$ retains a Gamma distribution
\begin{align}
    q(\mathbm U) = \prod_{j_1=1}^{J_1} \ldots \prod_{j_D=1}^{J_D} \text{Gamma}(\mathbm U_{j_1\ldots j_D} | \hat{ \mathbm P }_{j_1\ldots j_D}, \hat{ \mathbm Q }_{j_1\ldots j_D}), \nonumber
\end{align}
with updated parameters
\begin{align}
    \hat{ \mathbm P }_{j_1\ldots j_D} &= \mathbm P_{j_1\ldots j_D} + \frac{1}{2}  , \label{eq:outlier_gamma_P}\\
    \hat{ \mathbm Q }_{j_1\ldots j_D} &= \mathbm Q_{j_1\ldots j_D} + \frac{1}{2} \mathbb E \llbracket \mathbm E_{j_1\ldots j_D}^2 \rrbracket .\label{eq:outlier_gamma_Q}
\end{align}

\noindent \underline{\textbf{Update $\tau$}}

The variational distribution of $\tau$ follows a Gamma distribution
\begin{align*}
    q(\tau) = \text{Gamma}(\hat{\alpha}_{\tau},\hat{\beta}_{\tau}),
    % \label{eqn:tauupdate}
\end{align*}
with parameters
\begin{align}
    \hat{\alpha}_{\tau} = \alpha_{\tau} + \frac{|\Omega|}{2},
    \label{eqn:taualpha}
\end{align}
and
\begin{align}
    &\hat{\beta}_{\tau} = \beta_\tau + \frac{1}{2} \sum_{j_1=1}^{J_1} \ldots \sum_{j_D=1}^{J_D} \mathbm O_{j_1\ldots j_D} \bigg( (\mathbm Y_{j_1\ldots j_D} - \mathbb E\llbracket \mathbm E_{j_1\ldots j_D} \rrbracket)^2 \nonumber \\
    & + \mathbm V_{j_1\ldots j_D} + \mathbb{E}\llbracket\bm{\mathcal{G}}_{:,:,j_1}^{(1)}\otimes \bm{\mathcal{G}}_{:,:,j_1}^{(1)}\rrbracket \ldots \mathbb{E}\llbracket\bm{\mathcal{G}}_{:,:,j_D}^{(D)}\otimes \bm{\mathcal{G}}_{:,:,j_D}^{(D)}\rrbracket \nonumber \\
    & - 2   (\bm{\mathcal{Y}}_{j_1\ldots j_D} - \mathbb E \llbracket \mathbm E_{j_1 \ldots j_D} \rrbracket)\mathbb{E}\llbracket\bm{\mathcal{G}}_{:,:,j_1}^{(1)}\rrbracket\ldots \mathbb{E}\llbracket\bm{\mathcal{G}}_{:,:,j_D}^{(D)}\rrbracket\bigg).
    \label{eqn:taubeta}
\end{align}

As updating a certain $q(\bm{\Theta}_s)$ requires the statistics of other $\{\bm{\Theta}_k\}_{k\neq s}$, various variational distributions are updated iteratively. The proposed Bayesian algorithm is summarized in Algorithm \ref{alg:graphTTC_vi}, and the required expectations are given in Table. \ref{tab:expectation_cal}.

\subsection{Further discussions}
\label{sec:VIdiscussion}
\noindent
\textbf{Hyperparameter settting.} Proposition 1 reveals the sparsity-promoting property of the proposed model when all the hyperparameters tend to zero. Following Proposition 1, we set the values of $\{\bm{b}^{(d)}\}_{d=2}^{D}$, $\{\bm{c}^{(d)}\}_{d=2}^{D}$, $\{\bm{f}^{(d)}\}_{d=2}^{D}$, $\mathbm P$, $\mathbm Q$, $\alpha_\tau$, $\beta_\tau$ as $1e^{-6}$, and $\{\bm{\lambda}^{(d)}\}_{d=2}^{D}$ as $-1e^{-6}$. A justification for such configuration is provided through experiments in Supplemental Materials, which show the proposed GraphTT-VI is robust to the choice of hyperparameters as long as they are set as small values.

\vspace{2mm}

\noindent \textbf{Insights of the VI updates.} To give an insight into how the proposed algorithm works, we first see how (\ref{eqn:viupdate}) is expressed with respect to a TT core fiber, which follows a quadratic form
\begin{align}
   & \ln{q^*({\bm{G}_{:,p}^{(d)}})} =  \mathbb{E}_{\backslash{\bm{G}_{:,p}^{(d)}}} \bigg\llbracket \tau  \bigg\| \bm{O}_{(d)} \ast \bigg(\bm{Y}_{(d)} - \bm{G}_{(3)}^{(d)}\times (\bm{G}_{(1)}^{(>d)} \nonumber \\
   &  \otimes \bm{G}_{(d)}^{(<d)}) \bigg) \bigg\|_F^2 + \frac{1}{\bm{z}_{k_p}^{(d)}\bm{z}_{\ell_p}^{(d+1)}} \text{tr}({\bm{G}_{(3)}^{(d)}}_{:,p}^{T} \bm{L}^{(d)} {\bm{G}_{(3)}^{(d)}}_{:,p}) \bigg\rrbracket.
   \label{eqn:viupdateanalysis}
\end{align}
Since $q^*(\bm{G}_{:,p}^{(d)})$ follows a Gaussian distribution, the minimum value of $\ln{q^*({\bm{G}_{:,p}^{(d)}})}$ w.r.t. $\bm{G}_{:,p}^{(d)}$ is exactly attained when $\bm{G}_{:,p}^{(d)}=\bm{\nu}^{(d,p)}$. Furthermore, as $\{\bm{\nu}^{(d,p)}\}_{p=1,d=1}^{R_dR_{d+1},D}$ will be adopted to reconstruct the tensor after the convergence of Algorithm \ref{alg:graphTTC_vi}, the VI update using (\ref{eqn:viupdateanalysis}) is similar to the minimization problem (\ref{eqn:opt_wrtTTcore}) w.r.t. ${\bm{G}_{:,p}^{(d)}}$, except the expectation and different coefficients in (\ref{eqn:viupdateanalysis}). For the 'coefficients' in (\ref{eqn:viupdateanalysis}), they are actually modeled as variables and are updated adaptively as shown in Algorithm \ref{alg:graphTTC_vi}, which takes into consideration the noise level $\tau$ and the weighted TT core slice power $\{\bm{z}_d\}_{d=2}^{D}$, and thus in turns refines (\ref{eqn:viupdateanalysis}) to better model the observed data. In contrast, the coefficients in (\ref{eqn:opt_wrtTTcore}) are all fixed and do not have the ability to adapt to different types of data, e.g., different noise or missing ratio, as will be seen in various experiments in Section. \ref{sec:experiment}.

\vspace{2mm}

\noindent
\textbf{Rank Selection.} The proposed probabilistic TT model has the ability to introduce sparsity into the vertical and horizontal slices of the TT cores \cite{cheng2022towards,xu2021overfitting}. Even with graph Laplacian included, it has recently been proved that such model is also sparsity promoting \cite{CHEN2023108826}. After the iterative VI updates, the sparsity inducing variables $\{\bm{z}^{(d)}\}_{d=2}^{(D)}$ tend to have a variational distribution under which many $\mathbb{E}\llbracket{\bm{z}_{k}^{(d)}} \rrbracket$ are close to $0$, and thus the corresponding $\mathbb{E}\llbracket{1}/{\bm{z}_{k}^{(d)}} \rrbracket$ have very large values. On the other hand, it can be seen from (\ref{eqn:GcoreVar}) that a large $\mathbb{E}\llbracket{1}/{\bm{z}_{k_p}^{(d)}} \rrbracket$ would lead the elements of the corresponding $\bm{\Sigma}^{(d,p)}$ and $\bm{\Sigma}^{(d-1,p)}$ to be very small, and therefore lead all the elements in the expectation $\bm{\nu}^{(d,p)}$ and $\bm{\nu}^{(d-1,p)}$ close to zero. In this way, group sparsity is introduced and thus the TT-ranks can be automatically determined. In practice, if the power of $\mathbm{G}_{:,k,:}^{(d)}$ and $\mathbm{G}_{k,:,:}^{(d+1)}$ both tend to be $0$, e.g., less than $1e^{-7}$, these two slices can be discarded.

\vspace{2mm}

\noindent \textbf{Convergence Analysis.} The convergence of Algorithm \ref{alg:graphTTC_vi} is guaranteed, as it has been proved that (\ref{eqn:KLdivergence}) is convex with respect to each variable set $\bm{\Theta}_s$ under the mean-field approximation \cite[pp.~466]{bishop2006pattern}. As (\ref{eqn:viupdate}) is the optimal solution to (\ref{eqn:KLdivergence}) w.r.t. $\bm{\Theta}_s$, the KL divergence between the true posterior and the variational distribution is non-increasing after each update. Moreover, the rank pruning can be performed after every iteration with the convergence property preserved, since every time a slice is deleted, it is equivalent to restarting the VI algorithm with a smaller model size and with the current variational distribution serving as a new initialization.

\vspace{2mm}

\noindent \textbf{Complexity Analysis.} Algorithm \ref{alg:graphTTC_vi} uses most of the time on updating the variational distributions of the TT cores. For the updates of other variables $\{\bm{z}^{(d)}\}_{d=2}^D$, $\{\bm{a}^{(d)}\}_{d=2}^D$, {$\mathbm E$, $\mathbm U$} and $\tau$, they either are with simple expressions or can re-use computation results required for the TT core update. For each TT core fiber, it takes $\bm{O}({J_d^3}+{R^4|\Omega|})$ to obtain (\ref{eqn:GcoreVar}) and (\ref{eqn:Gcoremean}). By noticing that there are unchanged factors for different fibers in (\ref{eqn:GcoreVar}) and (\ref{eqn:Gcoremean}), the total complexity for updating one TT core is $\bm{O}\left(R^2|\Omega|+R^4J_d^3\right)$. Then for an iteration of Algorithm \ref{alg:graphTTC_vi}, it takes computational complexity $\bm{O}\left(DR^2|\Omega|+DR^4J_d^3\right)$.

\section{Experiments}
\label{sec:experiment}
In this section, the performance of the proposed algorithms will be tested on both synthetic and real-world data. In experiments on synthetic data, the effects of the parameters like initial ranks\footnote{For GraphTT-opt, as the TT ranks cannot be learned, the initial ranks are the assumed ranks in the model and will not change during the algorithm.} and regularization parameters will be tested under different noise and missing rates. The convergence performance of GraphTT-opt with fiber update is also compared to that with core update as in (\ref{eqn:opt_wrtTTcore}). In the real-world experiment, different kinds of datasets, including images and videos, are tested under different noise and missing patterns\footnote{The codes for Algorithm \ref{alg:graphTTC_opt} and \ref{alg:graphTTC_vi} are available at \url{https://github.com/xumaomao94/GraphTTC.git}}. For comparison, the performance of some state-of-the-art methods will also be provided.

For the initialization of the proposed methods, we first fill in the missing entries through i.i.d. Gaussian distributed variables with mean and variance obtained from the observed data. Then TT-SVD is performed with truncated TT-ranks set as the initialized ranks, and the initial TT cores $\{\mathbm{G}_0^{(d)}\}_{d=1}^{D}$ are obtained. For the regularization parameters, we choose to set only $\beta_0$ to better illustrate its effects on the optimization-based algorithm. To balance the regularization on each mode, $\beta_d$ is accordingly set as $\beta_0/\text{tr}({{\bm{G}_0}_{(3)}^{(d)}}^T{\bm{G}_0}_{(3)}^{(d)})$, where ${\bm{G}_0}_{(3)}^{(d)}$ is the mode-$3$ unfolding of the $d$-th initialized TT core $\mathbm{G}_0^{(d)}$. For the initialization of the compared methods, they are fine-tuned around the parameter setting introduced in the original works to obtain the best performance.

\subsection{Comparing fiber update vs. core update in synthetic data}
\label{subsec:synthetic}
In this subsection, we use synthetic data to test the performance of the proposed algorithms in terms of fiber update versus the cores update as in (\ref{eqn:opt_wrtTTcore}). The synthetic data is with size $[20,20,20,20]$ and TT-ranks $[1,5,5,5,1]$. To generate the synthetic data, we first generate $4$ TT cores according to (\ref{eqn:TTdefinition}). To make the synthetic data embedded with graph information, for each unfolded TT core ${\bm{G}}_{(3)}^{(d)}$, we generate it with its columns from $\mathcal{N}(\bm{0},\bm{\Sigma})$, where $\bm{\Sigma} \in \mathbb{R}^{J_d\times J_d}$ with $\bm{\Sigma}_{i,j} =  \text{exp}( \frac{1}{5}|i - j|^2)$.
{After generating the ground truth low TT-rank tensor $\mathbm X_\sharp$, additive white Gaussian noise $\mathbm W$ is added, with signal-to-noise ratio (SNR) defined as \[\text{SNR} = 20\log_{10}(\|\mathbm X_\sharp\|_F/\|\mathbm W\|_F).\]
Outliers are modeled as i.i.d. Gaussian variables with zero mean and variance equal to $\eta$ times the variance of $\mathbm X_\sharp$, where $\eta$ is typically set to a large value (e.g., $\eta=100$) to simulate high-magnitude corruptions.
}

For the graph adopted in the tested algorithms, we generate it as in (\ref{eqn:graphdef}), from the weighting matrix $\bm{A} \in \mathbb{R}^{J_d\times J_d}$ with $\bm{A}_{i,j} =  \text{exp}(|i - j|^2)$, which is different from $\bm{\Sigma}$ since for real-world data it is not very likely we have access to the ground truth weighting matrix. 
The relative square error (RSE) is adopted as the evaluation metric, which is defined as \[\text{RSE} = \|\bm{X}_\sharp-\hat{\mathbm{X}}\|_F/\|\mathbm{X}_\sharp\|_F,\] with $\hat{\mathbm{X}}$ denoting the recovered TT-format tensor.
For the tested algorithms, their performance under different SNRs and missing rates are tested. Especially, the effects of different parameters are evaluated for GraphTT-opt, i.e., performance under different TT ranks $[1,R,R,R,1]$ and regularization parameters $\beta_0$.
For each case, the experiments are conducted for $20$ Monte Carlo runs, and the average result is presented.

\begin{figure}[!tb]
    \centering
    \begin{subfigure}{.98\linewidth}
      \includegraphics[width=1\linewidth]{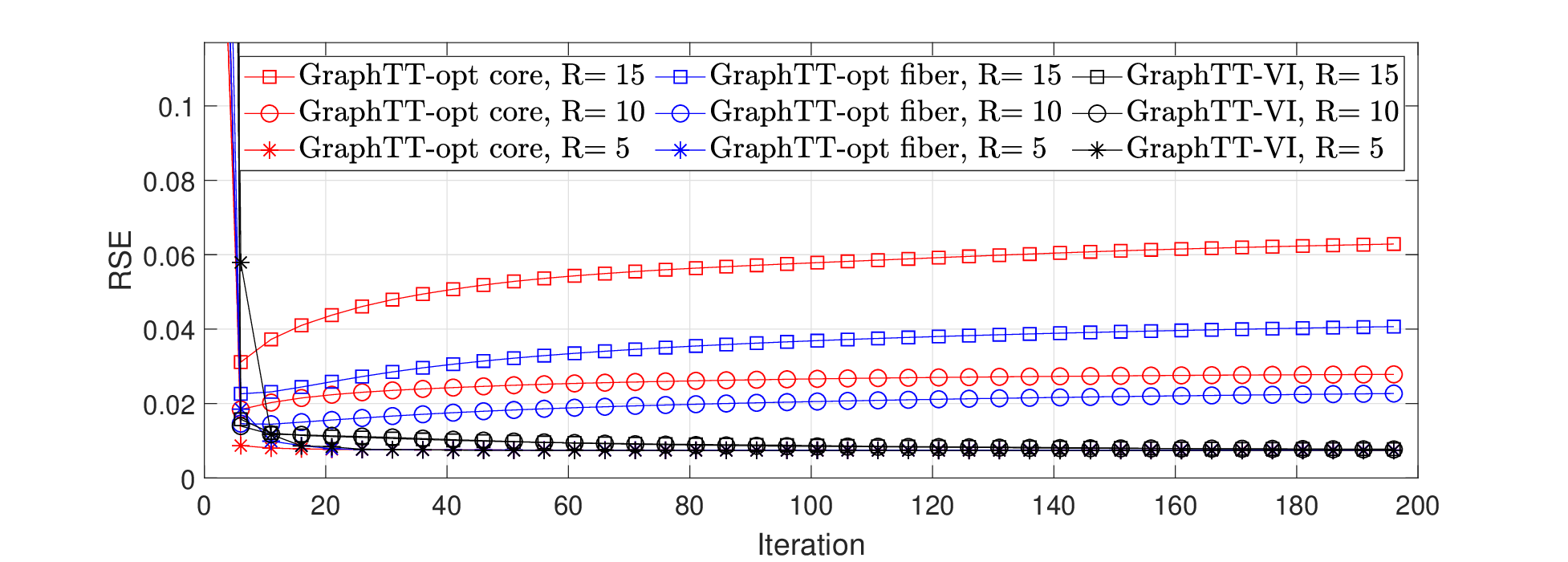}
      \caption{RSE at each iteration ($\beta_0=0.5$).}
      \label{subfig:converge_torank}
    \end{subfigure}
    \begin{subfigure}{.98\linewidth}
      \includegraphics[width=1\linewidth]{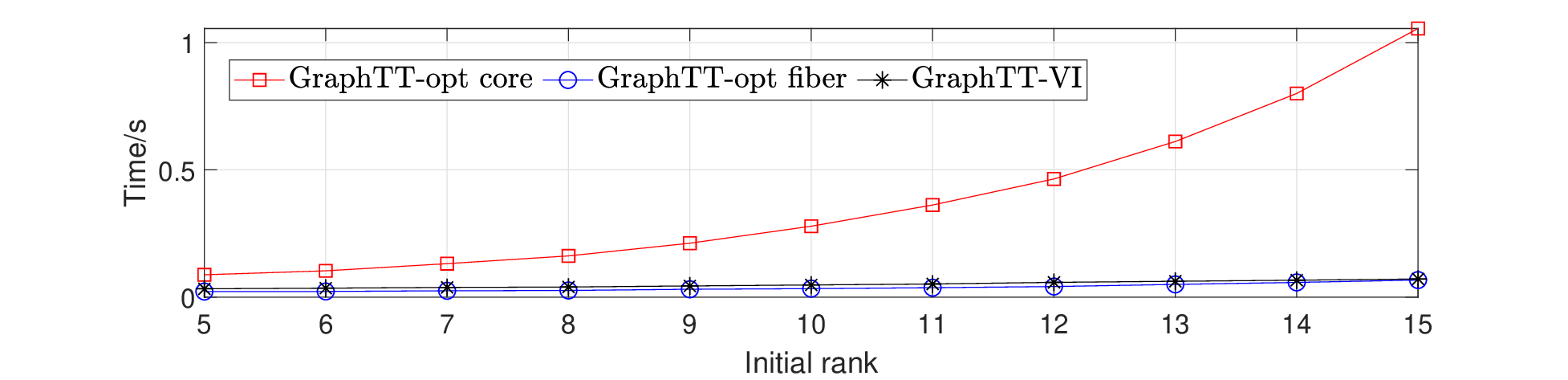}
      \caption{Time cost for each iteration}
      \label{subfig:timecost}
    \end{subfigure}
    \caption{Convergence of the proposed methods\\ (SNR = $10$dB, missing rate = $90\%$, no outlier).}
    \label{fig:convergence}
\end{figure}

Fig. \ref{fig:convergence} illustrates the convergence performance and time required by the compared algorithms under different rank initializations. The experiments are conducted with an SNR of $10$dB and a missing rate of $90\%$. From Fig. \ref{subfig:converge_torank}, it is evident that both GraphTT-VI and GraphTT-opt, employing fiber update and core update, achieve convergence across all tested configurations. However, careless choice of initial ranks would lead to slower convergence, especially for the core update, e.g., an inappropriate choice of $R = 15$ as shown in Fig. \ref{subfig:converge_torank}, which is far away from the true rank $R=5$ and results in a larger problem size. In addition, the recovery performance deteriorates when the initial ranks become larger, which is because with larger ranks the model tends to overfit the noise. In contrast, the proposed GraphTT-VI provides similar performance under different initialized TT ranks, which highlights its capability to automatically estimate the appropriate TT ranks.

Fig. \ref{subfig:timecost} shows the time consumption per iteration for the proposed algorithms. As can be seen, with larger initial TT ranks, the time cost grows, which is especially obvious for the GraphTT-opt with core update. Each iteration of it involves $D$ matrix inverses, each of which is with a complexity of $\bm{O}(J_d^3R^6)$. In comparison, the corresponding operations in the other two methods are with a complexity of $\bm{O}(J_d^3R^2)$.

\begin{figure}[!tb]
    \centering
    \includegraphics[width=1\linewidth]{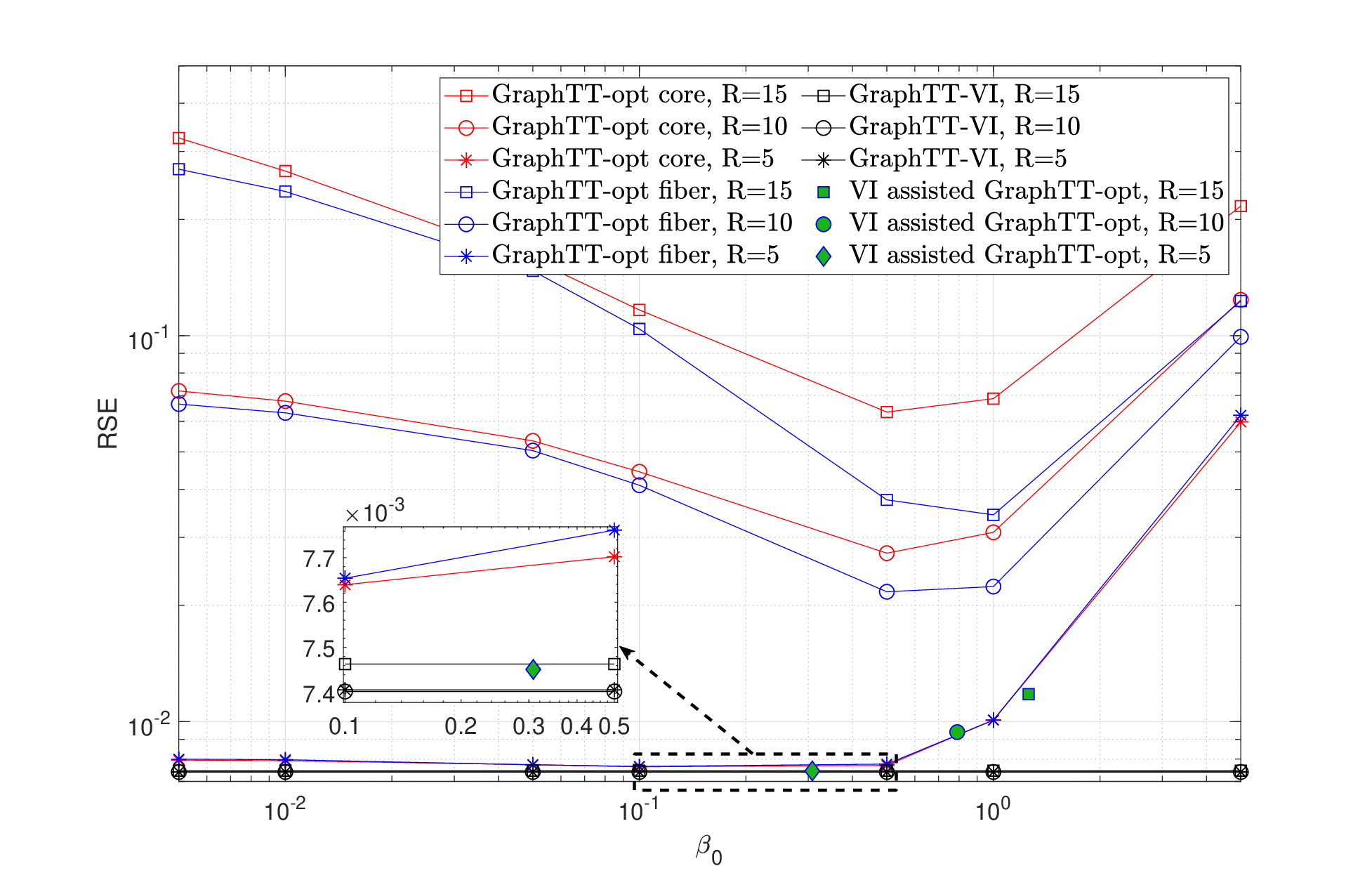}
    \caption{RSE w.r.t. different regularization parameters\\ (SNR = $10$dB, missing rate = $90\%$, no outlier).}
    \label{fig:perf_to_initPar}
\end{figure}

Fig. \ref{fig:perf_to_initPar} presents the performance of compared algorithms under different initial ranks (R set as $5$, $10$ and $15$) and regularization parameters ($\beta_0$ set as $0.005$, $0.01$, $0.05$, $0.1$, $0.5$, $1$ and $5$). Additionally, in order to showcase the advantage of GraphTT-VI in automatic parameters selection, we further evaluate GraphTT-opt with fiber update, in which the ranks are set as the estimated ranks from GraphTT-VI, and the regularization parameters are approximated by $\beta_d = \frac{1}{R_d R_{d+1}}\sum_{p=1}^{R_dR_{d+1}}{\mathbb{E}\llbracket 1/({\bm{z}_{k_p}^{(d)}\bm{z}_{\ell_p}^{(d+1)}})\rrbracket}$, by noticing that $1/({\bm{z}_{k_p}^{(d)}\bm{z}_{\ell_p}^{(d+1)}})$ in (\ref{eqn:viupdateanalysis}) plays a role similar to the regularization parameter in (\ref{eqn:opt_wrtTTcore}).

From Fig. \ref{fig:perf_to_initPar}, it is observed that with different regularization parameters and TT ranks, the performance of GraphTT-opt varies significantly, no matter with core update or fiber update. For example, when $R$ is set to $10$ or $15$, the performance of GraphTT-opt improves as $\beta_0$ increases from $0.005$ to $0.5$, but deteriorates when $\beta_0=5$. The reason is that, with $\beta_0$ smaller than $0.5$, the graph information is not fully used, while with $\beta_0=5$, the graph-regularized terms are excessively emphasized over the reconstruction error in the objective function. Furthermore, with the same choice of $\beta_0$, the performance of GraphTT-opt becomes worse when the ranks increase from $5$ to $15$, due to noise overfitting.

On the other hand, GraphTT-VI does not need a regularization parameter, and it consistently achieves the best performance under different initial ranks. Moreover, as can be seen from Fig. \ref{fig:perf_to_initPar}, under all initial ranks, VI-assisted GraphTT-opt outperforms GraphTT-opt with manually set regularization parameters, especially when $R=10$ or $15$. In general, the performance matches that of GraphTT-opt with $R=5$, which again shows the superiority of GraphTT-VI in automatic rank estimation.

Furthermore, it can be seen in both Fig. \ref{subfig:converge_torank} and Fig. \ref{fig:perf_to_initPar} that GraphTT-opt with fiber update exhibits similar or superior performance compared to core update across all parameter settings. Particularly, under unfavorable parameter settings like $\beta_0=5$ or $R=15$, the performance disparity between the two methods becomes more obvious. This is probably because of the greater flexibility of fiber update, enabling them to explore regions around specific local minima that core update cannot reach.

\begin{figure}[!tb]
    \centering
    \begin{subfigure}{.49\linewidth}
      \includegraphics[width=1\linewidth]{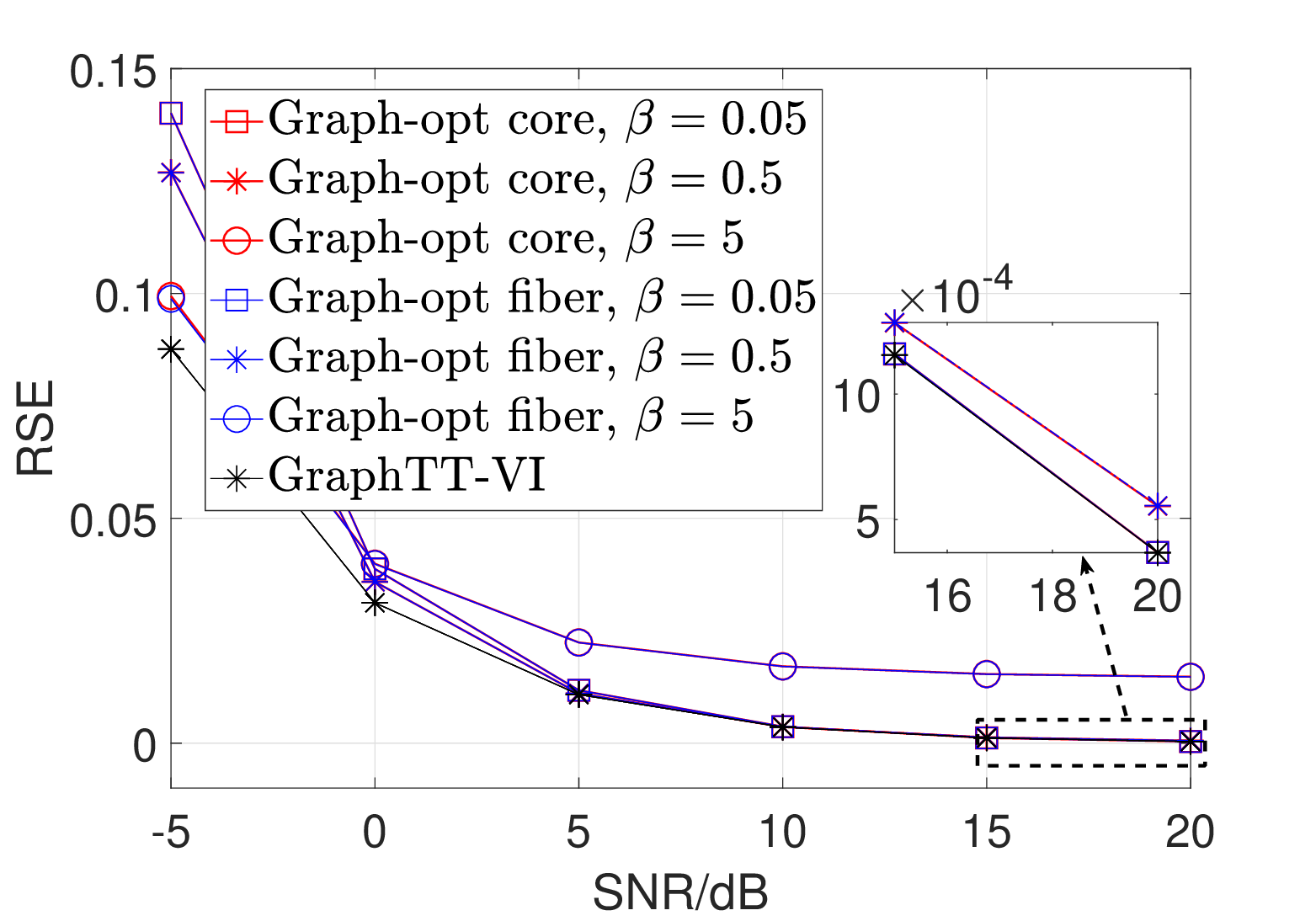}
      \caption{w.r.t SNR \\(missing rate = $80\%$)}
      \label{subfig:rsewrtsnr}
    \end{subfigure}
    \begin{subfigure}{.49\linewidth}
      \includegraphics[width=1\linewidth]{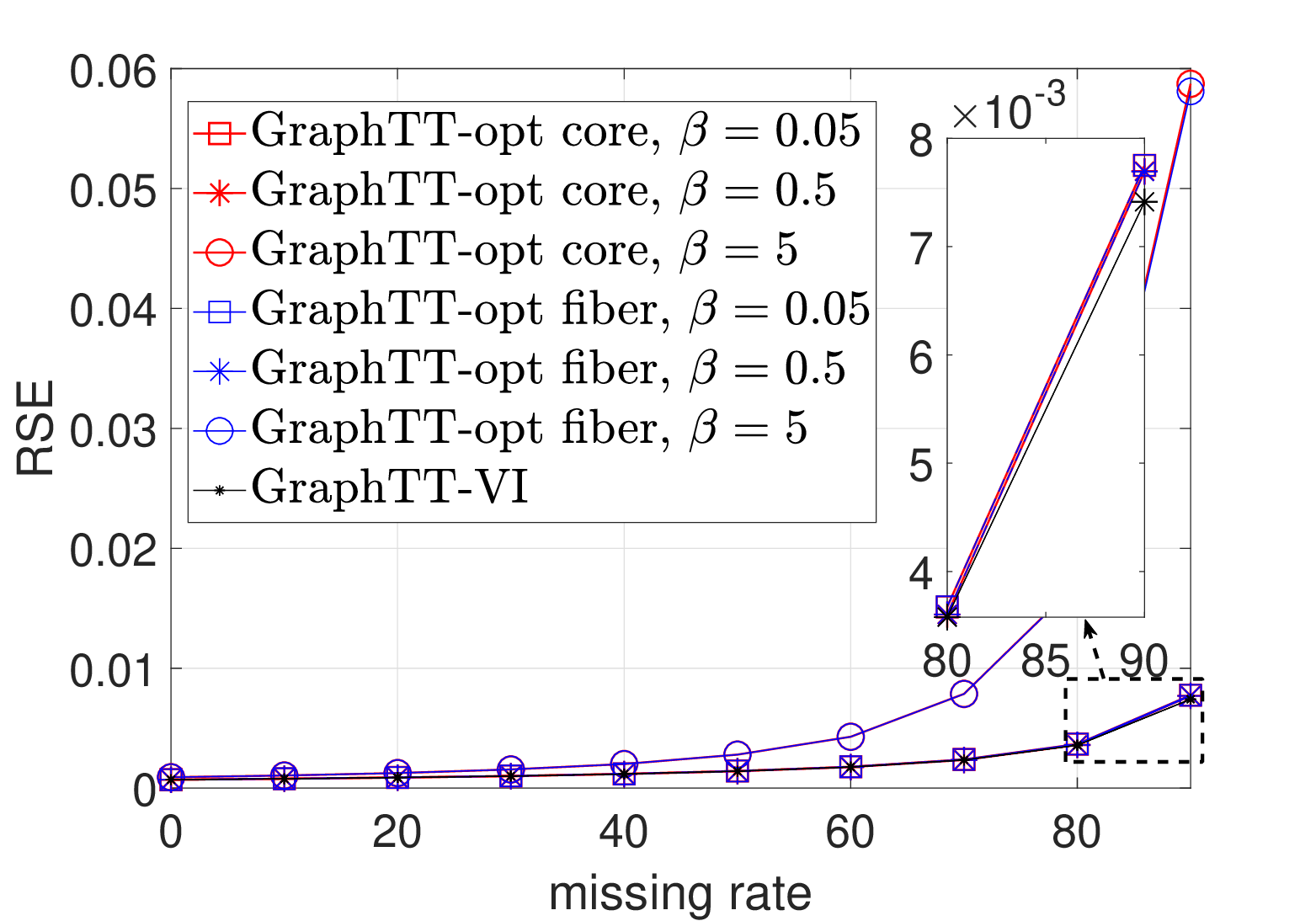}
      \caption{w.r.t missing rate\\ (SNR = $10$dB)}
      \label{subfig:rsewrtmr}
    \end{subfigure}
    \caption{RSE w.r.t. different SNRs and missing rates\\ ($R=5$, no outlier).}
    \label{fig:synthe_SNRMR}
\end{figure}

Fig. \ref{fig:synthe_SNRMR} shows the performance of the proposed algorithms under different SNR and missing rates with initial ranks set as the true TT-ranks, and the detailed settings of other parameters are labeled in the figure. In all settings, GraphTT-VI obtains the best performance, and GraphTT-opt under fiber and core update performs similarly. An interesting observation from Fig. \ref{subfig:rsewrtsnr} is that different $\beta_0$ lead to totally different performance under different SNRs, i.e., GraphTT-opt with $\beta_0=5$ performs the best under $-5\text{dB}$ but worst under $20\text{dB}$, and in contrast, with $\beta_0=0.05$ it performs the worst under $-5\text{dB}$ but the best under $20\text{dB}$. That is because with large noise, the graph regularization should be considered more important as the observed data are contaminated and not reliable, but with small noise, we can rely more on the observed data and lower the importance of the regularization terms.

From Fig. \ref{subfig:rsewrtmr} it can be seen that with moderate SNR ($10\text{dB}$) and relatively low missing rates, all methods perform similarly. However, as the missing rate goes higher, the effects of the choice of parameters become more obvious. With missing rate from $80\%$ to $90\%$, even with the TT-ranks initialized as the true ones, $\beta_0$ has a significant influence on the performance, e.g., $\beta_0 = 0.5$ provides the best performance, $\beta=0.05$ performs slightly worse but still close to that of $\beta_0=0.5$, and $\beta=5$ leads to unmistakably worse performance.

\begin{figure}[!tb]
    \centering
    \begin{subfigure}{.49\linewidth}
      \includegraphics[width=1\linewidth]{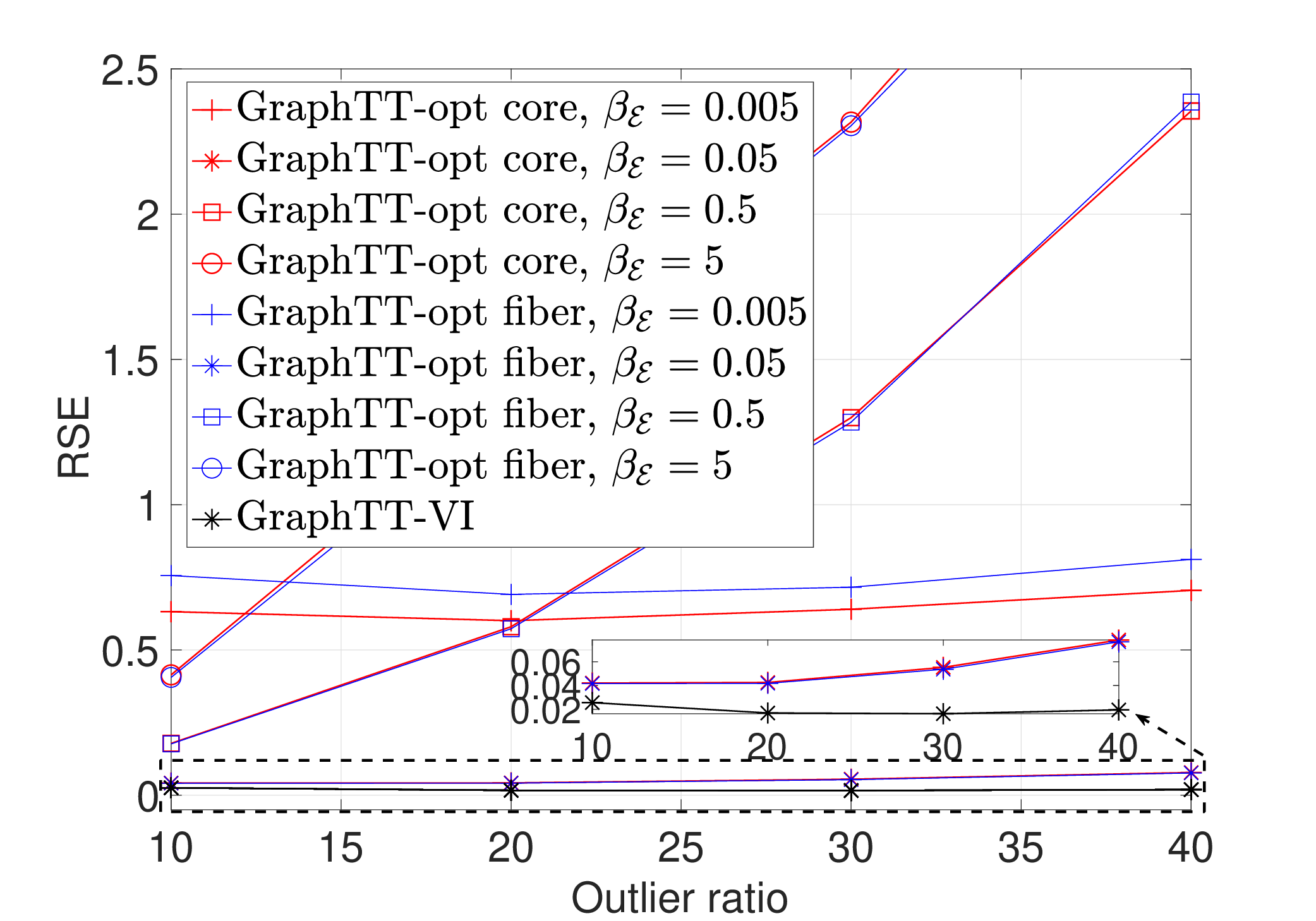}
      \caption{{w.r.t ratio of outliers \\ ($\eta=100$)}}
      \label{subfig:rsewrt_ratio}
    \end{subfigure}
    \begin{subfigure}{.49\linewidth}
      \includegraphics[width=1\linewidth]{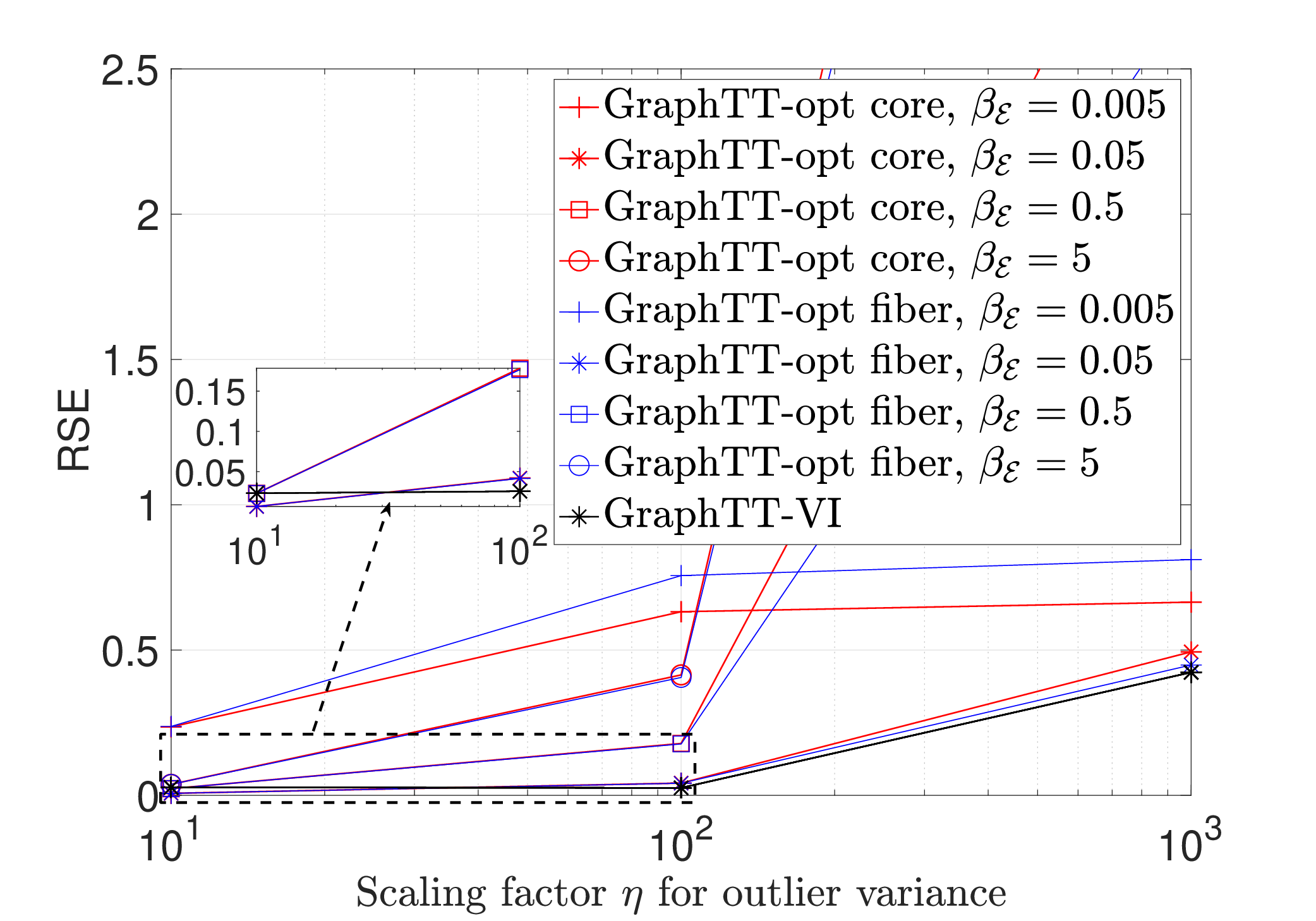}
      \caption{{w.r.t the outlier variance scaling factor $\eta$ ($10\%$ outliers)}}
      \label{subfig:rsewrt_scale}
    \end{subfigure}
    \caption{{RSE w.r.t. different outlier settings\\ (SNR = $10$dB, missing rate = $80\%$, $R = 5$, $\beta_0 = 0.5$).}}
    \label{fig:synthe_outlier}
\end{figure}

{Fig. \ref{fig:synthe_outlier} shows the performance of the compared methods under various outlier settings, with other parameters specified in the caption. This task is particularly challenging---for example, even with a relatively low outlier ratio of $10\%$ and a moderate variance scaling factor $\eta=100$, the resulting $\mathbm X_\sharp + \mathbm W + \mathbm E$ yields an SNR of around 1dB. This challenge is further compounded by a high missing rate of $80\%$.
In Fig. \ref{fig:synthe_outlier}, GraphTT-VI consistently achieves the best overall performance across different outlier ratios and values of $\eta$, reaching an RSE of approximately 0.04 for $\eta\leq 100$ under all tested outlier ratios. GraphTT-opt also performs well, but only when the regularization parameter $\beta_{\mathbm{E}}$ is carefully tuned. In this case, it achieves the best performance at $\beta_{\mathbm{E}}=0.05$, but degrades significantly for other values. These results highlight the effectiveness of the proposed methods in handling outliers, as well as the key advantage of GraphTT-VI---it requires no manual parameter tuning.}

\subsection{RGB images completion}
\label{subsec:rgb}
Next, we will test the performance of the proposed methods on real-world data. Noisy and incomplete images/videos with different missing patterns will be tested. Without loss of generality, all tested data are normalized such that their entries are valued from $0$ to $1$. The results of the following state-of-the-art methods are also presented as a comparison, with the parameters fine-tuned to their best performance.
\begin{itemize}[leftmargin=*]
  \item Simple low-rank tensor completion via TT (SiLRTC-TT) \cite{bengua2017efficient}, which adopts the TT nuclear norm as a regularization for the TT completion;
  \item Tensor completion by parallel matrix factorization via TT (TMAC-TT) \cite{bengua2017efficient}, which minimizes the reconstruction error using parallel matrix factorization;
  \item Tensor train completion with total variation regularizations (TTC-TV) \cite{ko2020fast}, which adopts the total variation as the regularization for the TT completion;
  \item Probabilistic tensor train completion (PTTC) \cite{xu2021overfitting}, which uses the Gaussian-Gamma sparsity promoting prior for the traditional TT format and solves it using variational inference. An improved folding strategy is also introduced by duplicating the folding edges.
  \item Tensor ring completion based on the variational Bayesian framework (TR-VBI) \cite{long2021bayesian}, which builds a probabilistic model from the Gaussian-Gamma prior for the tensor ring completion and learning through VI.
  \item Sparse tensor train optimization (STTO) \cite{yuan2018high}, which minimizes the square error between the completed TT tensor and the observed tensor by considering only the observed entries;
  \item Fully Bayesian Canonical Polyadic Decomposition (FBCP) \cite{twothreezhao2015bayesian}, which builds a probabilistic model for tensor CPD and learns it through VI methods.
  \item Fast low-rank tensor completion (FaLRTC) \cite{liu2013tensor}, which adopts the tensor trace norm as the regularization for tensor completion;
  \item {Diffusion posterior sampling (DPS) \cite{chung2023diffusion}, an inverse problem solver that samples from the posterior distribution using Langevin dynamics \cite{song2021scorebased}, guided by a pretrained diffusion model \cite{dhariwal2021diffusion} and the measurement likelihood. In this subsection, the adopted diffusion model is pretrained on the ImageNet $256\times 256$ dataset \cite{deng2009imagenet}, which contains over 1 million images across 1000 categories. The likelihood follows a Gaussian measurement model, corresponding to the inpainting task setting described in \cite{chung2023diffusion}.}
\end{itemize}
{We do not compare with other generative models, such as transformer-based methods \cite{he2022masked,Li2022MAT}, as they are not directly applicable to our setting with random missing elements. Furthermore, their objectives differ fundamentally from ours: they are designed to generate visually plausible images, whereas our methods aim to ensure data fidelity.
}

To investigate the effect of folding the image under graph regularizations, we evaluate the performance of GraphTT-VI under different folding strategies and present the best results, which is denoted as `GraphTT-fold'. The tested folding strategies and the performance can be found in the supplemental materials. For SiLRTC-TT, TMAC-TT, TTC-TV, PTTC, TR-VBI, and STTO, tensor folding is also performed before TT completion, and the way a tensor is folded follows that in the original work. For the detailed folding strategies and parameter settings for the compared methods, please see the supplemental materials.

The performance of these methods is evaluated by the peak signal-to-noise ratio (PSNR) which is defined as
\begin{align}
    \text{PSNR} = 20 \log_{10} \max(\mathbm{X}) - 20 \log_{10}(\text{MSE}),
\end{align}
where $\max(\mathbm{X})$ is the maximum value of the original data tensor $\mathbm{X}$, and MSE denotes the mean square error between the completed and original images. The structural similarity index measure (SSIM) \cite{wang2004image} is also tested, which takes more image information like luminance masking and contrast masking terms.

In this subsection, 12 RGB images with size $256\times 256 \times 3$ are tested. {All tensor-based methods are implemented on an Intel Core i7-8700K CPU, while DPS runs on an Intel Xeon Platinum 8168 CPU with a Tesla V100 GPU.}

\begin{table*}[!tb]
\caption{Performance of image completion with 90\% random missing entries without noise.}
\footnotesize
\centering
\scalebox{0.7}{
\begin{tabular}{ l |m{1.5em} m{2em} |m{1.5em} m{2em} |m{1.5em} m{2em} |m{1.5em} m{2em} |m{1.5em} m{2em} |m{1.5em} m{2em} |m{1.5em} m{2em} |m{1.5em} m{2em} |m{1.5em} m{2em} |m{1.5em} m{2em} |m{1.5em} m{2em} |m{2em} m{2.5em}}
 \hline
 \quad &  \multicolumn{2}{c|}{SiLRTC-TT} & \multicolumn{2}{c|}{TMAC-TT}  & \multicolumn{2}{c|}{TTC-TV} &\multicolumn{2}{c|}{PTTC} &\multicolumn{2}{c|}{TR-VBI} & \multicolumn{2}{c|}{STTO} &  \multicolumn{2}{c|}{FBCP}  & \multicolumn{2}{c|}{FaLRTC} & \multicolumn{2}{c|}{{DPS}}   & \multicolumn{2}{c|}{GraphTT-opt} & \multicolumn{2}{c|}{GraphTT-VI} & \multicolumn{2}{c}{GraphTT-fold} \\  
\textbf{} & PSNR & SSIM & PSNR & SSIM & PSNR & SSIM & PSNR & SSIM & PSNR & SSIM & PSNR & SSIM & PSNR & SSIM & PSNR & SSIM & PSNR & SSIM & PSNR & SSIM & PSNR & SSIM & PSNR & SSIM \\
\hline
airplane & 19.57 & 0.593 & 21.34 & 0.689 & 20.78 & 0.595 & {22.44} & \underline{0.709} & 21.44 & 0.541 & 20.67 & 0.533 & 19.53 & 0.448 & 18.97 & 0.496 & \underline{23.31} & 0.694  & 22.25 & 0.660 & \textbf{23.34} & \textbf{0.715} & 22.14 & 0.687\\
baboon & 19.11 & 0.375 & 18.77 & 0.414 & 20.16 & \underline{0.416} & 20.26 & 0.389 & 19.99 & 0.310 & 19.45 & 0.370 & 18.46 & 0.270 & 18.51 & 0.348 & \underline{20.65} & 0.311  & 19.61 & \textbf{0.446} & \textbf{20.85} & 0.380 & 20.20 & 0.324\\
barbara & 20.31 & 0.547 & 22.46 & 0.674 & 21.35 & 0.564 & 23.41 & 0.699 & 22.50 & 0.583 & 22.12 & 0.590 & 19.08 & 0.416 & 18.95 & 0.466 & {23.47} & {0.590}  & \textbf{24.38} & \textbf{0.722} & \underline{24.29} & \underline{0.701} & 22.90 & 0.639\\
couple & 23.20 & 0.625 & 21.89 & 0.595 & 24.99 & 0.666 & 26.33 & 0.745 & 25.85 & 0.658 & 25.37 & 0.665 & 24.05 & 0.560 & 23.28 & 0.615 & 26.51 & 0.720 & \textbf{28.16} & \textbf{0.837} & \underline{27.71} & \underline{0.780} & 26.14 & 0.713\\
facade & 19.34 & 0.424 & 22.05 & 0.653 & 22.30 & 0.638 & 22.25 & 0.636 & 25.56 & 0.782 & 20.81 & 0.560 & 25.56 & 0.799 & 24.74 & 0.788 & 23.98 & 0.520  & \underline{26.10} & \underline{0.828} & \textbf{27.14} & \textbf{0.839} & 20.38 & 0.396\\
goldhill & 20.74 & 0.477 & 23.17 & 0.621 & 21.83 & 0.539 & 23.55 & 0.615 & 22.86 & 0.522 & 22.04 & 0.535 & 20.46 & 0.427 & 20.43 & 0.481 & 22.73 & 0.425  & \underline{24.35} & \textbf{0.675} & \textbf{24.63} & \underline{0.650} & 23.13 & 0.533\\
house & 21.18 & 0.653 & 22.81 & 0.716 & 22.97 & 0.649 & {26.10} & {0.748} & 24.85 & 0.637 & 22.86 & 0.608 & 20.99 & 0.515 & 20.78 & 0.574 & \textbf{28.47} & \underline{0.758} & 25.66 & 0.716 & \underline{26.40} & \textbf{0.771} & 24.12 & 0.690\\
jellybeans & 21.94 & 0.801 & 23.82 & 0.849 & 23.78 & 0.848 & 26.63 & {0.892} & 20.79 & 0.780 & 23.35 & 0.604 & 20.51 & 0.748 & 21.86 & 0.807 & \textbf{29.55} & \underline{0.908} & {26.79} & 0.887 & \underline{27.07} & \textbf{0.910} & 24.68 & 0.866\\
peppers & 19.05 & 0.570 & 21.10 & 0.660 & 20.06 & 0.567 & 22.41 & 0.688 & 20.96 & 0.541 & 20.82 & 0.582 & 17.68 & 0.360 & 17.00 & 0.384 & \textbf{23.67} & 0.685 &\underline{23.32} & \underline{0.709} & {22.87} & \textbf{0.726} & 22.05 & 0.693\\
sailboat & 18.05 & 0.483 & 19.85 & 0.586 & 19.58 & 0.529 & 20.94 & 0.615 & 19.88 & 0.481 & 19.66 & 0.497 & 18.43 & 0.400 & 17.91 & 0.444 & 20.72 & 0.544 & \underline{21.46} & \underline{0.641} & \textbf{21.85} & \textbf{0.647} & 20.42 & 0.581\\
splash & 21.24 & 0.662 & 23.81 & 0.729 & 23.17 & 0.681 & 25.74 & 0.743 & 22.75 & 0.647 & 23.10 & 0.667 & 21.47 & 0.605 & 21.93 & 0.663 & \textbf{27.18} & \textbf{0.801} & {26.08} & {0.757} & \underline{26.66} & \underline{0.768} & 24.12 & 0.715\\
tree & 17.95 & 0.471 & 20.56 & 0.597 & 19.12 & 0.501 & 21.05 & 0.598 & 20.10 & 0.469 & 19.84 & 0.490 & 17.43 & 0.342 & 16.95 & 0.369 & \underline{21.48} & 0.561 & {21.39} & \underline{0.599} & \textbf{21.89} & \textbf{0.631} & 20.55 & 0.562\\
\hline
\end{tabular}}
\label{table:90mrclean}
\end{table*}

\begin{table*}[!tb]
\caption{Performance of image completion with 90\% random missing with $10\%$ salt and pepper noise.}
\footnotesize
\centering
\scalebox{0.7}{
\begin{tabular}{ l |m{1.5em} m{2em} |m{1.5em} m{2em} |m{1.5em} m{2em} |m{1.5em} m{2em} |m{1.5em} m{2em} |m{1.5em} m{2em} |m{1.5em} m{2em} |m{1.5em} m{2em} |m{1.5em} m{2em}  |m{1.5em} m{2em} |m{1.5em} m{2em} |m{2em} m{2.5em}}
 \hline
 \quad &  \multicolumn{2}{c|}{SiLRTC-TT} & \multicolumn{2}{c|}{TMAC-TT}  & \multicolumn{2}{c|}{TTC-TV} &\multicolumn{2}{c|}{PTTC} &\multicolumn{2}{c|}{TR-VBI} & \multicolumn{2}{c|}{STTO} &  \multicolumn{2}{c|}{FBCP}  & \multicolumn{2}{c|}{FaLRTC} &  \multicolumn{2}{c|}{DPS} & \multicolumn{2}{c|}{GraphTT-opt} & \multicolumn{2}{c|}{GraphTT-VI} & \multicolumn{2}{c}{GraphTT-fold} \\   %  \multicolumn{2}{c|}{\begin{tabular}{@{}c@{}}Proposed \\ Algorithm\end{tabular}} &\\
\textbf{} & PSNR & SSIM & PSNR & SSIM & PSNR & SSIM & PSNR & SSIM & PSNR & SSIM & PSNR & SSIM & PSNR & SSIM & PSNR & SSIM & PSNR & SSIM & PSNR & SSIM & PSNR & SSIM & PSNR & SSIM \\
\hline
airplane & 15.55 & 0.282 & 8.48 & 0.091 & 16.84 & 0.329 & 18.79 & 0.456 & 17.87 & 0.409 & 16.08 & 0.234 & 16.82 & 0.221 & 14.83 & 0.212  &18.28 & 0.467 & \textbf{22.14} & \textbf{0.691} & \underline{21.48} & \underline{0.681} & 19.65 & 0.519\\
baboon & 16.60 & 0.251 & 11.63 & 0.095 & 17.24 & 0.281 & 18.73 & 0.271 & 18.58 & 0.251 & 16.08 & 0.206 & 16.49 & 0.181 & 15.62 & 0.214 & 18.17 & 0.191  & \underline{19.15} & \textbf{0.409} & \textbf{20.41} & \underline{0.359}& 19.05 & 0.264\\
barbara & 16.68 & 0.309 & 11.92 & 0.092 & 17.82 & 0.367 & 19.23 & 0.435 & 18.32 & 0.364 & 16.86 & 0.265 & 16.66 & 0.235 & 15.39 & 0.236 &  17.58 &  0.258  & \textbf{23.51} & \textbf{0.689} & \underline{23.08} & \underline{0.670}  & 20.46 & 0.482\\
couple & 19.01 & 0.314 & 13.63 & 0.066 & 18.40 & 0.323 & 20.90 & 0.409 & 18.99 & 0.369 & 16.39 & 0.160 & 19.46 & 0.232 & 18.33 & 0.247 & 19.97 & 0.501  & \textbf{26.17} & \textbf{0.777} & \underline{25.53} & \underline{0.753}  & 23.58 & 0.553\\
facade & 16.97 & 0.279 & 13.54 & 0.131 & 18.07 & 0.407 & 18.39 & 0.213 & 20.42 & 0.409 & 16.65 & 0.292 & 20.03 & 0.501 & 18.34 & 0.475 & 18.52 & 0.212 & \underline{24.76} & \underline{0.772} & \textbf{25.93} & \textbf{0.814}  & 18.91 & 0.269\\
goldhill & 17.64 & 0.294 & 12.95 & 0.106 & 18.20 & 0.353 & 19.84 & 0.367 & 19.47 & 0.321 & 17.45 & 0.261 & 18.04 & 0.259 & 16.77 & 0.260  & 18.52 & 0.212 & \textbf{23.90} & \textbf{0.632} & \underline{23.66} & \underline{0.613}  & 20.82 & 0.380\\
house & 17.14 & 0.332 & 11.55 & 0.075 & 17.91 & 0.338 & 20.55 & 0.506 & 19.37 & 0.463 & 17.35 & 0.257 & 17.34 & 0.243 & 16.09 & 0.236 & 20.12 &0.532  & \textbf{24.49} & \underline{0.715} & \underline{24.02} & \textbf{0.743}& 21.16 & 0.582\\
jellybeans & 16.55 & 0.344 & 8.60 & 0.061 & 17.59 & 0.363 & 20.81 & 0.667 & 20.30 & 0.644 & 16.78 & 0.230 & 17.84 & 0.286 & 15.71 & 0.231 & 20.42&0.639 & \textbf{25.51} & \underline{0.865} & \underline{25.03} & \textbf{0.879}& 21.92 & 0.763 \\
peppers & 15.32 & 0.296 & 10.39 & 0.059 & 16.77 & 0.354 & 17.57 & 0.396 & 16.95 & 0.378 & 16.19 & 0.277 & 14.83 & 0.183 & 13.86 & 0.198 & 17.22& 0.408 & \textbf{21.94} & \underline{0.702} & \underline{21.31} & \textbf{0.704} & 18.87 & 0.528\\
sailboat & 15.23 & 0.272 & 9.66 & 0.067 & 16.73 & 0.354 & 18.00 & 0.389 & 17.01 & 0.364 & 15.93 & 0.260 & 15.77 & 0.226 & 14.79 & 0.250 & 17.09 &0.381  & \textbf{20.96} & \textbf{0.640} & \underline{20.40} & \underline{0.612}& 18.50 & 0.436\\
splash & 16.49 & 0.338 & 11.02 & 0.062 & 17.96 & 0.360 & 19.65 & 0.444 & 19.81 & 0.553 & 17.15 & 0.289 & 17.89 & 0.253 & 16.62 & 0.293 & 20.12 & 0.640 & \textbf{24.28} & \underline{0.740} & \underline{23.15} & \textbf{0.749} & 21.75 & 0.640\\
tree & 15.05 & 0.272 & 9.78 & 0.067 & 16.64 & 0.351 & 17.97 & 0.387 & 16.25 & 0.274 & 15.93 & 0.274 & 15.26 & 0.183 & 14.11 & 0.206  & 16.26 & 0.284  & \textbf{20.87} & \textbf{0.599} & \underline{20.39} & \underline{0.576}& 18.33 & 0.409\\
\hline
\end{tabular}}
\label{table:90mrsaltpepper}
\end{table*}

\begin{figure*}[!tb]
\centering

\begin{subfigure}{.066\textwidth}
  \includegraphics[width=1\linewidth]{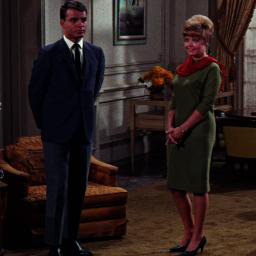}
\end{subfigure}
\begin{subfigure}{.066\textwidth}
  \includegraphics[width=1\linewidth]{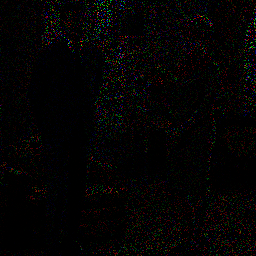}
\end{subfigure}
\begin{subfigure}{.066\textwidth}
  \includegraphics[width=1\linewidth]{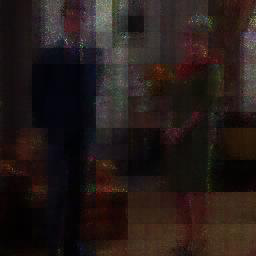}
\end{subfigure}
\begin{subfigure}{.066\textwidth}
  \includegraphics[width=1\linewidth]{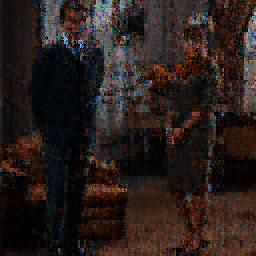}
\end{subfigure}
\begin{subfigure}{.066\textwidth}
  \includegraphics[width=1\linewidth]{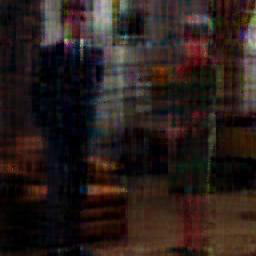}
\end{subfigure}
\begin{subfigure}{.066\textwidth}
  \includegraphics[width=1\linewidth]{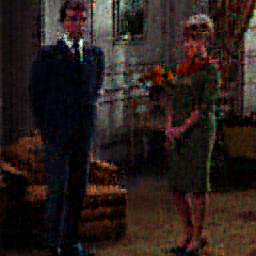}
\end{subfigure}
\begin{subfigure}{.066\textwidth}
  \includegraphics[width=1\linewidth]{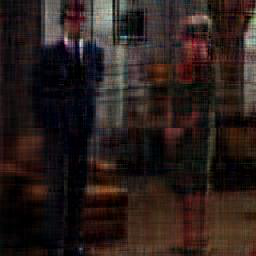}
\end{subfigure}
\begin{subfigure}{.066\textwidth}
  \includegraphics[width=1\linewidth]{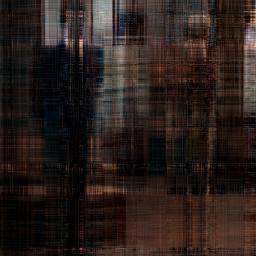}
\end{subfigure}
\begin{subfigure}{.066\textwidth}
  \includegraphics[width=1\linewidth]{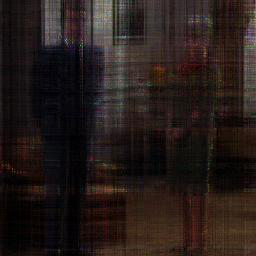}
\end{subfigure}
\begin{subfigure}{.066\textwidth}
  \includegraphics[width=1\linewidth]{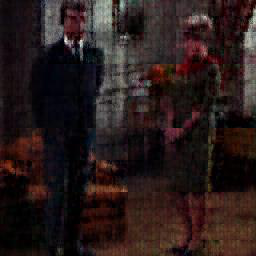}
\end{subfigure}
\begin{subfigure}{.066\textwidth}
  \includegraphics[width=1\linewidth]{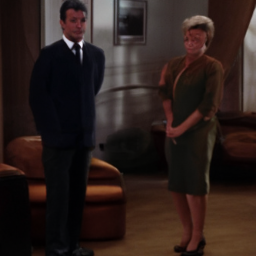}
\end{subfigure}
\begin{subfigure}{.066\textwidth}
  \includegraphics[width=1\linewidth]{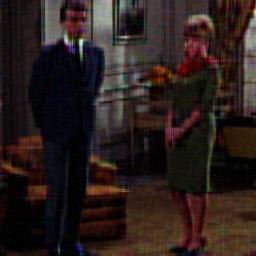}
\end{subfigure}
\begin{subfigure}{.066\textwidth}
  \includegraphics[width=1\linewidth]{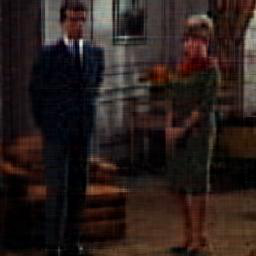}
\end{subfigure}
\begin{subfigure}{.066\textwidth}
  \includegraphics[width=1\linewidth]{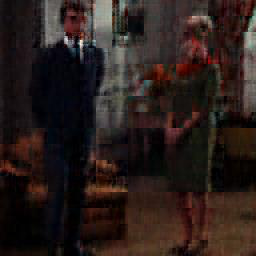}
\end{subfigure}

\vspace{1mm}

\begin{subfigure}{.066\textwidth}
  \includegraphics[width=1\linewidth]{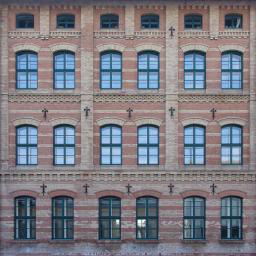}
\end{subfigure}
\begin{subfigure}{.066\textwidth}
  \includegraphics[width=1\linewidth]{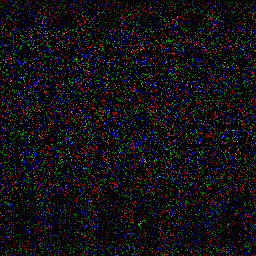}
\end{subfigure}
\begin{subfigure}{.066\textwidth}
  \includegraphics[width=1\linewidth]{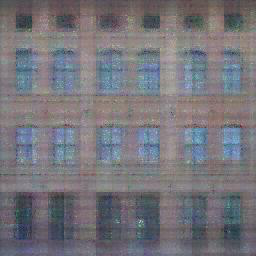}
\end{subfigure}
\begin{subfigure}{.066\textwidth}
  \includegraphics[width=1\linewidth]{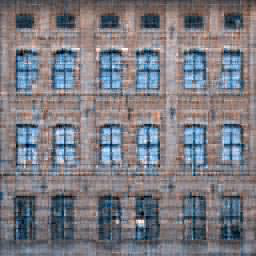}
\end{subfigure}
\begin{subfigure}{.066\textwidth}
  \includegraphics[width=1\linewidth]{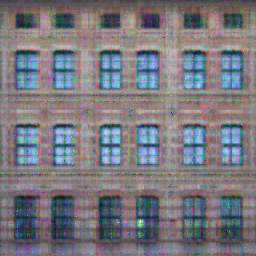}
\end{subfigure}
\begin{subfigure}{.066\textwidth}
  \includegraphics[width=1\linewidth]{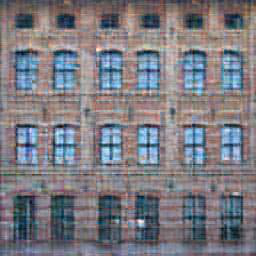}
\end{subfigure}
\begin{subfigure}{.066\textwidth}
  \includegraphics[width=1\linewidth]{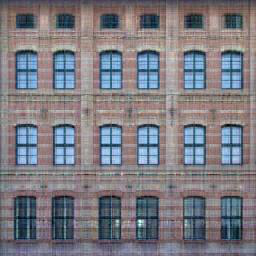}
\end{subfigure}
\begin{subfigure}{.066\textwidth}
  \includegraphics[width=1\linewidth]{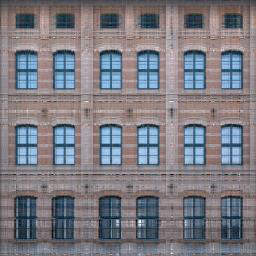}
\end{subfigure}
\begin{subfigure}{.066\textwidth}
  \includegraphics[width=1\linewidth]{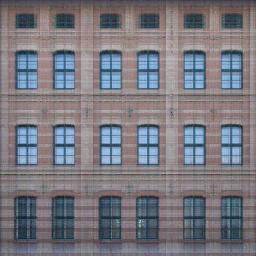}
\end{subfigure}
\begin{subfigure}{.066\textwidth}
  \includegraphics[width=1\linewidth]{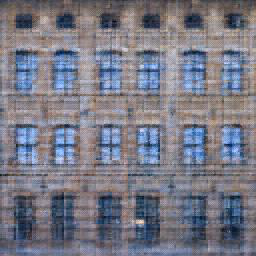}
\end{subfigure}
\begin{subfigure}{.066\textwidth}
  \includegraphics[width=1\linewidth]{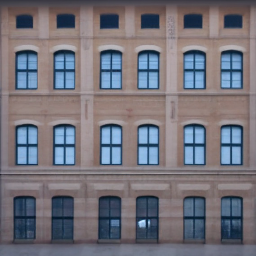}
\end{subfigure}
\begin{subfigure}{.066\textwidth}
  \includegraphics[width=1\linewidth]{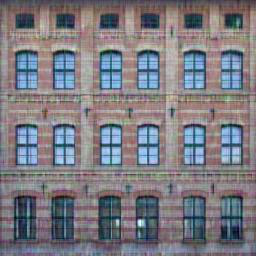}
\end{subfigure}
\begin{subfigure}{.066\textwidth}
  \includegraphics[width=1\linewidth]{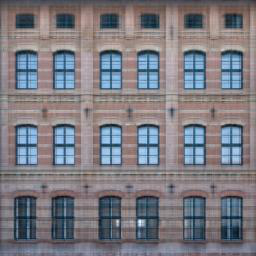}
\end{subfigure}
\begin{subfigure}{.066\textwidth}
  \includegraphics[width=1\linewidth]{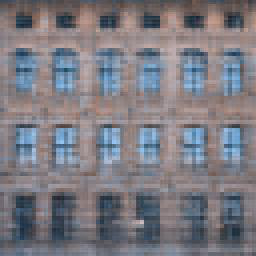}
\end{subfigure}

\vspace{1mm}

\begin{subfigure}{.066\textwidth}
  \includegraphics[width=1\linewidth]{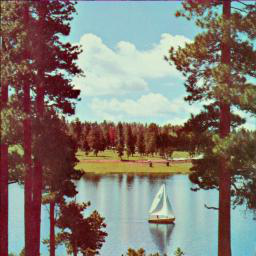}
\end{subfigure}
\begin{subfigure}{.066\textwidth}
  \includegraphics[width=1\linewidth]{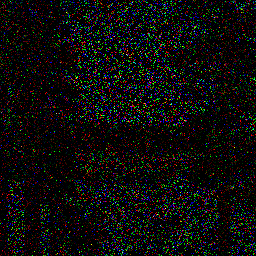}
\end{subfigure}
\begin{subfigure}{.066\textwidth}
  \includegraphics[width=1\linewidth]{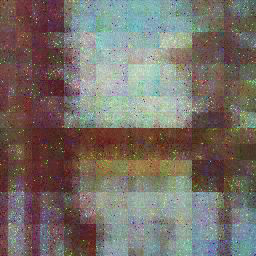}
\end{subfigure}
\begin{subfigure}{.066\textwidth}
  \includegraphics[width=1\linewidth]{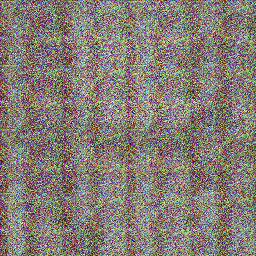}
\end{subfigure}
\begin{subfigure}{.066\textwidth}
  \includegraphics[width=1\linewidth]{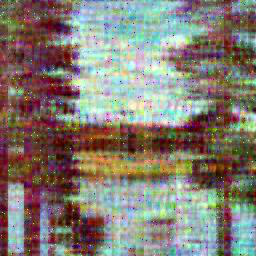}
\end{subfigure}
\begin{subfigure}{.066\textwidth}
  \includegraphics[width=1\linewidth]{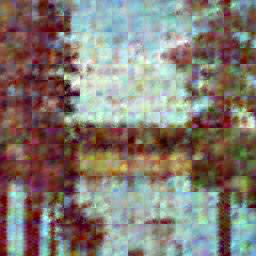}
\end{subfigure}
\begin{subfigure}{.066\textwidth}
  \includegraphics[width=1\linewidth]{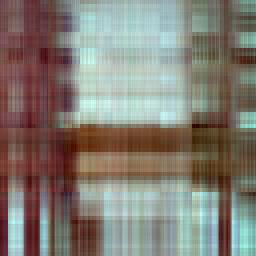}
\end{subfigure}
\begin{subfigure}{.066\textwidth}
  \includegraphics[width=1\linewidth]{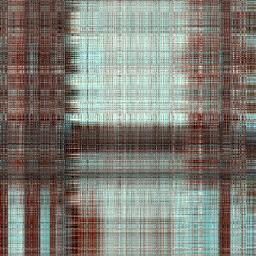}
\end{subfigure}
\begin{subfigure}{.066\textwidth}
  \includegraphics[width=1\linewidth]{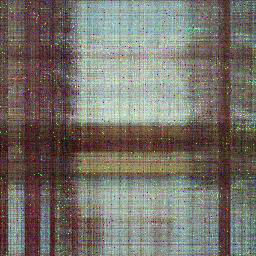}
\end{subfigure}
\begin{subfigure}{.066\textwidth}
  \includegraphics[width=1\linewidth]{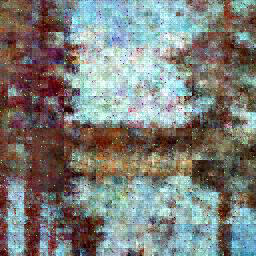}
\end{subfigure}
\begin{subfigure}{.066\textwidth}
  \includegraphics[width=1\linewidth]{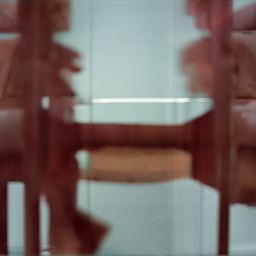}
\end{subfigure}
\begin{subfigure}{.066\textwidth}
  \includegraphics[width=1\linewidth]{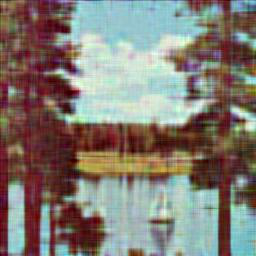}
\end{subfigure}
\begin{subfigure}{.066\textwidth}
  \includegraphics[width=1\linewidth]{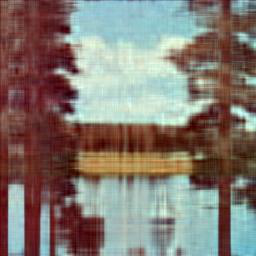}
\end{subfigure}
\begin{subfigure}{.066\textwidth}
  \includegraphics[width=1\linewidth]{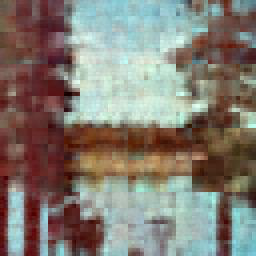}
\end{subfigure}

\vspace{1mm}

\begin{subfigure}{.066\textwidth}
  \includegraphics[width=1\linewidth]{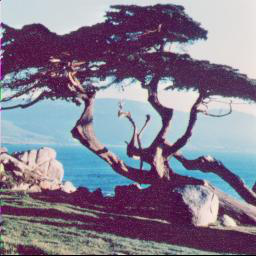}
\end{subfigure}
\begin{subfigure}{.066\textwidth}
  \includegraphics[width=1\linewidth]{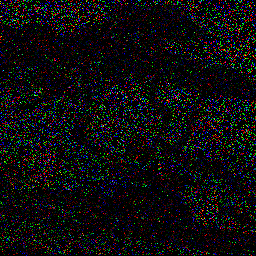}
\end{subfigure}
\begin{subfigure}{.066\textwidth}
  \includegraphics[width=1\linewidth]{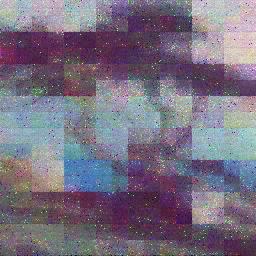}
\end{subfigure}
\begin{subfigure}{.066\textwidth}
  \includegraphics[width=1\linewidth]{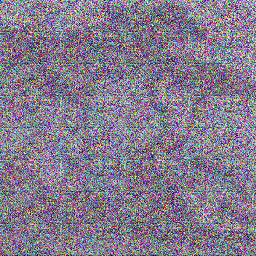}
\end{subfigure}
\begin{subfigure}{.066\textwidth}
  \includegraphics[width=1\linewidth]{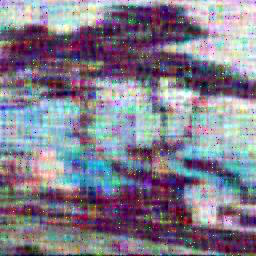}
\end{subfigure}
\begin{subfigure}{.066\textwidth}
  \includegraphics[width=1\linewidth]{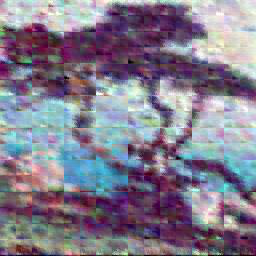}
\end{subfigure}
\begin{subfigure}{.066\textwidth}
  \includegraphics[width=1\linewidth]{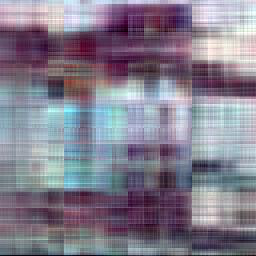}
\end{subfigure}
\begin{subfigure}{.066\textwidth}
  \includegraphics[width=1\linewidth]{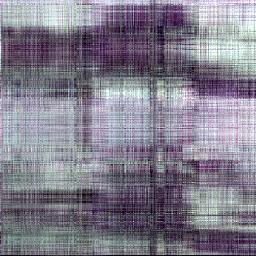}
\end{subfigure}
\begin{subfigure}{.066\textwidth}
  \includegraphics[width=1\linewidth]{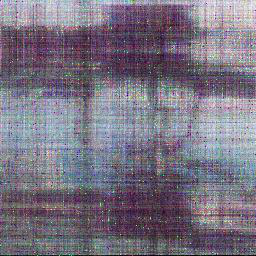}
\end{subfigure}
\begin{subfigure}{.066\textwidth}
  \includegraphics[width=1\linewidth]{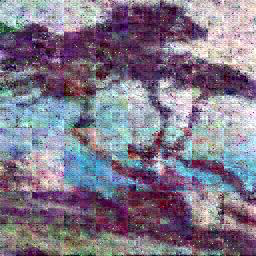}
\end{subfigure}
\begin{subfigure}{.066\textwidth}
  \includegraphics[width=1\linewidth]{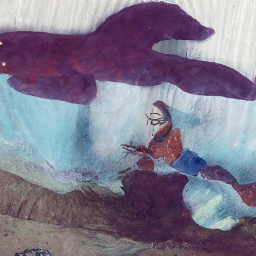}
\end{subfigure}
\begin{subfigure}{.066\textwidth}
  \includegraphics[width=1\linewidth]{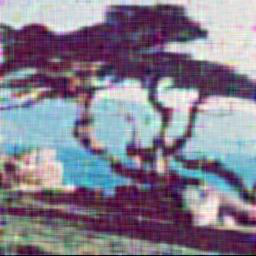}
\end{subfigure}
\begin{subfigure}{.066\textwidth}
  \includegraphics[width=1\linewidth]{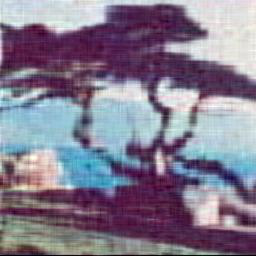}
\end{subfigure}
\begin{subfigure}{.066\textwidth}
  \includegraphics[width=1\linewidth]{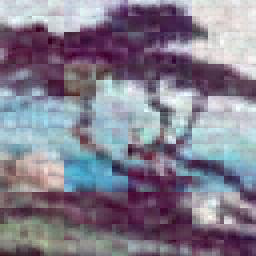}
\end{subfigure}
\caption{Visual effects of the image completion experiments, from top to bottom: recovered 'couple' and 'facade' images under $90\%$ missing rate and no noise, recovered 'sailboat' and 'tree' images under $90\%$ missing rate and {$10\%$ salt-and-pepper noise}; from left to right: original images, observed images, recovered images by SiLRTC-TT, TMAC-TT, TTC-TV, PTTC, TR-VBI, STTO, FBCP, FaLRTC, {DPS,} GraphTT-opt, GraphTT-VI, and GrphTT-fold, respectively.}
\label{fig:imgcompletion}
\end{figure*}

\subsubsection{Random missing elements}
Firstly, images with 90 percent random missing entries are tested. {Two cases are considered: one without noise and the other with $10\%$ salt-and-pepper noise \cite{ng2006salt}, which is used to model outliers.} {These constitute challenging conditions for recovery; in particular, even without missing entries, the noise alone reduces the image SNR to approximately 14dB.} Some original and observed images in no noise case can be seen in the left two columns of Fig. \ref{fig:imgcompletion}. For the proposed algorithms, the Laplacian is generated using (\ref{eqn:graphdef}). The first two weighting matrices are with elements $\bm{A}_{i,j}^{(d)} = \text{exp}(|i-j|^2)$, and the third one is set as an identity matrix. The reason is that the spatial smoothness only exhibits in the columns and rows of an image, but can be barely found among the RGB layers. Similarly, for GraphTT-fold, the same weighting matrix is applied on the first two modes only, as folding brings pixels from different regions into different dimensions, making it challenging to establish correlations between pixels in higher dimensions.
The initial ranks are set as $[1,64,3,1]$ for both GraphTT-opt and GraphTT-VI. For GraphTT-opt, $\beta_0$ is set as $2$ for the clean data and $100$ for the noisy data.

\begin{table*}[!tb]
\caption{Performance of image completion under character mask with noise variance 0.01.}
\footnotesize
\centering
\scalebox{0.7}{
\begin{tabular}{ l |m{1.5em} m{2em} |m{1.5em} m{2em} |m{1.5em} m{2em} |m{1.5em} m{2em} |m{1.5em} m{2em} |m{1.5em} m{2em} |m{1.5em} m{2em} |m{1.5em} m{2em} |m{1.5em} m{2em} |m{1.5em} m{2em} |m{1.5em} m{2em} |m{2em} m{2.5em}}
 \hline
 \quad &  \multicolumn{2}{c|}{SiLRTC-TT} & \multicolumn{2}{c|}{TMAC-TT}  & \multicolumn{2}{c|}{TTC-TV} &\multicolumn{2}{c|}{PTTC} &\multicolumn{2}{c|}{TR-VBI} & \multicolumn{2}{c|}{STTO} &  \multicolumn{2}{c|}{FBCP}  & \multicolumn{2}{c|}{FaLRTC}  & \multicolumn{2}{c|}{DPS}  & \multicolumn{2}{c|}{GraphTT-opt} & \multicolumn{2}{c|}{GraphTT-VI} & \multicolumn{2}{c}{GraphTT-fold} \\   %  \multicolumn{2}{c|}{\begin{tabular}{@{}c@{}}Proposed \\ Algorithm\end{tabular}} &\\
\textbf{} & PSNR & SSIM & PSNR & SSIM & PSNR & SSIM & PSNR & SSIM & PSNR & SSIM & PSNR & SSIM & PSNR & SSIM & PSNR & SSIM & PSNR & SSIM & PSNR & SSIM & PSNR & SSIM & PSNR & SSIM \\
\hline
airplane & 19.50 & 0.473 & 19.37 & 0.471 & 19.48 & 0.471 & 25.43 & 0.722 & 24.41 & 0.607 & 19.31 & 0.459 & 24.41 & 0.613 & 19.41 & 0.465 & \underline{25.89} & 0.739 & 25.22 & 0.674 & \textbf{26.05} & \underline{0.735} & {25.82} & \textbf{0.760}\\
baboon & 20.11 & \underline{0.636} & 19.92 & 0.630 & 20.11 & 0.635 & 22.49 & 0.591 & 22.26 & 0.582 & 19.92 & 0.622 & 22.42 & 0.607 & 20.04 & 0.631 & 21.52 & 0.366 & \underline{23.00} & \textbf{0.637} & \textbf{23.16} & 0.623 & 22.74 & 0.580\\
barbara & 20.08 & 0.540 & 19.84 & 0.529 & 19.94 & 0.532 & 25.96 & 0.731 & 24.79 & 0.665 & 19.87 & 0.526 & 24.64 & 0.667 & 19.97 & 0.533 & 26.09 & 0.696 & \underline{26.35} & 0.737 & \textbf{26.40} & \underline{0.746} & 26.12 & \textbf{0.747}\\
couple & 21.44 & 0.427 & 21.48 & 0.430 & 21.45 & 0.427 & 26.68 & 0.644 & 26.89 & 0.640 & 21.41 & 0.426 & 26.49 & 0.612 & 21.45 & 0.427 & 27.43 & \textbf{0.739} & \underline{27.83} & \underline{0.720} & \textbf{27.92} & {0.709} & 27.42 & 0.690\\
facade & 20.09 & 0.679 & 19.83 & 0.664 & 20.30 & 0.696 & 25.22 & 0.787 & 26.78 & 0.824 & 19.92 & 0.666 & \underline{27.36} & 0.850 & 20.38 & 0.705 & 25.72 & 0.636 & 27.19 & \underline{0.859} & \textbf{28.24} & \textbf{0.869} & 25.31 & 0.770\\
goldhill & 20.03 & 0.559 & 19.89 & 0.554 & 19.97 & 0.554 & 25.37 & 0.707 & 24.77 & 0.664 & 19.96 & 0.552 & 24.76 & 0.667 & 19.97 & 0.556 & 24.69 & 0.548 & \underline{26.24} & \textbf{0.742} & \textbf{26.34} & \underline{0.734} & 25.83 & 0.707\\
house & 20.07 & 0.408 & 19.91 & 0.399 & 20.05 & 0.406 & 27.50 & 0.727 & 26.25 & 0.633 & 19.94 & 0.395 & 25.81 & 0.620 & 20.02 & 0.406 & \textbf{28.86} & \textbf{0.764} & 27.20 & 0.677 & \underline{28.16} & {0.758} & {27.53} & \underline{0.759}\\
jellybeans & 19.06 & 0.328 & 18.91 & 0.324 & 19.10 & 0.333 & 28.09 & 0.798 & 27.36 & 0.795 & 18.91 & 0.315 & 25.97 & 0.629 & 19.04 & 0.327 & \textbf{31.63} & \textbf{0.921} & 26.54 & 0.641 & \underline{29.06} & {0.857} & {28.46} & \underline{0.884}\\
peppers & 19.66 & 0.488 & 19.31 & 0.473 & 19.46 & 0.480 & 25.53 & 0.746 & 24.13 & 0.664 & 19.46 & 0.477 & 23.44 & 0.635 & 19.45 & 0.477 & \textbf{26.48} & 0.756 & {25.60} & 0.730 & \underline{25.71} & \underline{0.766} & 25.37 & \textbf{0.779}\\
sailboat & 19.83 & 0.557 & 19.63 & 0.547 & 19.78 & 0.553 & 24.22 & 0.711 & 23.61 & 0.651 & 19.75 & 0.550 & 23.64 & 0.661 & 19.86 & 0.557 & 23.47 & 0.664 & \underline{24.65} & 0.722 & \textbf{24.86} & \textbf{0.738} & 24.31 & \underline{0.733}\\
splash & 20.50 & 0.419 & 20.36 & 0.414 & 20.50 & 0.421 & 27.17 & 0.705 & 26.79 & 0.665 & 20.46 & 0.418 & 26.42 & 0.653 & 20.52 & 0.421 & \textbf{29.27} & \textbf{0.819} & 27.12 & 0.706 & \underline{28.26} & \underline{0.771} & {27.52} & {0.757}\\
tree & 19.91 & 0.568 & 19.55 & 0.557 & 19.70 & 0.560 & 24.54 & 0.705 & 23.58 & 0.639 & 19.78 & 0.557 & 23.31 & 0.636 & 19.76 & 0.560 & 23.88 & 0.660 & \underline{24.64} & 0.704 & \textbf{24.98} & \underline{0.728} & 24.58 & \textbf{0.730}\\
\hline
\end{tabular}}
\label{table:lettermissing}
\end{table*}

\begin{figure*}[!tb]
    \centering
    \begin{subfigure}{0.135\linewidth}
        \centering
        \captionsetup{justification=centering}
        \includegraphics[width=0.98\linewidth]{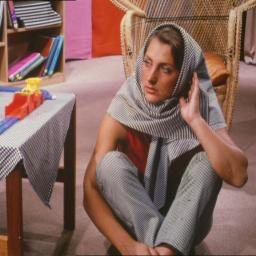}
        \caption{Original}
        \label{subfig:originalbarbara}
    \end{subfigure}
    \begin{subfigure}{0.135\linewidth}
        \centering
        \captionsetup{justification=centering}
        \includegraphics[width=0.98\linewidth]{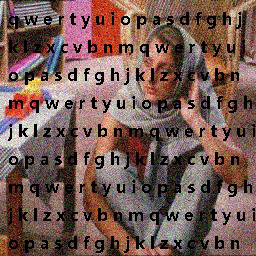}  
        \caption{Observed}
        \label{subfig:wordmissingbar}
    \end{subfigure}
    \begin{subfigure}{0.135\linewidth}
        \centering
        \captionsetup{justification=centering}
        \includegraphics[width=0.98\linewidth]{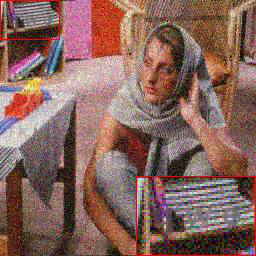}  
        \caption{SiLRTC-TT}
        % \label{subfig:reclettervittg}
    \end{subfigure}
    \begin{subfigure}{0.135\linewidth}
        \centering
        \captionsetup{justification=centering}
        \includegraphics[width=0.98\linewidth]{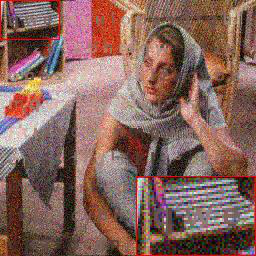}  
        \caption{TMAC-TT}
        % \label{subfig:reclettervittg}
    \end{subfigure}
    \begin{subfigure}{0.135\linewidth}
        \centering
        \captionsetup{justification=centering}
        \includegraphics[width=0.98\linewidth]{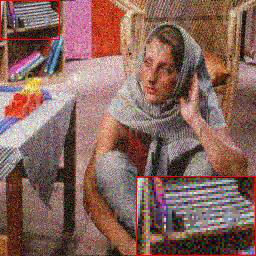}  
        \caption{TTC-TV}
        % \label{subfig:reclettervittg}
    \end{subfigure}
    \begin{subfigure}{0.135\linewidth}
        \centering
        \captionsetup{justification=centering}
        \includegraphics[width=0.98\linewidth]{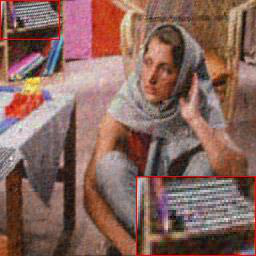}  
        \caption{PTTC}
        % \label{subfig:reclettervittg}
    \end{subfigure}
    \begin{subfigure}{0.135\linewidth}
        \centering
        \captionsetup{justification=centering}
        \includegraphics[width=0.98\linewidth]{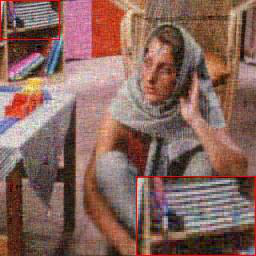}  
        \caption{TR-VBI}
        % \label{subfig:reclettervittg}
    \end{subfigure}
    
    \begin{subfigure}{0.135\linewidth}
        \centering
        \captionsetup{justification=centering}
        \includegraphics[width=0.98\linewidth]{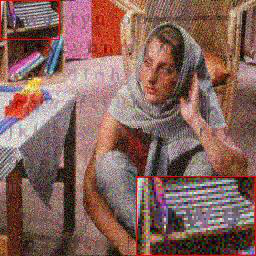}  
        \caption{STTO}
        % \label{subfig:reclettervittg}
    \end{subfigure}
    \begin{subfigure}{0.135\linewidth}
        \centering
        \captionsetup{justification=centering}
        \includegraphics[width=0.98\linewidth]{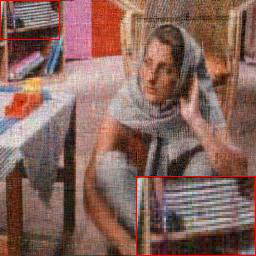}
        \caption{FBCP}
        % \label{subfig:reclettervittg}
    \end{subfigure}
    \begin{subfigure}{0.135\linewidth}
        \centering
        \captionsetup{justification=centering}
        \includegraphics[width=0.98\linewidth]{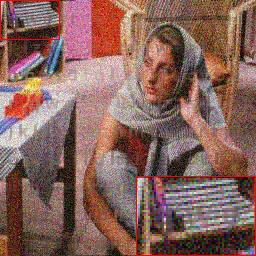}  
        \caption{FaLRTC}
        % \label{subfig:reclettervittg}
    \end{subfigure}
    \begin{subfigure}{0.135\linewidth}
        \centering
        \captionsetup{justification=centering}
        \includegraphics[width=0.98\linewidth]{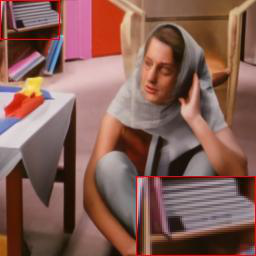}  
        \caption{DPS}
        % \label{subfig:reclettervittg}
    \end{subfigure}
    \begin{subfigure}{0.135\linewidth}
        \centering
        \captionsetup{justification=centering}
        \includegraphics[width=0.98\linewidth]{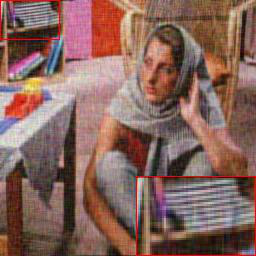}  
        \caption{GraphTT-opt}
        % \label{subfig:reclettervittg}
    \end{subfigure}
    \begin{subfigure}{0.135\linewidth}
        \centering
        \captionsetup{justification=centering}
        \includegraphics[width=0.98\linewidth]{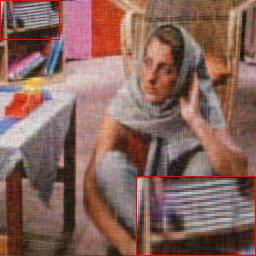}  
        \caption{GraphTT-VI}
        % \label{subfig:reclettervittg}
    \end{subfigure}
    \begin{subfigure}{0.135\linewidth}
        \centering
        \captionsetup{justification=centering}
        \includegraphics[width=0.98\linewidth]{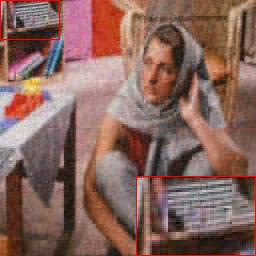}  
        \caption{{GraphTT-fold}}
        \label{subfig:reclettervittfold}
    \end{subfigure}
    \caption{Recovered 'barbara' images with character missing and noise variance $0.01$.}
    \label{fig:lettermissingbar}
\end{figure*}

The PSNR and SSIM of the recovered images without noise are listed in Table \ref{table:90mrclean}. {As can be seen, both GraphTT-opt and GraphTT-VI achieve comparable performance to the deep learning (DL) based method---DPS, and rank the top two among tensor-based methods.}
In general, the two proposed algorithms perform similarly, and GraphTT-VI achieves an average {$0.43\text{dB}$} higher PSNR and {$0.004$} higher SSIM than GraphTT-opt. In comparison, GraphTT-VI achieves an average {$1.13\text{dB}$} higher PSNR and {$0.037$} higher SSIM than PTTC, which is the third best {among tensor-based methods}. 
For the 'couple' and 'facade' image, GraphTT-VI achieves significantly better performance, surpassing the third best methods---PTTC and FBCP---by $1.38/1.57\text{dB}$ in PSNR and $0.035/0.040$ in SSIM, respectively. The superior performance of the proposed algorithms on these two images can also be visualized in the top two rows of Fig. \ref{fig:imgcompletion}. {In particular, while DPS produces images that appear sharp at first glance, it tends to generate details that not necessarily appear in the original images. For example, in the 'couple' image, the lady appears stronger and is dressed in brown, whereas in the ground truth, she is wearing dark green.} For GraphTT-fold, it performs worse than GraphTT-VI without folding. This is because the graph information cannot be fully utilized under folding. Such disadvantage is clearly illustrated in the recovered 'facade' image at the end of the second row in Fig. \ref{fig:imgcompletion}.

{Table \ref{table:90mrsaltpepper} presents the results under the challenging setting of $10\%$ salt-and-pepper noise. While all methods experience performance degradation, the proposed GraphTT-based methods are notably more robust. Specifically, GraphTT-opt and GraphTT-VI show only modest drops of 0.99/1.69dB in PSNR and 0.020/0.031 in SSIM, respectively. In contrast, non-GraphTT baselines suffer at least 3.10dB and 0.179 losses in PSNR and SSIM, respectively. This robustness is further illustrated in the visually cleaner results on the 'sailboat' and 'tree' images in Fig. \ref{fig:imgcompletion}. For GraphTT-fold, while it also exhibits some resistance to outliers, its inability to fully utilize graph structure across spatial dimensions leads to obvious block artifacts. One thing worth noting is that DPS performs poorly under outlier corruption; the noise severely disrupts its generative process, leading to unrealistic results---for example, hallucinating a person in the reconstructed 'tree' image.}

The average runtimes of all the compared methods on the $13$ images are listed in the first two rows in Table \ref{table:RGBtimerecord}. As can be seen, the proposed GraphTT-opt and GraphTT-VI achieve the overall best performance mentioned above at the cost of a moderate runtime. Specifically, GraphTT-VI takes twice to third times longer than GraphTT-opt, mainly due to the more complicated expectations in the VI update. However, it should be recognized that GraphTT-VI does not require any parameter tuning, which is practically helpful since there would not be any ground-truth images for computing the PSNR or SSIM. Even if a tuning strategy could be adopted without the ground truth, the exhaustive tuning may eventually end up with a longer runtime.

\begin{table*}[!tb]
\caption{Average runtime/s of all the compared methods in experiments on RGB images.}
\footnotesize
\centering
\scalebox{0.7}{
\begin{tabular}{ l m{5em}m{5em}m{4.5em}m{4.5em}m{4.5em}m{4.5em}m{4.5em}m{4.5em}m{4.5em}m{5.9em}m{5.5em}m{6.3em}}
 \hline
 &SiLRTC-TT & TMAC-TT & TTC-TV &PTTC & TR-VBI  & STTO &FBCP & FaLRTC & DPS & GraphTT-opt &GraphTT-VI & GraphTT-fold\\  
\hline
Random missing, clean & 32.62 & 63.75 & 27.97 & 641.22 & 148.40 & 315.66 & 24.87  & 22.46 & 269.32 & 43.30 & 94.50 & 82.73\\
Random missing, outlier  & 107.12 & 33.89 & 50.74 & 495.72 & 142.29 & 560.80 & 19.17 & 62.02 & 270.82 & 46.40 & 117.08 & 96.31 \\
Character missing, noisy  & 14.27 & 1.37 & 32.14 & 1405.28 & 717.65 & 311.48 & 46.51 & 2.86 & 270.17 &  13.00 & 118.13 & 91.87\\
\hline
\end{tabular}}
\label{table:RGBtimerecord}
\end{table*}
\label{subsec:yale}
\begin{figure*}[!tb]
    \centering
    \begin{subfigure}{0.14\linewidth}
        \centering
        \captionsetup{justification=centering}
        \includegraphics[width=0.98\linewidth]{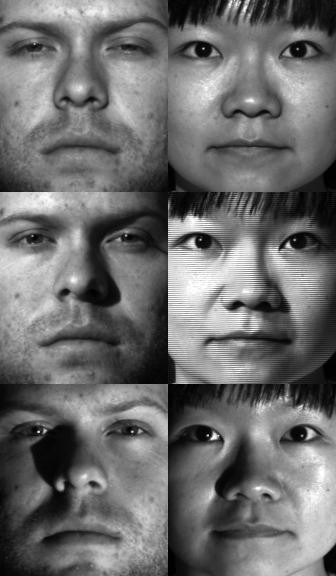}
        \caption{Original}
    \end{subfigure}
    \begin{subfigure}{0.14\linewidth}
        \centering
        \captionsetup{justification=centering}
        \includegraphics[width=0.98\linewidth]{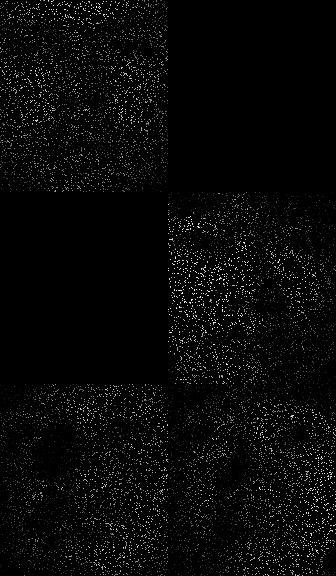}
        \caption{Observed}
        \label{subfig:yaleobserved}
    \end{subfigure}
    \begin{subfigure}{0.14\linewidth}
        \centering
        \captionsetup{justification=centering}
        \includegraphics[width=0.98\linewidth]{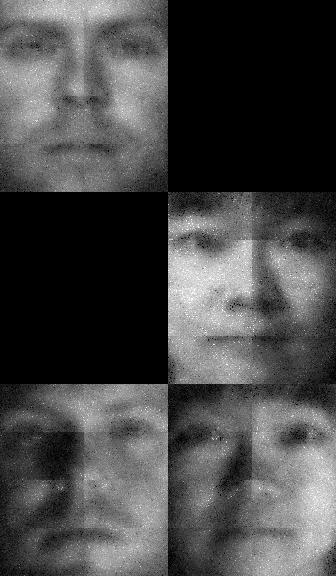}
        \caption{SiLRTC-TT}
        \label{subfig:yalefacesilrtctt}
    \end{subfigure}
    \begin{subfigure}{0.14\linewidth}
        \centering
        \captionsetup{justification=centering}
        \includegraphics[width=0.98\linewidth]{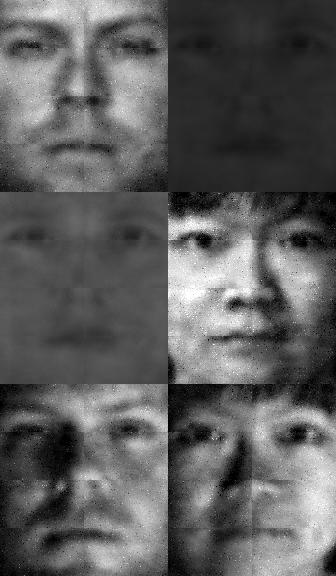}
        \caption{TMAC-TT}
    \end{subfigure}
    \begin{subfigure}{0.14\linewidth}
        \centering
        \captionsetup{justification=centering}
        \includegraphics[width=0.98\linewidth]{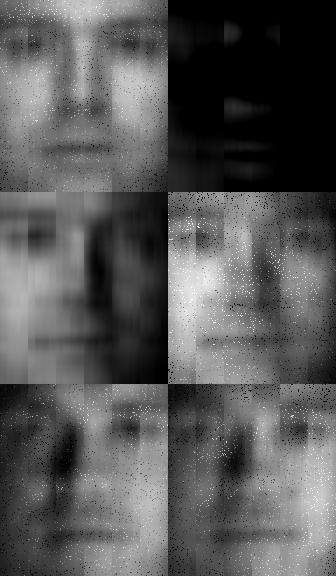}
        \caption{TTC-TV}
        \label{subfig:yalefacettctv}
    \end{subfigure}
    \begin{subfigure}{0.14\linewidth}
        \centering
        \captionsetup{justification=centering}
        \includegraphics[width=0.98\linewidth]{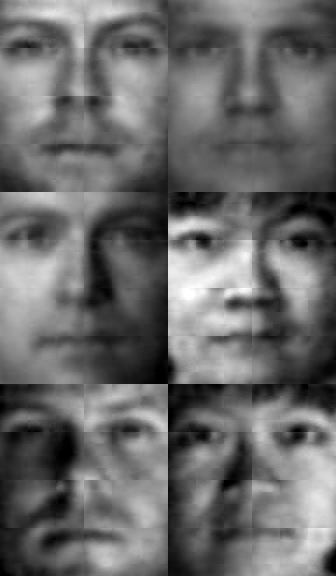}
        \caption{PTTC}
    \end{subfigure}
    
    \begin{subfigure}{0.14\linewidth}
        \centering
        \captionsetup{justification=centering}
        \includegraphics[width=0.98\linewidth]{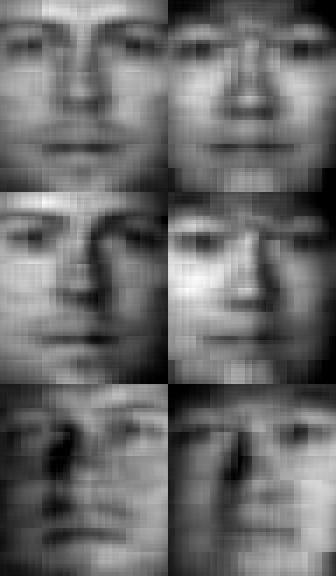}
        \caption{TR-VBI}
        \label{subfig:yalefacetrvbi}
    \end{subfigure}
    \begin{subfigure}{0.14\linewidth}
        \centering
        \captionsetup{justification=centering}
        \includegraphics[width=0.98\linewidth]{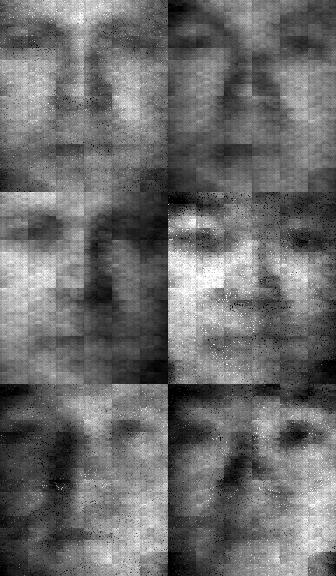}
        \caption{STTO}
        \label{subfig:yalefacestto}
    \end{subfigure}
    \begin{subfigure}{0.14\linewidth}
        \centering
        \captionsetup{justification=centering}
        \includegraphics[width=0.98\linewidth]{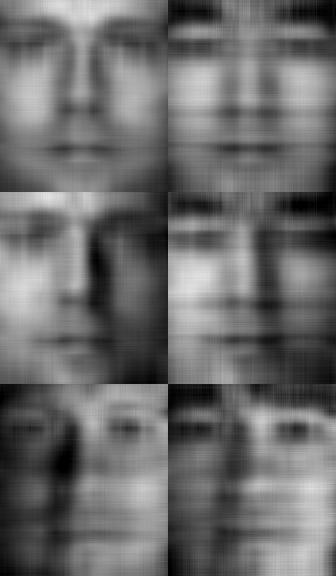}
        \caption{FBCP}
        \label{subfig:yalefacefbcp}
    \end{subfigure}
    \begin{subfigure}{0.14\linewidth}
        \centering
        \captionsetup{justification=centering}
        \includegraphics[width=0.98\linewidth]{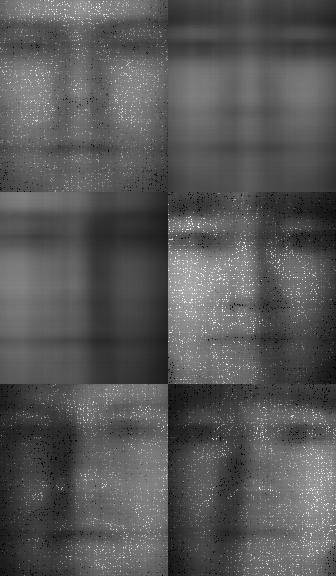}
        \caption{FaLRTC}
    \end{subfigure}
    \begin{subfigure}{0.14\linewidth}
        \centering
        \captionsetup{justification=centering}
        \includegraphics[width=0.98\linewidth]{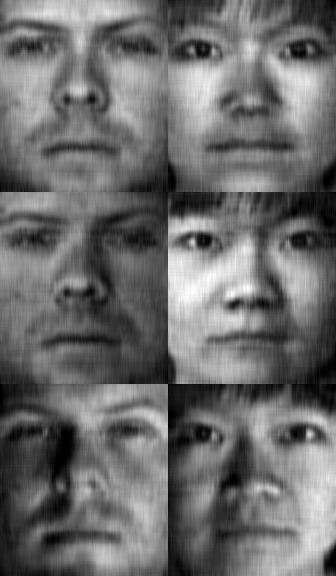}
        \caption{GraphTT-opt}
        \label{subfig:yaleGraphTTOPT}
    \end{subfigure}
    \begin{subfigure}{0.14\linewidth}
        \centering
        \captionsetup{justification=centering}
        \includegraphics[width=0.98\linewidth]{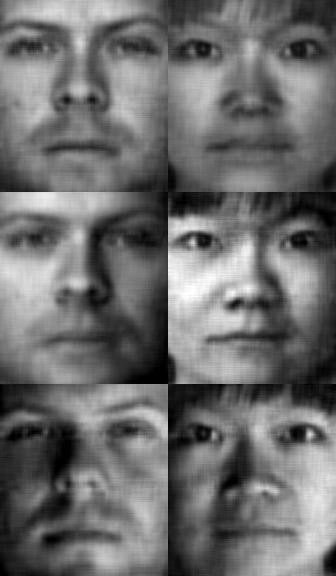}
        \caption{GraphTT-VI}
        \label{subfig:yaleGraphttvi}
    \end{subfigure}
    \caption{Recovered face data under 90\% random element missing rate, 20\% random pose missing rate and 0.01 noise variance.}
    \label{fig:YalefaceVisual}
\end{figure*}
\begin{figure}[!tb]
    \centering
    \includegraphics[width=0.95\linewidth]{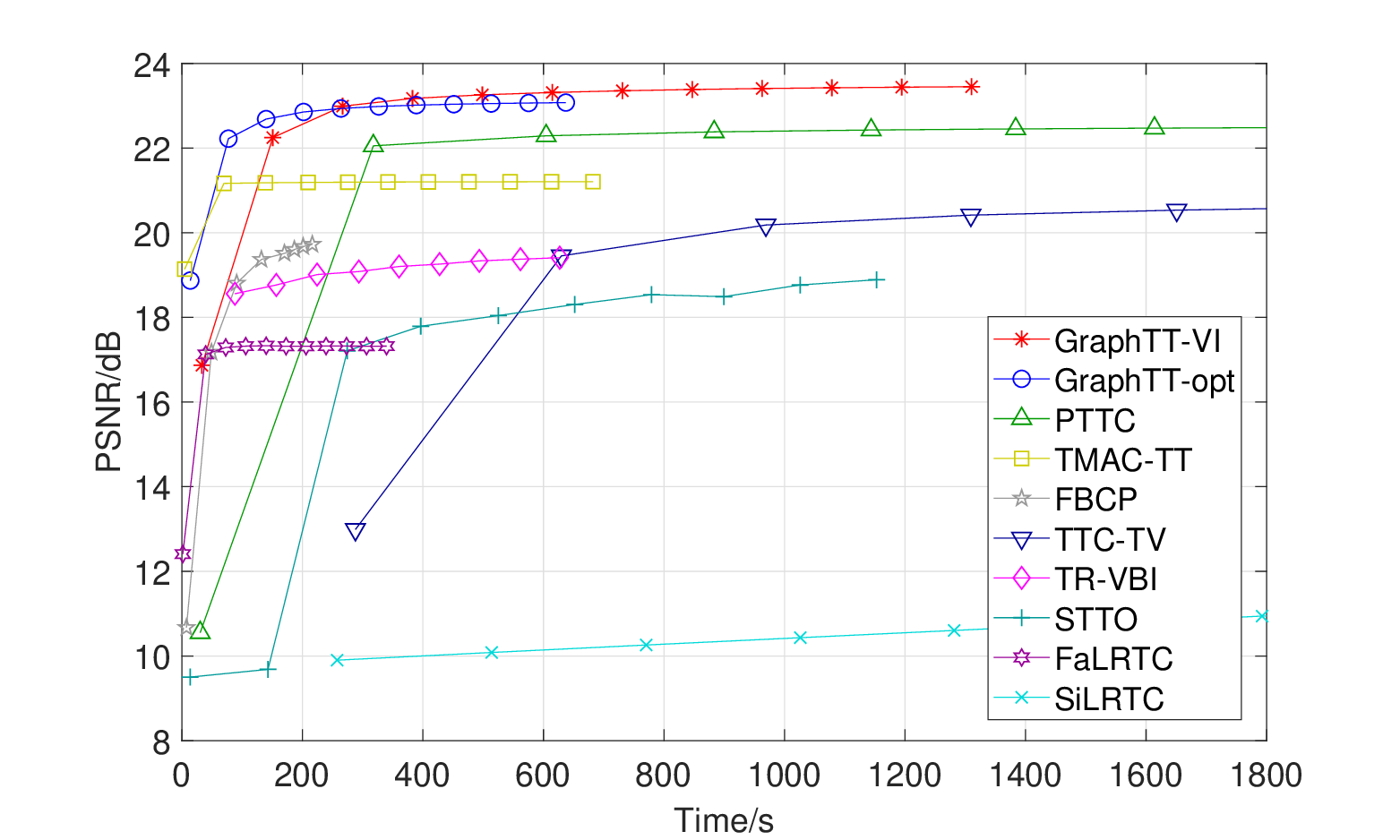}
    \caption{Performance of completion on YaleFace Dataset under 90\% random entry missing and 20\% random pose missing with noise variance 0.01.}
    \label{fig:yalecompletion}
\end{figure}

\subsubsection{Character missing patterns}
Character missing patterns are considered in this subsection. Every character corrupted image is further added with Gaussian noise with zero mean and variance $0.01$. An example of the observed image is shown in Fig. \ref{subfig:wordmissingbar}. The Laplacian matrices and initial ranks are set the same as in the previous experiments for both GraphTT-VI and GraphTT-opt, and $\beta_0$ is set as $100$ for GraphTT-opt, the same as that in the salt-and-pepper noise case.

Table. \ref{table:lettermissing} summarizes the performance of the compared methods. GraphTT-VI achieves the best overall performance, {with an PSNR $0.36$dB higher than the second-best in PSNR---DPS, and an SSIM $0.012$ higher than the second-best in SSIM---GraphTT-fold. In addition, GraphTT-opt, GraphTT-fold and DPS rank second to fourth overall, with their relative rankings varying across different images.}

Fig. \ref{fig:lettermissingbar} presents the visual effects of the recovered 'barbara' images. As seen in the bottom-right corner of each figure, {DPS, GraphTT-opt and GraphTT-VI are more effective at removing the overlaid character patterns.} {While DPS produces visually clear reconstructions at first glance, closer inspection reveals inconsistencies; e.g., slight facial distortions and overly smoothed textures in the background chair.}
For GraphTT-fold, though it reports competitive performance metrics, the recovered 'barbara' image exhibits noticeable inconsistency along the edges of the books. This is due to the block effects induced by tensor folding, which highlights the drawback of tensor folding even when enhanced with graph regularization.

The average runtimes of the compared algorithms are presented in the third row of Table \ref{table:RGBtimerecord}. As can be seen, the proposed methods cost moderate times among all competing algorithms. Specifically, SiLRTC-TT, TMAC-TT, FaLRTC and GraphTT-opt obviously take less time than that in the random missing cases, mainly because they converge faster due to more observed entries.

\subsection{YaleFace dataset}

In this subsection, the YaleFace dataset, which contains gray images of 38 people under 64 illumination conditions, each with size $192\times 168$, is adopted. Without loss of generality, images of 10 people are chosen, resulting in a data tensor with size 192×168×64×10. $90\%$ elements are randomly removed, and Gaussian noise with mean $0$ and variance $0.01$ is added to the dataset. Apart from that, $20\%$ of poses are further randomly removed. For the proposed algorithm, the Laplacian as in (\ref{eqn:graphdef}) is adopted. For the first $3$ TT cores, the weighting matrix is with element $\bm{A}_{i,j}=\text{exp}(|i-j|^2)$, and for the last TT core, the weighting matrix is set as an identity matrix. The reason why such a Laplacian matrix is adopted for the $3$rd TT core is that the pose image of the same person will not change much, even under different illuminations. The initial ranks for both the proposed algorithms are $[1,32,32,10,1]$, and $\beta_0$ is set as $100$ for GraphTT-opt. The visual effects of the $7$th, $16$th and $48$th poses of the $1$st and $5$th person are shown in Fig. \ref{subfig:yaleobserved}, in which the second pose of the man and the first pose of the woman are totally missing.

\begin{figure*}[!tb]
    \centering
    \begin{subfigure}{0.48\linewidth}
        \centering
        \captionsetup{justification=centering}
        \includegraphics[width=0.98\linewidth]{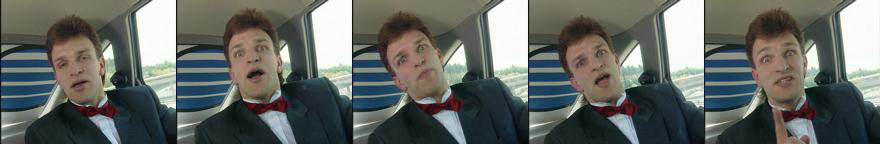}
    \end{subfigure}
    \begin{subfigure}{0.48\linewidth}
        \centering
        \captionsetup{justification=centering}
        \includegraphics[width=0.98\linewidth]{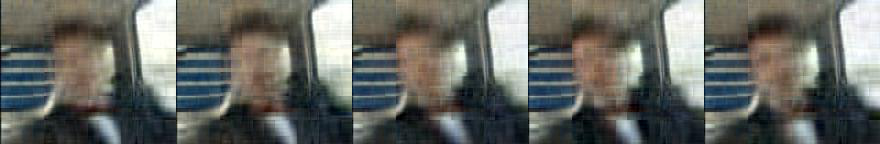}
    \end{subfigure}
    
    \vspace{1mm}
    
    \begin{subfigure}{0.48\linewidth}
        \centering
        \captionsetup{justification=centering}
        \includegraphics[width=0.98\linewidth]{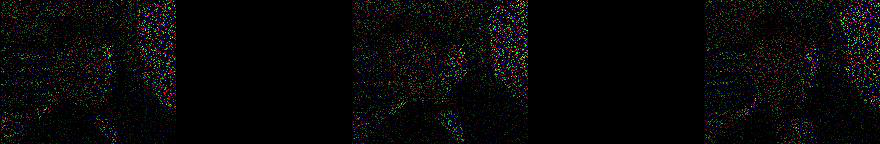}
    \end{subfigure}
    \begin{subfigure}{0.48\linewidth}
        \centering
        \captionsetup{justification=centering}
        \includegraphics[width=0.98\linewidth]{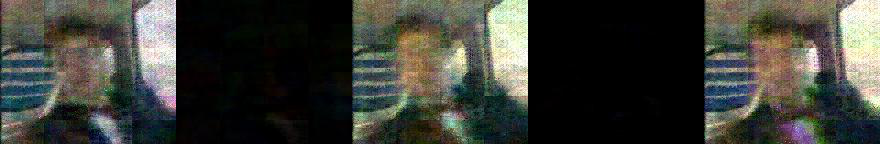}
    \end{subfigure}
    
    \vspace{1mm}
    
    \begin{subfigure}{0.48\linewidth}
        \centering
        \captionsetup{justification=centering}
        \includegraphics[width=0.98\linewidth]{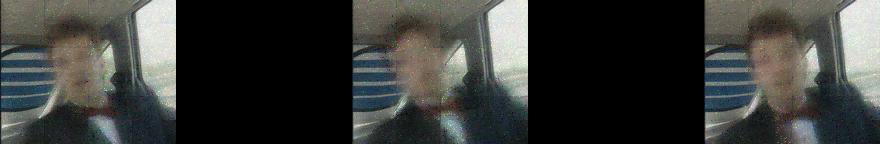}
    \end{subfigure}
    \begin{subfigure}{0.48\linewidth}
        \centering
        \captionsetup{justification=centering}
        \includegraphics[width=0.98\linewidth]{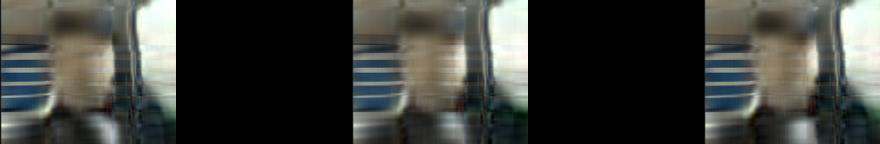}
    \end{subfigure}
    
    \vspace{1mm}
    
    \begin{subfigure}{0.48\linewidth}
        \centering
        \captionsetup{justification=centering}
        \includegraphics[width=0.98\linewidth]{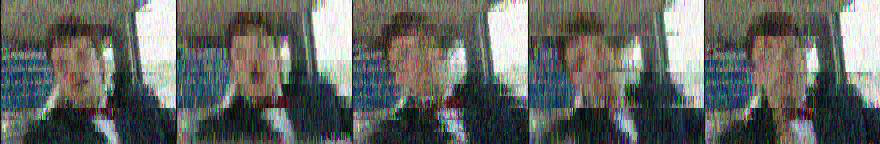}
    \end{subfigure}
    \begin{subfigure}{0.48\linewidth}
        \centering
        \captionsetup{justification=centering}
        \includegraphics[width=0.98\linewidth]{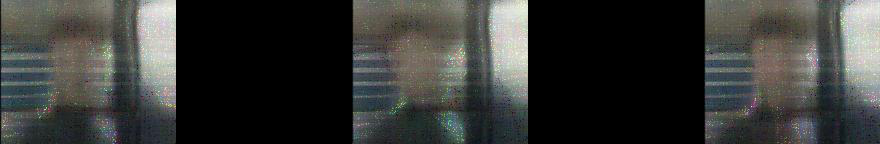}
    \end{subfigure}
    
    \vspace{1mm}
    
    \begin{subfigure}{0.48\linewidth}
        \centering
        \captionsetup{justification=centering}
        \includegraphics[width=0.98\linewidth]{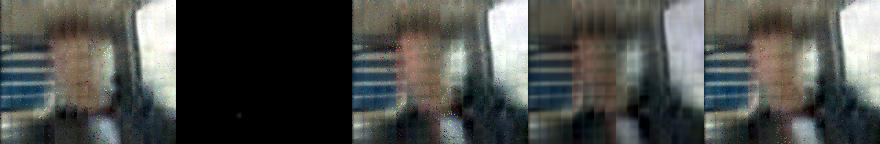}
    \end{subfigure}
    \begin{subfigure}{0.48\linewidth}
        \centering
        \captionsetup{justification=centering}
        \includegraphics[width=0.98\linewidth]{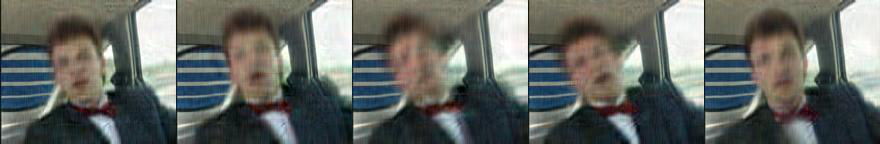}
    \end{subfigure}
    
    \vspace{1mm}
    
    \begin{subfigure}{0.48\linewidth}
        \centering
        \captionsetup{justification=centering}
        \includegraphics[width=0.98\linewidth]{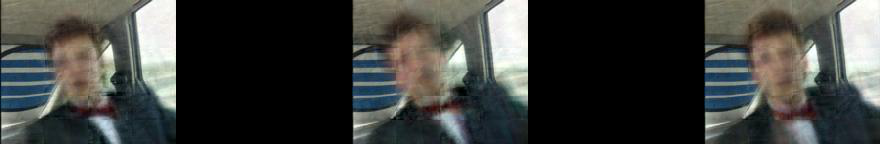}
    \end{subfigure}
    \begin{subfigure}{0.48\linewidth}
        \centering
        \captionsetup{justification=centering}
        \includegraphics[width=0.98\linewidth]{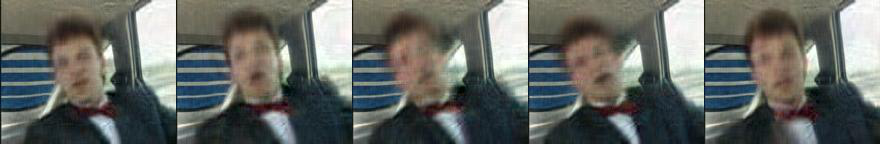}
    \end{subfigure}
    \caption{Recovered 'carphone' video under 90\% random element missing, 20\% random frame missing and noise variance 0.01. From top to bottom (left): the original images, the observed images, recovered images by SiLRTC-TT, TMAC-TT, TTC-TV and PTTC; (right): recovered images by TR-VBI, STTO, FBCP, FaLRTC, GraphTT-opt, and GraphTT-VI, respectively.}
    \label{fig:VideoVisual}
\end{figure*}
\begin{figure}[!tb]
    \centering
    \includegraphics[width=0.95\linewidth]{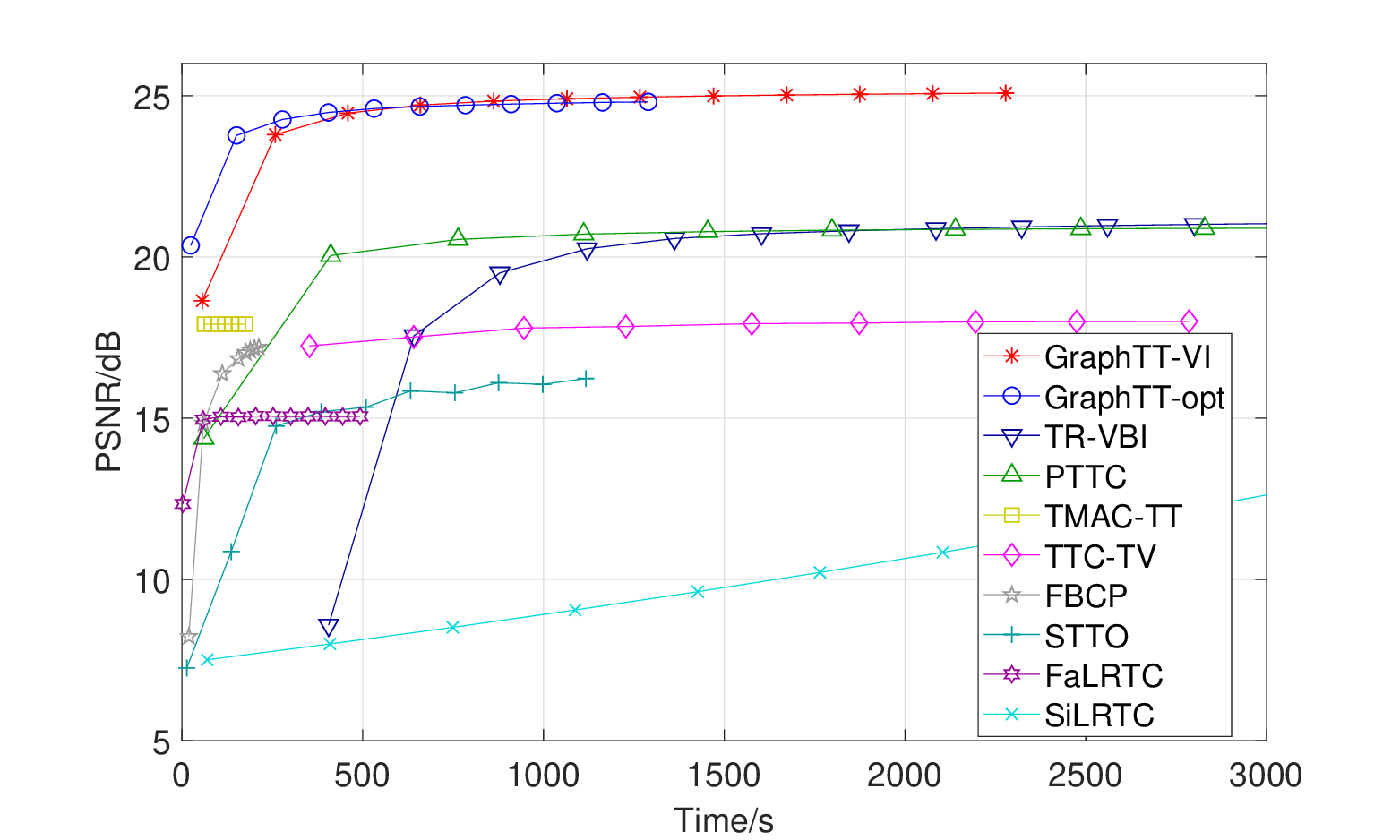}
    \caption{Performance of video completion under 90\% random entry missing and 20\% random pose missing with noise variance 0.01.}
    \label{fig:video}
\end{figure}

The PSNR of various methods w.r.t. runtime is presented in Fig. \ref{fig:yalecompletion}. As can be seen, since about $250s$, GraphTT-VI and GraphTT-opt keep the highest and second highest PSNR. Their good performance can also be observed from the visual effects in Fig. \ref{fig:YalefaceVisual}, which shows the recovered face images after the algorithms converging. From Fig. \ref{fig:YalefaceVisual} it can be seen that only TMAC-TT, PTTR and the proposed methods recovered recognizable images. For the pose images that are totally missing, TMAC-TT fails to recover them. For PTTR, even though it tries to recover the missing pose and achieves the third highest PSNR, it wrongly borrows information from other people, leading to its top right image look like a man. In particular, the block effects are obviously seen for the methods combined with tensor folding, as shown in Fig. \ref{subfig:yalefacesilrtctt}-\ref{subfig:yalefacestto}. Additionally, due to the heavy memory consumption caused by the update on a whole TT core, the initial ranks for TTC-TV and TR-VBI are bounded by $10$, which is the highest value possible for them to run without exceeding the memory limitation (32GB). However, the ranks are too small to recover the details of the data, leading to blurred face images shown in Fig. \ref{subfig:yalefacettctv} and \ref{subfig:yalefacetrvbi}.

\subsection{Video Completion}
\label{subsec:video}

In this subsection, we assess the performance of the proposed methods on video completion. A color video with size $144\times 176 \times 3 \times 382$ is tested with $90\%$ elements randomly missed, plus $20\%$ frames randomly missed, and Gaussian noise with mean $0$ and variance $0.01$ added. The parameter setting follows those in the YaleFace experiment, except that the $3$rd Laplacian matrix is set as an identity matrix while the $4$th is set as one that measures similarity between pixels. The reason is that there is no particular relations between RGB pixels, but for nearby time frames, they tend to be similar with each other. The $44$th, $64$th, $84$th, $104$th and $124$th frames of the video are presented in the second line of Fig. \ref{fig:VideoVisual}, among which the $64$th and $104$th frames are totally missing under observation.

The performance of the compared methods w.r.t. runtime is shown in Fig. \ref{fig:video}, with the visual effects of the corresponding recovered frames after algorithms converging shown in Fig. \ref{fig:VideoVisual}. As can be seen from Fig. \ref{fig:video}, graphTT-VI and graphTT-opt keep the highest two PSNRs all the time, and achieve about $4$dB higher PSNR than the third best after convergence. From Fig. \ref{fig:VideoVisual}, it can be seen that the proposed methods recover videos with recognizable faces and expressions, while most other compared methods cannot. Though PTTC also generate recognizable faces for normal frames, they cannot handle cases when a whole frame is missing. On the other hand, though TMAC-TT and TR-VBI provide estimations of the missing frames, the recovered frames are hard to recognize, as those recovered by TMAC-TT are highly corrupted with noise, while those recovered by TR-VBI are blurry.

\section{Conclusion}
In this paper, a graph-regularized TT completion method was proposed for visual data completion without the need to fold a tensor. To overcome the high computational burden introduced by graph regularization without tensor folding, tensor core fibers were updated as the basic blocks under the BCD framework. Based on that, a probabilistic graph regularized TT model, which has the ability to automatically learn the TT ranks and the regularization parameters, was further proposed. Experiments on synthetic data showed that the proposed optimization-based method with fiber update performs similarly to core update, but is much more computationally efficient. Further experiments on image and video data showed the superiority of the proposed algorithms, especially for GraphTT-VI, which achieves the overall best performance compared to other state-of-the-art methods under different settings without the need to finetune parameters.

This paper partially answered the question of to fold or not to fold in TT completion: if the graph information is provided and put along each mode of the TT-format tensor, then in general not to fold would give better performance. However, a further question might be if the graph information could be provided among every two elements like that in (\ref{eqn:graphregu_intro}) and if the heavy computational burden could be overcome, then would not folding a tensor still a better option? This is a good topic for future work. 

While the proposed methods have demonstrated advancements in visual data completion, there are several directions worth exploring.
Firstly, in this paper, we only consider the local similarity, and it would be valuable to investigate the incorporation of more robust and effective structural information into the TT completion problem. Secondly, the complexity analysis at the end of Section \ref{sec:graphTTopt} and \ref{sec:VIdiscussion} reveals that both methods have a cubic complexity with respect to the number of data samples. Therefore it is worthy to study methods to reduce the complexity of the proposed methods for large-scale datasets, e.g., using stochastic optimization methods \cite{hoffman2013stochastic} or using approximate message passing to replace the matrix inverse \cite{rangan2019vector}.

\bibliographystyle{IEEEtran}
\bibliography{main}

\begin{thebibliography}{10}
\providecommand{\url}[1]{#1}
\csname url@samestyle\endcsname
\providecommand{\newblock}{\relax}
\providecommand{\bibinfo}[2]{#2}
\providecommand{\BIBentrySTDinterwordspacing}{\spaceskip=0pt\relax}
\providecommand{\BIBentryALTinterwordstretchfactor}{4}
\providecommand{\BIBentryALTinterwordspacing}{\spaceskip=\fontdimen2\font plus
\BIBentryALTinterwordstretchfactor\fontdimen3\font minus \fontdimen4\font\relax}
\providecommand{\BIBforeignlanguage}[2]{{%
\expandafter\ifx\csname l@#1\endcsname\relax
\typeout{** WARNING: IEEEtran.bst: No hyphenation pattern has been}%
\typeout{** loaded for the language `#1'. Using the pattern for}%
\typeout{** the default language instead.}%
\else
\language=\csname l@#1\endcsname
\fi
#2}}
\providecommand{\BIBdecl}{\relax}
\BIBdecl

\bibitem{zhou2022optimal}
Y.~Zhou, A.~R. Zhang, L.~Zheng, and Y.~Wang, ``Optimal high-order tensor {SVD} via tensor-train orthogonal iteration,'' \emph{IEEE Transactions on Information Theory}, vol.~68, no.~6, pp. 3991--4019, 2022.

\bibitem{liu2020low}
Y.~Liu, J.~Liu, and C.~Zhu, ``Low-rank tensor train coefficient array estimation for tensor-on-tensor regression,'' \emph{IEEE transactions on neural networks and learning systems}, vol.~31, no.~12, pp. 5402--5411, 2020.

\bibitem{sofuoglu2021multi}
S.~E. Sofuoglu and S.~Aviyente, ``Multi-branch tensor network structure for tensor-train discriminant analysis,'' \emph{IEEE Transactions on Image Processing}, vol.~30, pp. 8926--8938, 2021.

\bibitem{liu2022efficient}
D.~Liu, M.~D. Sacchi, and W.~Chen, ``Efficient tensor completion methods for 5d seismic data reconstruction: Low-rank tensor train and tensor ring,'' \emph{IEEE Transactions on Geoscience and Remote Sensing}, 2022.

\bibitem{baust2016combined}
M.~Baust, A.~Weinmann, M.~Wieczorek, T.~Lasser, M.~Storath, and N.~Navab, ``Combined tensor fitting and tv regularization in diffusion tensor imaging based on a riemannian manifold approach,'' \emph{IEEE transactions on medical imaging}, vol.~35, no.~8, pp. 1972--1989, 2016.

\bibitem{jiang2020framelet}
T.-X. Jiang, M.~K. Ng, X.-L. Zhao, and T.-Z. Huang, ``Framelet representation of tensor nuclear norm for third-order tensor completion,'' \emph{IEEE Transactions on Image Processing}, vol.~29, pp. 7233--7244, 2020.

\bibitem{oseledets2012solution}
I.~V. Oseledets and S.~V. Dolgov, ``Solution of linear systems and matrix inversion in the tt-format,'' \emph{SIAM Journal on Scientific Computing}, vol.~34, no.~5, pp. A2718--A2739, 2012.

\bibitem{zhao2016tensor}
Q.~Zhao, G.~Zhou, S.~Xie, L.~Zhang, and A.~Cichocki, ``Tensor ring decomposition,'' \emph{arXiv preprint arXiv:1606.05535}, 2016.

\bibitem{yuan2018high}
L.~Yuan, Q.~Zhao, and J.~Cao, ``High-order tensor completion for data recovery via sparse tensor-train optimization,'' in \emph{2018 IEEE international conference on acoustics, speech and signal processing (ICASSP)}.\hskip 1em plus 0.5em minus 0.4em\relax IEEE, 2018, pp. 1258--1262.

\bibitem{wang2017efficient}
W.~Wang, V.~Aggarwal, and S.~Aeron, ``Efficient low rank tensor ring completion,'' in \emph{Proceedings of the IEEE International Conference on Computer Vision}, 2017, pp. 5697--5705.

\bibitem{bengua2017efficient}
J.~A. Bengua, H.~N. Phien, H.~D. Tuan, and M.~N. Do, ``Efficient tensor completion for color image and video recovery: Low-rank tensor train,'' \emph{IEEE Transactions on Image Processing}, vol.~26, no.~5, pp. 2466--2479, 2017.

\bibitem{yu2020low}
J.~Yu, G.~Zhou, C.~Li, Q.~Zhao, and S.~Xie, ``Low tensor-ring rank completion by parallel matrix factorization,'' \emph{IEEE transactions on neural networks and learning systems}, vol.~32, no.~7, pp. 3020--3033, 2020.

\bibitem{huang2020robust}
H.~Huang, Y.~Liu, Z.~Long, and C.~Zhu, ``Robust low-rank tensor ring completion,'' \emph{IEEE Transactions on Computational Imaging}, vol.~6, pp. 1117--1126, 2020.

\bibitem{xu2021overfitting}
L.~Xu, L.~Cheng, N.~Wong, and Y.-C. Wu, ``Overfitting avoidance in tensor train factorization and completion: Prior analysis and inference,'' in \emph{2021 IEEE International Conference on Data Mining (ICDM)}.\hskip 1em plus 0.5em minus 0.4em\relax IEEE, 2021, pp. 1439--1444.

\bibitem{long2021bayesian}
Z.~Long, C.~Zhu, J.~Liu, and Y.~Liu, ``Bayesian low rank tensor ring for image recovery,'' \emph{IEEE Transactions on Image Processing}, vol.~30, pp. 3568--3580, 2021.

\bibitem{latorre2005image}
J.~I. Latorre, ``Image compression and entanglement,'' \emph{arXiv preprint quant-ph/0510031}, 2005.

\bibitem{ko2020fast}
C.-Y. Ko, K.~Batselier, L.~Daniel, W.~Yu, and N.~Wong, ``Fast and accurate tensor completion with total variation regularized tensor trains,'' \emph{IEEE Transactions on Image Processing}, 2020.

\bibitem{jiang2013graph}
B.~Jiang, C.~Ding, B.~Luo, and J.~Tang, ``Graph-laplacian pca: Closed-form solution and robustness,'' in \emph{Proceedings of the IEEE Conference on Computer Vision and Pattern Recognition (CVPR)}, 2013, pp. 3492--3498.

\bibitem{strahl2020scalable}
J.~Strahl, J.~Peltonen, H.~Mamitsuka, and S.~Kaski, ``Scalable probabilistic matrix factorization with graph-based priors,'' in \emph{Proceedings of the AAAI Conference on Artificial Intelligence}, vol.~34, no.~04, 2020, pp. 5851--5858.

\bibitem{CHEN2023108826}
Y.~Chen, L.~Cheng, and Y.-C. Wu, ``Bayesian low-rank matrix completion with dual-graph embedding: Prior analysis and tuning-free inference,'' \emph{Signal Processing}, vol. 204, p. 108826, 2023.

\bibitem{holtz2012alternating}
S.~Holtz, T.~Rohwedder, and R.~Schneider, ``The alternating linear scheme for tensor optimization in the tensor train format,'' \emph{SIAM Journal on Scientific Computing}, vol.~34, no.~2, pp. A683--A713, 2012.

\bibitem{cichocki2017tensor}
A.~Cichocki, A.-H. Phan, Q.~Zhao, N.~Lee, I.~Oseledets, M.~Sugiyama, D.~P. Mandic \emph{et~al.}, ``Tensor networks for dimensionality reduction and large-scale optimization: Part 2 applications and future perspectives,'' \emph{Foundations and Trends{\textregistered} in Machine Learning}, vol.~9, no.~6, pp. 431--673, 2017.

\bibitem{grasedyck2015alternating}
L.~Grasedyck, M.~Kluge, and S.~Kr{\"a}mer, ``Alternating least squares tensor completion in the tt-format,'' \emph{arXiv preprint arXiv:1509.00311}, 2015.

\bibitem{yu2021robust}
J.~Yu, G.~Zhou, W.~Sun, and S.~Xie, ``Robust to rank selection: Low-rank sparse tensor-ring completion,'' \emph{IEEE Transactions on Neural Networks and Learning Systems}, 2021.

\bibitem{yu2022graph}
Y.~Yu, G.~Zhou, N.~Zheng, Y.~Qiu, S.~Xie, and Q.~Zhao, ``Graph-regularized non-negative tensor-ring decomposition for multiway representation learning,'' \emph{IEEE Transactions on Cybernetics}, 2022.

\bibitem{sixteenoseledets2011tensor}
I.~V. Oseledets, ``Tensor-train decomposition,'' \emph{SIAM Journal on Scientific Computing}, vol.~33, no.~5, pp. 2295--2317, 2011.

\bibitem{chen2013simultaneous}
Y.-L. Chen, C.-T. Hsu, and H.-Y.~M. Liao, ``Simultaneous tensor decomposition and completion using factor priors,'' \emph{IEEE transactions on pattern analysis and machine intelligence}, vol.~36, no.~3, pp. 577--591, 2013.

\bibitem{shahid2016fast}
N.~Shahid, N.~Perraudin, V.~Kalofolias, G.~Puy, and P.~Vandergheynst, ``Fast robust pca on graphs,'' \emph{IEEE Journal of Selected Topics in Signal Processing}, vol.~10, no.~4, pp. 740--756, 2016.

\bibitem{paradkar2017graph}
M.~Paradkar and M.~Udell, ``Graph-regularized generalized low-rank models,'' in \emph{Proceedings of the IEEE Conference on Computer Vision and Pattern Recognition Workshops}, 2017, pp. 7--12.

\bibitem{tibshirani1996regression}
R.~Tibshirani, ``Regression shrinkage and selection via the lasso,'' \emph{Journal of the Royal Statistical Society Series B: Statistical Methodology}, vol.~58, no.~1, pp. 267--288, 1996.

\bibitem{candes2011robust}
E.~J. Cand{\`e}s, X.~Li, Y.~Ma, and J.~Wright, ``Robust principal component analysis?'' \emph{Journal of the ACM}, vol.~58, no.~3, pp. 1--37, 2011.

\bibitem{goldfarb2014robust}
D.~Goldfarb and Z.~Qin, ``Robust low-rank tensor recovery: Models and algorithms,'' \emph{SIAM Journal on Matrix Analysis and Applications}, vol.~35, no.~1, pp. 225--253, 2014.

\bibitem{xu2013block}
Y.~Xu and W.~Yin, ``A block coordinate descent method for regularized multiconvex optimization with applications to nonnegative tensor factorization and completion,'' \emph{SIAM Journal on imaging sciences}, vol.~6, no.~3, pp. 1758--1789, 2013.

\bibitem{wen2019nonconvex}
F.~Wen, R.~Ying, P.~Liu, and T.-K. Truong, ``Nonconvex regularized robust pca using the proximal block coordinate descent algorithm,'' \emph{IEEE Transactions on Signal Processing}, vol.~67, no.~20, pp. 5402--5416, 2019.

\bibitem{bolte2014proximal}
J.~Bolte, S.~Sabach, and M.~Teboulle, ``Proximal alternating linearized minimization for nonconvex and nonsmooth problems,'' \emph{Mathematical Programming}, vol. 146, no.~1, pp. 459--494, 2014.

\bibitem{deng2010graph}
D.~Cai, X.~He, J.~Han, and T.~S. Huang, ``Graph regularized nonnegative matrix factorization for data representation,'' \emph{IEEE Transactions on Pattern Analysis and Machine Intelligence}, vol.~33, no.~8, pp. 1548--1560, 2011.

\bibitem{twofourcheng2017probabilistic}
L.~Cheng, Y.-C. Wu, and H.~V. Poor, ``Probabilistic tensor canonical polyadic decomposition with orthogonal factors.'' \emph{IEEE Trans. Signal Processing}, vol.~65, no.~3, pp. 663--676, 2017.

\bibitem{cheng2020learning}
L.~Cheng, X.~Tong, S.~Wang, Y.-C. Wu, and H.~V. Poor, ``Learning nonnegative factors from tensor data: Probabilistic modeling and inference algorithm,'' \emph{IEEE Transactions on Signal Processing}, vol.~68, pp. 1792--1806, 2020.

\bibitem{twofivezhao2015bayesian}
Q.~Zhao, L.~Zhang, and A.~Cichocki, ``Bayesian sparse tucker models for dimension reduction and tensor completion,'' \emph{arXiv preprint arXiv:1505.02343}, 2015.

\bibitem{babacan2014bayesian}
S.~D. Babacan, S.~Nakajima, and M.~N. Do, ``Bayesian group-sparse modeling and variational inference,'' \emph{IEEE transactions on signal processing}, vol.~62, no.~11, pp. 2906--2921, 2014.

\bibitem{holtz2012manifolds}
S.~Holtz, T.~Rohwedder, and R.~Schneider, ``On manifolds of tensors of fixed tt-rank,'' \emph{Numerische Mathematik}, vol. 120, no.~4, pp. 701--731, 2012.

\bibitem{west1987scale}
M.~West, ``On scale mixtures of normal distributions,'' \emph{Biometrika}, vol.~74, no.~3, pp. 646--648, 1987.

\bibitem{murphy2012probabilistic}
K.~P. Murphy, \emph{Machine learning: a probabilistic perspective}.\hskip 1em plus 0.5em minus 0.4em\relax MIT press, 2012.

\bibitem{cheng2022towards}
L.~Cheng, Z.~Chen, Q.~Shi, Y.-C. Wu, and S.~Theodoridis, ``Towards flexible sparsity-aware modeling: Automatic tensor rank learning using the generalized hyperbolic prior,'' \emph{IEEE Transactions on Signal Processing}, vol.~70, pp. 1834--1849, 2022.

\bibitem{bishop2006pattern}
C.~M. Bishop, \emph{Pattern recognition and machine learning}.\hskip 1em plus 0.5em minus 0.4em\relax springer, 2006.

\bibitem{twothreezhao2015bayesian}
Q.~Zhao, L.~Zhang, and A.~Cichocki, ``Bayesian cp factorization of incomplete tensors with automatic rank determination,'' \emph{IEEE transactions on pattern analysis and machine intelligence}, vol.~37, no.~9, pp. 1751--1763, 2015.

\bibitem{liu2013tensor}
J.~Liu, P.~Musialski, P.~Wonka, and J.~Ye, ``Tensor completion for estimating missing values in visual data,'' \emph{IEEE Transactions on Pattern Analysis and Machine Intelligence}, vol.~35, no.~1, pp. 208--220, 2013.

\bibitem{chung2023diffusion}
H.~Chung, J.~Kim, M.~T. Mccann, M.~L. Klasky, and J.~C. Ye, ``Diffusion posterior sampling for general noisy inverse problems,'' in \emph{The Proceedings of the International Conference on Learning Representations (ICLR)}, 2023.

\bibitem{song2021scorebased}
Y.~Song, J.~Sohl-Dickstein, D.~Kingma, A.~Kumar, S.~Ermon, and B.~Poole, ``Score-based generative modeling through stochastic differential equations,'' \emph{International Conference on Learning Representations (ICLR)}, 2021.

\bibitem{dhariwal2021diffusion}
P.~Dhariwal and A.~Q. Nichol, ``Diffusion models beat {GAN}s on image synthesis,'' in \emph{Advances in Neural Information Processing Systems}, A.~Beygelzimer, Y.~Dauphin, P.~Liang, and J.~W. Vaughan, Eds., 2021.

\bibitem{deng2009imagenet}
J.~Deng, W.~Dong, R.~Socher, L.-J. Li, K.~Li, and L.~Fei-Fei, ``Imagenet: A large-scale hierarchical image database,'' in \emph{Proceedings of the IEEE Conference on Computer Vision and Pattern Recognition (CVPR)}.\hskip 1em plus 0.5em minus 0.4em\relax IEEE, 2009, pp. 248--255.

\bibitem{he2022masked}
K.~He, X.~Chen, S.~Xie, Y.~Li, P.~Doll{\'a}r, and R.~Girshick, ``Masked autoencoders are scalable vision learners,'' in \emph{Proceedings of the IEEE/CVF conference on computer vision and pattern recognition}, 2022, pp. 16\,000--16\,009.

\bibitem{Li2022MAT}
W.~Li, Z.~Lin, K.~Zhou, L.~Qi, Y.~Wang, and J.~Jia, ``Mat: Mask-aware transformer for large hole image inpainting,'' in \emph{2022 IEEE/CVF Conference on Computer Vision and Pattern Recognition (CVPR)}, 2022, pp. 10\,748--10\,758.

\bibitem{wang2004image}
Z.~Wang, A.~C. Bovik, H.~R. Sheikh, and E.~P. Simoncelli, ``Image quality assessment: from error visibility to structural similarity,'' \emph{IEEE transactions on image processing}, vol.~13, no.~4, pp. 600--612, 2004.

\bibitem{ng2006salt}
P.-E. Ng and K.-K. Ma, ``A switching median filter with boundary discriminative noise detection for extremely corrupted images,'' \emph{IEEE Transactions on Image Processing}, pp. 1506--1516, 2006.

\bibitem{hoffman2013stochastic}
M.~D. Hoffman, D.~M. Blei, C.~Wang, and J.~Paisley, ``Stochastic variational inference,'' \emph{Journal of Machine Learning Research}, 2013.

\bibitem{rangan2019vector}
S.~Rangan, P.~Schniter, and A.~K. Fletcher, ``Vector approximate message passing,'' \emph{IEEE Transactions on Information Theory}, vol.~65, no.~10, pp. 6664--6684, 2019.

\end{thebibliography}

\begin{IEEEbiography}[{\includegraphics[width=1in,height=1.25in,clip,keepaspectratio]{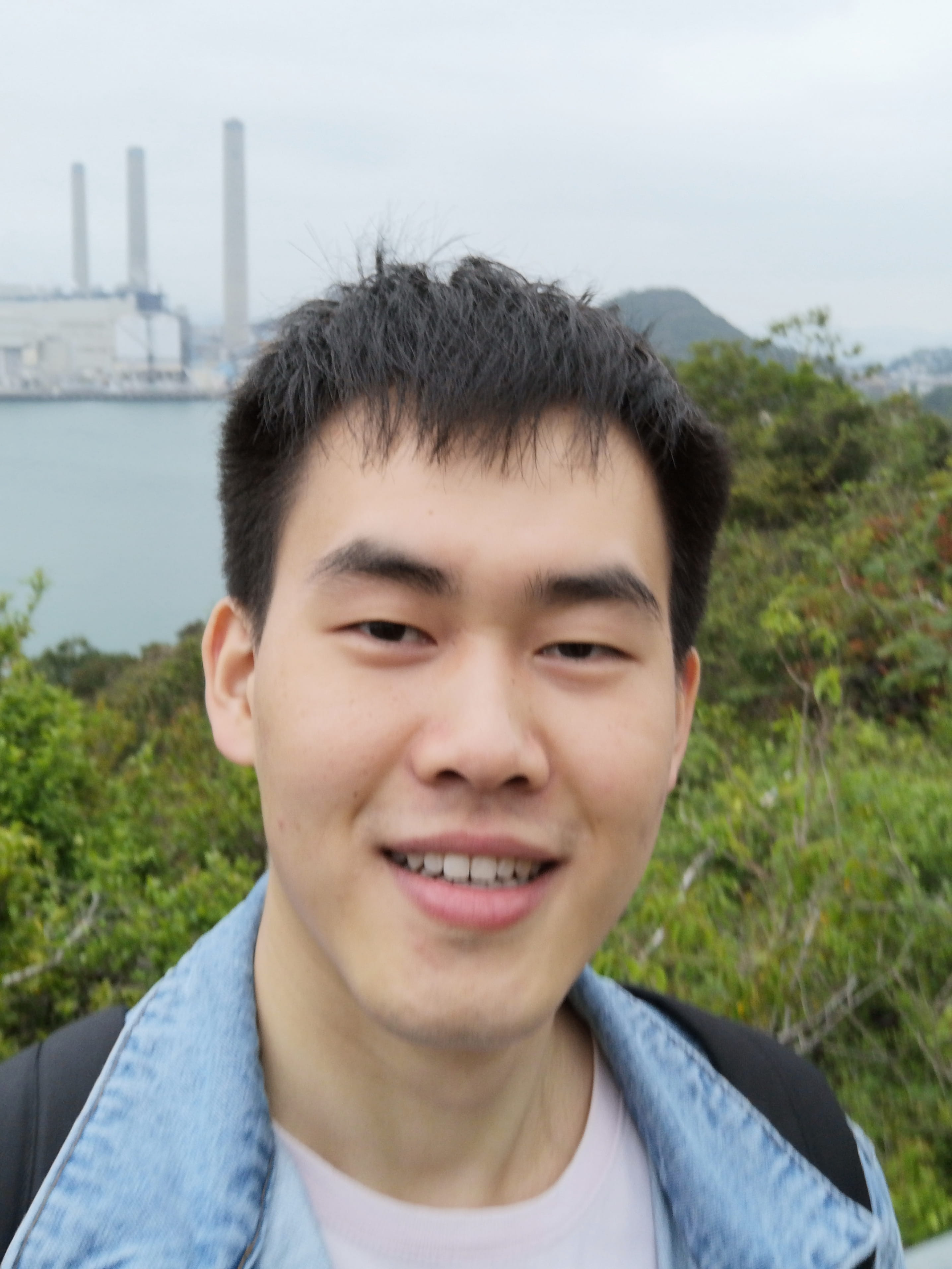}}]{Le Xu}
received the B.Eng. degree from Southeast University, Nanjing, China, in 2017. He is currently pursuing the Ph.D. degree at the University of Hong Kong. His research interests include tensor decomposition, Bayesian inference, and their applications in machine learning and wireless communication.
\end{IEEEbiography}

\begin{IEEEbiography}[{\includegraphics[width=1in,height=1.25in,clip,keepaspectratio]{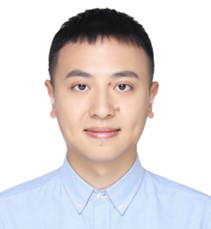}}]{Lei Cheng}
is an Assistant Professor (ZJU Young Professor) in the College of Information Science and Electronic Engineering at Zhejiang University, Hangzhou, China. He received the B.Eng. degree from Zhejiang University in 2013, and the Ph.D. degree from the University of Hong Kong in 2018. He was a research scientist in Shenzhen Research Institute of Big Data from 2018 to 2021. His research interests are in Bayesian machine learning for tensor data analytics, and interpretable machine learning for information systems.
\end{IEEEbiography}

\begin{IEEEbiography} 
[{\includegraphics[width=1in,height=1.25in,clip,keepaspectratio]{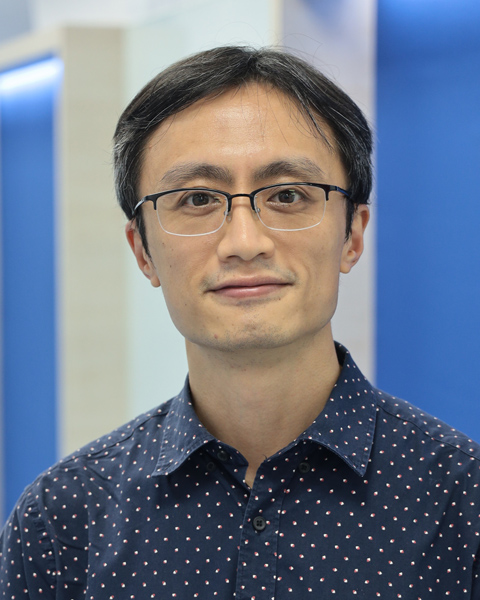}}]{Ngai Wong}  (SM, IEEE) received his B.Eng and Ph.D. in EEE from The University of Hong Kong (HKU), and he was a visiting scholar with Purdue University, West Lafayette, IN, in 2003. He is currently an Associate Professor with the Department of Electrical and Electronic Engineering at HKU. His research interests include electronic design automation (EDA), model order reduction, tensor algebra, linear and nonlinear modeling \& simulation, and compact neural network 
design.
\end{IEEEbiography}

\begin{IEEEbiography}[{\includegraphics[width=1in,height=1.25in,clip,keepaspectratio]{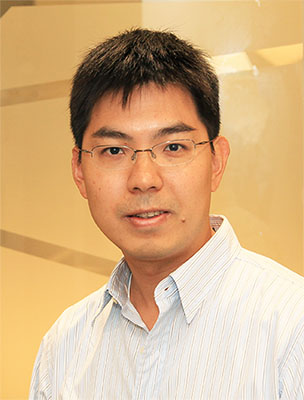}}]{Yik-Chung Wu} (SM, IEEE)
received the B.Eng. (EEE) and M.Phil. degrees from The University of Hong Kong (HKU) in 1998 and 2001, respectively, and the Ph.D. degree from Texas A\&M University, College Station, in 2005. From 2005 to 2006, he was with Thomson Corporate Research, Princeton, NJ, USA, as a Member of Technical Staff. Since 2006, he has been with HKU, where he is currently as an Associate Professor. He was a Visiting Scholar at Princeton University in Summers of 2015 and 2017. His research interests include signal processing, machine learning, and communication systems. He served as an Editor for IEEE COMMUNICATIONS LETTERS and IEEE TRANSACTIONS ON COMMUNICATIONS. He is currently an Editor for IEEE TRANSACTIONS ON SIGNAL PROCESSING and Journal of Communications and Networks.
\end{IEEEbiography}

\newpage

\allowdisplaybreaks

\begin{appendices}

\section{Derivation of \eqref{eqn:opt_solfiber}}
\label{app:derivation_opt}
Since $\| \bm{A}+\bm{B} \|_F^2 = \| \bm{A} \|_F^2 + \| \bm{B} \|_F^2 + 2\text{tr}(\bm{A}^T \bm{B})$, \eqref{eqn:opt_wrtTTcorefiber} can be re-written as
\begin{align}
    & \min_{{\bm{G}_{(3)}^{(d)}}_{:,p}}  \bigg\| \bm{O}_{(d)}\ast \big({\bm{G}_{(3)}^{(d)}}_{:,p} \Big[ \bm{G}_{(1)}^{(>d)}\otimes \bm{G}_{(d)}^{(<d)} \Big]_{p,:}\big) \bigg\|_F^2 \nonumber \\
    & + \beta_d {\bm{G}_{(3)}^{(d)}}_{:,p}^T \bm{L}^{(d)} {\bm{G}_{(3)}^{(d)}}_{:,p} - 2 \text{tr}\bigg(  \bm{\Xi} ^T \Big( \bm{O}_{(d)}\ast ({\bm{G}_{(3)}^{(d)}}_{:,p} \nonumber \\
    & \times \Big[ \bm{G}_{(1)}^{(>d)}\otimes \bm{G}_{(d)}^{(<d)} \Big]_{p,:}) \Big) \bigg).
     \label{eqn:opt_wrtTTcorefiber_app}
\end{align}

It is clearly that (\ref{eqn:opt_wrtTTcorefiber_app}) is quadratic with respect to each TT core fiber ${\bm{G}_{(3)}^{(d)}}_{:,p}$. In order to obtain the solution of (\ref{eqn:opt_wrtTTcorefiber_app}), we put the objective function of (\ref{eqn:opt_wrtTTcorefiber_app}) into a standard form ${\bm{G}_{(3)}^{(d)} }_{:,p}^T \bm{\Upsilon} {\bm{G}_{(3)}^{(d)} }_{:,p} + \bm{\mu}^T {\bm{G}_{(3)}^{(d)} }_{:,p}$. For $\bm{\Upsilon}$, it comes from the Frobenius norm and the graph regularization term in (\ref{eqn:opt_wrtTTcorefiber_app}), the latter of which is obvious. Since
\begin{align}
    \bigg\| \bm{O} \ast (\bm{a} \bm{b}^T) \bigg\|_F^2 &= \sum_{i=1} \sum_{j=1} \bm{O}_{ij} \bm{a}_i^2 \bm{b}_j^2 = \sum_i \bm{a}_i^2 (\sum_j \bm{O}_{ij} \bm{b}_j^2) \nonumber\\
    & =\sum_i \bm{a}_i^2 \bm{O}_{i,:}(\bm{b}\ast \bm{b})
    = \bm{a}^T \text{diag}\Big( \bm{O}(\bm{b}\ast \bm{b}) \Big) \bm{a}\nonumber,
\end{align}
in which $\bm{O}$ is a boolean matrix, the Frobenius norm in (\ref{eqn:opt_wrtTTcorefiber_app}) can be written as
\begin{align}
    & {\bm{G}_{(3)}^{(d)}}_{:,p}^T \text{diag}\bigg( \bm{O}_{(d)} (\Big[ \bm{G}_{(1)}^{(>d)}\otimes \bm{G}_{(d)}^{(<d)} \Big]_{p,:}^T \nonumber \\
    & \quad \ast \Big[ \bm{G}_{(1)}^{(>d)}\otimes \bm{G}_{(d)}^{(<d)} \Big]_{p,:}^T) \bigg)  {\bm{G}_{(3)}^{(d)}}_{:,p}.
\end{align}

For the coefficient $\bm{\mu}$, since
\begin{align}
    &\text{tr}\bigg(\bm{Y}^T \Big( \bm{O} \ast (\bm{a} \bm{b}^T)  \Big) \bigg) = \sum_i \sum_j \bm{O}_{ij} \bm{Y}_{ij} \bm{a}_i \bm{b}_j \nonumber \\
    & = \sum_i \bm{a}_i (\sum_j \bm{O}_{ij} \bm{Y}_{ij} \bm{b}_j) = \Big((\bm{O}\ast \bm{Y})\bm{b} \Big)^T\bm{a}, \nonumber
\end{align}
the trace term in (\ref{eqn:opt_wrtTTcorefiber_app}) can be written as
\begin{align}
    -2 \bigg( \Big(\bm{O}_{(d)} \ast \bm{\Xi}\Big) \Big[ \bm{G}_{(1)}^{(>d)}\otimes \bm{G}_{(d)}^{(<d)} \Big]_{p,:}^T  \bigg) {\bm{G}_{(3)}^{(d)}}_{:,p}.
    \label{eqn:traceterm}
\end{align}

Therefore, (\ref{eqn:opt_wrtTTcorefiber_app}) can be formulated as
\begin{align}
    \min_{\mathbm{G}_{r_d,r_{d+1},:}^{(d)}} {\bm{G}_{(3)}^{(d)}}_{:,p}^T \bm{\Upsilon} {\bm{G}_{(3)}^{(d)}}_{:,p} - 2 \bm{\mu}^T {\bm{G}_{(3)}^{(d)}}_{:,p},
    \label{eqn:opt_fiberstandardform}
\end{align}
with
\begin{align}
    \bm{\Upsilon} &= \text{diag}\bigg( \bm{O}_{(d)} (\Big[ \bm{G}_{(1)}^{(>d)}\otimes \bm{G}_{(d)}^{(<d)} \Big]_{p,:}^T \nonumber \\
    & \ast \Big[ \bm{G}_{(1)}^{(>d)}\otimes \bm{G}_{(d)}^{(<d)} \Big]_{p,:}^T) \bigg) + \beta_d \bm{L}^{(d)}, \\
    \bm{\mu} &=\Big(\bm{O}_{(d)} \ast \bm{\Xi}\Big) \Big[ \bm{G}_{(1)}^{(>d)}\otimes \bm{G}_{(d)}^{(<d)} \Big]_{p,:}^T,
\end{align}
and the solution of (\ref{eqn:opt_fiberstandardform}) is given by ${\bm{G}_{(3)}^{(d)}}_{:,p} = {\bm{\Upsilon}}^{-1} \bm{\mu}$.

\section{\textcolor{black}{Proof of Proposition 1}}
% \textcolor{blue}{\section{Proof of Proposition 1}}
\label{sec:apx-prioranalysis}
\textcolor{black}{We take the marginal distribution of $p(\mathbm{G}_{k,:,:}^{(d)})$ as an example, and the results are similar for $p(\mathbm{G}_{:,\ell,:}^{(d+1)})$. Firstly, notice that when $\bm{a}_\ell^{(d+1)}$ tends to $0$ and $\bm{\lambda}_\ell^{(d+1)}<0$, the distribution of $\bm{z}_\ell^{(d+1)}$ \eqref{eqn:zprior} becomes an inverse Gamma distribution \cite{babacan2014bayesian}
\begin{align}
    &p(\bm{z}_\ell^{(d+1)}) = \frac{{(\frac{\bm{b}_\ell^{(d+1)}}{2})}^{-\bm{\lambda}_\ell^{(d+1)}}}{\Gamma(-\bm{\lambda}_\ell^{(d+1)}) }  {\bm{z}_\ell^{(d+1)}}^{\bm{\lambda}_\ell^{(d+1)}-1} \nonumber \\
    &\quad \times \text{exp}(-\frac{\bm{b}_\ell^{(d+1)}}{2} {\bm{z}_\ell^{(d+1)}}^{-1}).
\end{align}
Then the joint distribution of $\mathbm{G}_{k,:,:}^{(d)}$, $\bm{z}_k^{(d)}$ and $\bm{z}^{(d+1)}$ can be derived as
\begin{align}
    &p(\mathbm{G}_{k,:,:}^{(d)},\bm{z}_k^{(d)},\bm{z}^{(d+1)}) \propto \prod_{\ell=1}^{S_{d+1}}\bigg( {({\bm{z}_k^{(d)}} {\bm{z}_\ell^{(d+1)}})}^{-J_d}  \nonumber\\
    & \quad \times \text{exp} \Big(-\frac{1}{2{\bm{z}_k^{(d)}} {\bm{z}_k^{(d)}}}{\mathbm{G}_{k,\ell,:}^{(d)} }^T \bm{L}^{(d)} {\mathbm{G}_{k,\ell,:}^{(d)} } \Big) \bigg)\nonumber \\
    & \quad \times {\bm{z}_k^{(d)}}^{\bm{\lambda}_k^{(d)} -1} \text{exp} \Big(  -\frac{\bm{a}_k^{(d)}}{2}  {\bm{z}_{k}^{(d)}} -\frac{\bm{b}_k^{(d)}}{2}  {\bm{z}_{k}^{(d)}}^{-1}\Big) \nonumber \\
    & \quad \times \prod_{\ell=1}^{S_{d+1}}\bigg( {\bm{z}_\ell^{(d+1)}}^{\bm{\lambda}_\ell^{(d+1)} -1} \text{exp} \Big(   -\frac{\bm{b}_\ell^{(d+1)}}{2}  {\bm{z}_{\ell}^{(d+1)}}^{-1}\Big)  \bigg). \nonumber
\end{align}
Extracting terms related to ${\bm{z}_\ell^{(d+1)}}$, the above equation can be reformulated as
\begin{align}
    &p(\mathbm{G}_{k,:,:}^{(d)},\bm{z}_k^{(d)},\bm{z}^{(d+1)}) \propto {\bm{z}_k^{(d)}}^{\bm{\lambda}_k^{(d)} -J_d -1} \text{exp} \Big(  -\frac{\bm{a}_k^{(d)}}{2}  {\bm{z}_{k}^{(d)}}\nonumber\\
    & \quad -\frac{\bm{b}_k^{(d)}}{2}  {\bm{z}_{k}^{(d)}}^{-1}\Big) \prod_{\ell=1}^{S_{d+1}}\bigg( {\bm{z}_\ell^{(d+1)}}^{\bm{\lambda}_\ell^{(d+1)} -J_d - 1} \nonumber \\
    & \quad  \times \text{exp} \Big( \frac{{\bm{z}_{\ell}^{(d+1)}}^{-1}}{2}  ({\bm{z}_{k}^{(d)}}^{-1}  {\mathbm{G}_{k,\ell,:}^{(d)} }^T \bm{L}^{(d)} {\mathbm{G}_{k,\ell,:}^{(d)} } + \bm{b}_\ell^{(d+1)})\Big)\bigg),
    \label{eqn:apx-jointG_z}
\end{align}
which reveals that the marginal distribution of ${\bm{z}_{\ell}^{(d+1)}}^{-1}$ also follows a inverse Gamma distribution. Then the joint distribution of $\mathbm{G}_{k,:,:}^{(d)}$ and $\bm{z}_k^{(d)}$ can be obtained by integrating out $\bm{z}_{\ell}^{(d+1)}$ for all $\ell$, as
\begin{align}
    &p(\mathbm{G}_{k,:,:}^{(d)},\bm{z}_k^{(d)}) \propto  {\bm{z}_k^{(d)}}^{\bm{\lambda}_k^{(d)} -J_d -1} \text{exp} \Big(  -\frac{\bm{a}_k^{(d)}}{2}  {\bm{z}_{k}^{(d)}}\nonumber\\
    & \quad -\frac{\bm{b}_k^{(d)}}{2}  {\bm{z}_{k}^{(d)}}^{-1}\Big) \prod_{\ell=1}^{S_{d+1}} \bigg( \frac{1}{2} ({\bm{z}_{k}^{(d)}}^{-1}  {\mathbm{G}_{k,\ell,:}^{(d)} }^T \bm{L}^{(d)} {\mathbm{G}_{k,\ell,:}^{(d)} } \nonumber \\
    & \quad + \bm{b}_\ell^{(d+1)}) \bigg)^{\bm{\lambda}_{\ell}^{(d+1)} - J_d}.
\end{align}
With $\bm{b}_\ell^{(d+1)}$ and $\bm{\lambda}_{\ell}^{(d+1)}$ tending to $0$ for all $\ell$, it can be observed that $\mathbm{G}_{k,:,:}^{(d)}$ and $\bm{z}_k^{(d)}$ become independent with each other, and then \eqref{eqn:apx-proposition1} is obtained. \hfill $\square$
}

\begin{figure*}[th!]
\normalsize
\setcounter{equation}{56}
% Set the equation number to one less than the one
% desired for the first equation here.
% The value here will have to changed if equations
% are added or removed prior to the place these
% equations are referenced in the main text.
\begin{align}
     \ln{q({\bm{G}_{(3)}^{(d)}}_{:,p})}  &= - \mathbb{E}_{\bm{\Theta} \setminus {\bm{G}_{(3)}^{(d)}}_{:,p}} \bigg\llbracket \frac{\tau}{2} \left\|\bm{\mathcal{O}} \ast \big(\bm{ \mathcal{Y}} {- \mathbm E} - \ll \bm{\mathcal{G}}^{(1)},\bm{\mathcal{G}}^{(2)},\ldots  ,\bm{\mathcal{G}}^{(D)} \gg \big) \right\|_F^2 +  \frac{{\bm{G}_{(3)}^{(d)}}_{:,p}^T \bm{L}^{(d)} {\bm{G}_{(3)}^{(d)}}_{:,p}}{\bm{z}_{k_p}^{(d)}\bm{z}_{\ell_p}^{(d+1)}} \bigg\rrbracket + \text{const} \nonumber \\
    & = -\frac{1}{2} {\bm{G}_{(3)}^{(d)}}_{:,p}^T\Bigg(\mathbb{E}{\big\llbracket \tau\big\rrbracket}\text{diag}\bigg( \bm{O}_{(d)} \mathbb{E}\bigg\llbracket\Big[ \bm{G}_{(1)}^{(>d)}\otimes \bm{G}_{(d)}^{(<d)} \Big]_{p,:}^T \ast \Big[ \bm{G}_{(1)}^{(>d)}\otimes \bm{G}_{(d)}^{(<d)} \Big]_{p,:}^T\bigg\rrbracket  \bigg) + \mathbb{E}\bigg\llbracket\frac{1}{\bm{z}_{k_p}^{(d)}\bm{z}_{\ell_p}^{(d+1)}}\bigg\rrbracket\bm{L}^{(d)}\Bigg){\bm{G}_{(3)}^{(d)}}_{:,p} \nonumber \\
    & + \mathbb{E}{\big\llbracket \tau\big\rrbracket} \Bigg(\Big(\bm{O}_{(d)} \ast {(\bm{Y}_{(d)} - \mathbb{E}\llbracket \bm E_{(d)}\rrbracket)}\Big) \mathbb{E}\bigg\llbracket \Big[ \bm{G}_{(1)}^{(>d)}\otimes \bm{G}_{(d)}^{(<d)} \Big]_{p,:}^T\bigg\rrbracket \nonumber \\
    & - \mathbb{E}\bigg\llbracket \underbrace{\bm{O}_{(d)} \ast \Big(\sum_{q=1,q\neq p }^{S_dS_{d+1}} {\bm{G}_{(3)}^{(d)}}_{:,q} \Big[ \bm{G}_{(1)}^{(>d)}\otimes \bm{G}_{(d)}^{(<d)} \Big]_{q,:} \Big)\Big[ \bm{G}_{(1)}^{(>d)}\otimes \bm{G}_{(d)}^{(<d)} \Big]_{p,:}^T}_{\bm{\phi}}\bigg\rrbracket\Bigg)^T {\bm{G}_{(3)}^{(d)}}_{:,p} + \text{const},
    \label{eqn:q_fiberapx}
\end{align}
% Restore the current equation number.
% IEEE uses as a separator
\setcounter{equation}{55}
\end{figure*}

\section{Derivation of VI TT completion with graph regularization}
\label{app:vi_updatederivation}
Firstly, based on the proposed probabilistic model \eqref{eqn:ttlikelihood}-\eqref{eqn:aprior} and \eqref{eqn:outlierprior}, the logarithm of the joint distribution of the observed tensor and all the variables is derived as
\begin{align}
    & \ln{ \left( {p(\bm{\mathcal{Y}}, \bm{\Theta})} \right)}\nonumber \\
    & = \frac{|\Omega|}{2}\ln{\tau} -\frac{\tau}{2} \left\|\bm{\mathcal{O}} \ast \big(\bm{ \mathcal{Y}} - \ll \bm{\mathcal{G}}^{(1)},\bm{\mathcal{G}}^{(2)},\ldots  ,\bm{\mathcal{G}}^{(D)} \gg - {\mathbm E }\big) \right\|_F^2\nonumber \\
    & - \frac{1}{2}\sum_{d=1}^{D} \sum_{k}^{S_d} \sum_{\ell}^{S_{d+1}} \Big( J_d\ln(\bm{z}_{k}^{(d)}\bm{z}_{\ell}^{(d+1)}) + \frac{{{\bm{\mathcal{G}}_{k,\ell,:}^{(d)}}}^T \bm{L}^{(d)} {{\bm{\mathcal{G}}_{k,\ell,:}^{(d)}}}}{\bm{z}_k^{(d)}\bm{z}_\ell^{(d+1)}}\Big)\nonumber \\
    &  + \sum_{d=2}^{D} \sum_{k=1}^{S_d} \Big( \big( \bm{\lambda}_{k}^{(d)}-1 \big){\ln{\bm{z}_{k}^{(d)}}} - \frac{1}{2} (\bm{a}_k^{(d)}\bm{z}_k^{(d)} + \bm{b}_k^{(d)}\frac{1}{\bm{z}_k^{(d)}}) \nonumber \\
    & + \frac{\bm{\lambda_k^{(d)}}}{2} \ln \bm{a}_k^{(d)} + (\bm{c}_d-1) \ln \bm{a}_k^{(d)} - \bm{f}_d \bm{a}_k^{(d)} \Big)  + \left( a_{\tau}-1 \right)\ln{\tau}\nonumber\\
    & - b_{\tau}\tau {+ \sum_{j_1=1}^{J_1}\ldots \sum_{j_1=D}^{J_D} \Big( - \mathbm U_{j_1\ldots j_D} (\frac{1}{2}\mathbm E_{j_1\ldots j_D}^2 + \mathbm Q_{j_1\ldots j_D})} \nonumber \\
    & {+ (\mathbm P_{j_1\ldots j_D} - \frac{1}{2}) \ln \mathbm U_{j_1\ldots j_D}  \Big) }+ \text{const}.
\label{eqn:TTjointapdx}
\end{align}
\setcounter{equation}{57}
To make the equations of VI update be expressed using notations in deterministic optimization algorithm in Section 3, we notice that $\mathbm{G}_{k,\ell,:}^{(d)}$ with $k$ from $1$ to $S_d$ and $\ell$ from $1$ to $S_{d+1}$ is equivalent to ${\bm{G}_{(3)}^{(d)}}_{:,p}$ for $p$ from $1$ to $S_dS_{d+1}$, under the bijection $p = (\ell_p-1)S_d + k_p$.

Then, according to the optimal variational distribution \eqref{eqn:viupdate}, $q({\bm{G}_{(3)}^{(d)}}_{:,p})$ is obtained by taking expectation on (\ref{eqn:TTjointapdx}) and focusing on terms that are only related to ${\bm{G}_{(3)}^{(d)}}_{:,p}$, in which previous results \eqref{eqn:opt_wrtTTcorefiber}, (\ref{eqn:opt_wrtTTcorefiber_app})-(\ref{eqn:traceterm}) are used. It can be seen that (\ref{eqn:q_fiberapx}) is quadratic with respect to ${\bm{G}_{(3)}^{(d)}}_{:,p}$, and therefore it follows a Gaussian distribution with covariance matrix and mean
\begin{align}
    & \bm{\Sigma}^{(d,p)} = \Bigg(\mathbb{E}\bigg\llbracket\frac{1}{\bm{z}_{k_p}^{(d)}}\bigg\rrbracket  \mathbb{E}\bigg\llbracket\frac{1}{\bm{z}_{\ell_p}^{(d+1)}}\bigg\rrbracket\bm{L}^{(d)} + \mathbb{E}{\big\llbracket \tau\big\rrbracket}\text{diag}\bigg( \bm{O}_{(d)}\nonumber \\
    & \times \mathbb{E}\bigg\llbracket \underbrace{\Big[ \bm{G}_{(1)}^{(>d)}\otimes \bm{G}_{(d)}^{(<d)} \Big]_{p,:}^T \ast \Big[ \bm{G}_{(1)}^{(>d)}\otimes \bm{G}_{(d)}^{(<d)} \Big]_{p,:}^T}_{\mathbm{KG}_{p,p,:}^{(d)}} \bigg\rrbracket  \bigg)\Bigg)^{-1},
    \label{eqn:Sigma_fiberdp_apx}
\end{align}
\begin{align}
    & \bm{\nu}^{(d,p)} =\mathbb{E}{\big\llbracket \tau\big\rrbracket} \bm{\Sigma}^{(d,p)}  \bigg(\Big(\bm{O}_{(d)} \ast (\bm{Y}_{(d)} - {\mathbb E \llbracket\bm E_{(d)} \rrbracket})\Big)\nonumber \\
    & \times \mathbb{E}\bigg\llbracket \Big[ \bm{G}_{(1)}^{(>d)}\otimes \bm{G}_{(d)}^{(<d)} \Big]_{p,:}^T\bigg\rrbracket - \mathbb{E}\big\llbracket \bm{\phi}\big\rrbracket \bigg),
    \label{eqn:mean_fiberdp_nuapx}
\end{align}
respectively, where $\mathbm{KG}_{q,p,:}^{(d)}\in \mathbb{R}^{J_1\ldots J_{d-1}J_{d+1}\ldots J_{D}}$ is defined as $\Big[ \bm{G}_{(1)}^{(>d)}\otimes \bm{G}_{(d)}^{(<d)} \Big]_{q,:}^T \ast \Big[ \bm{G}_{(1)}^{(>d)}\otimes \bm{G}_{(d)}^{(<d)} \Big]_{p,:}^T$ for any $q$ and $p$ from $1$ to $S_dS_{d+1}$. The difficulty of calculating (\ref{eqn:Sigma_fiberdp_apx}) and (\ref{eqn:mean_fiberdp_nuapx}) comes from the expectation of $\mathbm{KG}_{p,p,:}^{(d)}$ and $\bm{\phi}$, in which the TT cores are heavily coupled and contains square terms. Below we will first rewriting $\bm{\phi}$, which turns out is related to $\mathbb{E}\llbracket\mathbm{KG}_{q,p,:}^{(d)} \rrbracket$.

Since
\begin{align}
    \Big[\big(\bm{A} \ast (\bm{b} \bm{c}^T) \big) \bm{d}\Big]_i = \bm{b}_i\sum_{j} \bm{A}_{ij} \bm{c}_j \bm{d_j},\nonumber
\end{align}
it can be verified that $\big(\bm{A} \ast (\bm{b} \bm{c}^T) \big) \bm{d} = \text{diag}(\bm{b})\bm{A}(\bm{c}\ast \bm{d})$. Using this result, we obtain
\begin{align}
    & \mathbb{E}\big\llbracket\bm{\phi}\big\rrbracket = \sum_{q=1,q\neq p}^{S_dS_{d+1}} \text{diag}\bigg( \mathbb{E}\Big\llbracket {\bm{G}_{(3)}^{(d)}}_{:,q} \Big\rrbracket\bigg) \bm{O}_{(d)}\nonumber\\
    & \times \mathbb{E}\bigg \llbracket \underbrace{\Big[ \bm{G}_{(1)}^{(>d)}\otimes \bm{G}_{(d)}^{(<d)} \Big]_{q,:}^T \ast \Big[ \bm{G}_{(1)}^{(>d)}\otimes \bm{G}_{(d)}^{(<d)} \Big]_{p,:}^T}_{\mathbm{KG}_{q,p,:}^{(d)}}\bigg \rrbracket.
\end{align}

According to \textit{Definition 2} and the definition of $\bm{G}^{(<d)}$ and $\bm{G}^{(>d)}$ in \textit{Property 1},
\begin{align}
    &\mathbm{KG}_{q,p,i}^{(d)} = \bigg( \mathbm{G}_{1,:,j_1}^{(1)}\ldots \mathbm{G}_{:,m,j_{d-1}}^{(d-1)} \mathbm{G}_{n,:,j_d}^{(d+1)}\ldots \mathbm{G}_{:,1,j_D}^{(D)} \bigg) \nonumber\\
    &\quad  \times \bigg( \mathbm{G}_{1,:,j_1}^{(1)}\ldots \mathbm{G}_{:,k,j_{d-1}}^{(d-1)} \mathbm{G}_{\ell,:,j_d}^{(d+1)}\ldots \mathbm{G}_{:,1,j_D}^{(D)} \bigg) \nonumber\\
    & = \Big( \mathbm{G}_{1,:,j_1}^{(1)}\otimes \mathbm{G}_{1,:,j_1}^{(1)}\Big)\ldots \Big( \mathbm{G}_{:,m,j_{d-1}}^{(d-1)}\otimes \mathbm{G}_{:,k,j_{d-1}}^{(d-1)}\Big) \nonumber \\
    & \quad \times \Big( \mathbm{G}_{n,:,j_{d+1}}^{(d+1)}\otimes \mathbm{G}_{\ell,:,j_{d+1}}^{(d+1)}\Big)\ldots\Big( \mathbm{G}_{:,1,j_D}^{(D)}\otimes \mathbm{G}_{:,1,j_D}^{(D)}\Big),\
    \label{eqn:KG2sperate}
\end{align}
with bijections $i = j_1 + \prod_{s=2 , s\neq d}^D \Big( (j_s -1) \prod_{t=1 , t \neq d}^{s-1} J_{t} \Big)$, $q = (n-1)R_{d}+m$ and $p = (\ell-1)S_d + k$. In the last line of (\ref{eqn:KG2sperate}), since the TT cores are separated, expectation of $\mathbm{KG}$ can be obtained by the product of the expectations on the kronecker product of the TT core frontal slices
\begin{align}
    &\mathbb{E}\big \llbracket \mathbm{G}_{:,:,j_t}^{(t)}\otimes \mathbm{G}_{:,:,j_t}^{(t)} \big \rrbracket = \mathbb{E}\big \llbracket \mathbm{G}_{:,:,j_t}^{(t)}\big \rrbracket \otimes \mathbb{E}\big \llbracket \mathbm{G}_{:,:,j_t}^{(t)}\big \rrbracket\nonumber \\
    &+ \underbrace{ \mathbb{E}\Big\llbracket \big(\mathbm{G}_{:,:,j_t}^{(t)} - \mathbb{E}\llbracket\mathbm{G}_{:
    ,:,j_t}^{(t)} \rrbracket\big)\otimes\big(\mathbm{G}_{:,:,j_t}^{(t)} - \mathbb{E}\llbracket\mathbm{G}_{:,:,j_t}^{(t)} \rrbracket\big) \Big\rrbracket}_{\bm{\mathrm{Var}}^{(t,j_t)}},
\end{align}
where $\mathbb{E}\big \llbracket \mathbm{G}_{k,\ell,j_t}^{(t)} \big \rrbracket= \bm{\nu}_{j_t}^{(t,(\ell-1)S_t+k)}$, and $\bm{\mathrm{Var}}^{(t,j_t)}$ comes from the covariance matrix of $\mathbm{G}_{:,:,j_t}^{(t)}$ but with elements permuted in another order. Since the mean-field approximation \eqref{eqn:meanfieldproposed} assumes that different mode-$3$ fibers of $\mathbm{G}^{(d)}$ are independent of each other, $\bm{\mathrm{Var}}^{(t,j_t)}$ would be a very sparse matrix, in which only elements with index pairs $[\{(k-1)R_t+k\}_{k=1}^{R_t},\{(\ell-1)R_{t+1}+\ell\}_{\ell=1}^{R_{t+1}}]$ are non-zero, with value
\begin{align}
    \quad \bm{\mathrm{Var}}_{(k-1)S_t+k,(\ell-1)S_{t+1}+\ell}^{(t,j_t)}  = \bm{\Sigma}_{j_t,j_t}^{(t,(\ell-1)S_t + k)}.
    \label{eqn:kronvarapd}
\end{align}

On the other hand, the variational distribution of $\bm{z}^{(d)}$ is obtained by taking expectations on (\ref{eqn:TTjointapdx}) and focusing only on the terms related to $\bm{z}^{(d)}$, it is obtained that
\begin{align}
    \ln q(\bm{z}^{(d)}) = \sum_{k=1}^{S_d}\ln q(\bm{z}_k^{(d)}) + \text{const}, \nonumber
\end{align}
with
\begin{align}
     & \ln q(\bm{z}_k^{(d)})= \bigg( \bm{\lambda}_k^{(d)} - \frac{J_d S_{d+1}}{2} - \frac{J_{d-1}S_{d-1}}{2}-1\bigg) \ln \bm{z}_k^{(d)} \nonumber \\
     &  -\frac{1}{2} \bigg( \mathbb{E}\llbracket\bm{a}_k^{(d)} \rrbracket\bigg) \bm{z}_k^{(d)} - \frac{1}{2}  \bigg( \bm{b}_k^{(d)} + \sum_{\ell=1}^{S_{d-1}} \mathbb{E} \big\llbracket \frac{1}{\bm{z}_\ell^{(d-1)}}\big\rrbracket \nonumber \\
    & \times  \mathbb{E} \big\llbracket {\bm{\mathcal{G}}_{\ell,k,:}^{(d-1)}}^T \bm{L}^{(d-1)} \bm{\mathcal{G}}_{\ell,k,:}^{(d-1)} \big\rrbracket \nonumber \\
    &  + \sum_{\ell=1}^{S_{d+1}} \mathbb{E} \big\llbracket \frac{1}{\bm{z}_\ell^{(d+1)}}\big\rrbracket\mathbb{E} \big\llbracket {\bm{\mathcal{G}}_{\ell,k,:}^{(d)}}^T \bm{L}^{(d)} \bm{\mathcal{G}}_{\ell,k,:}^{(d)} \big\rrbracket \bigg) \frac{1}{\bm{z}_k^{(d)}}.
    \label{eqn:PDFzapd}
\end{align}
Notice that in (\ref{eqn:PDFzapd}) there are only terms linear to $\ln\bm{z}_k^{(d)}$, $\bm{z}_k^{(d)}$ and ${1}/{\bm{z}_k^{(d)}}$. Comparing (\ref{eqn:PDFzapd}) to \eqref{eqn:gigdiscribe}, we obtain that $\bm{z}_k^{(d)}$ follows $\text{GIG}(\hat{\bm{a}_k^{(d)}},\hat{\bm{\lambda}}_{k}^{(d)},\hat{\bm{b}}_k^{(d)})$, with parameters
\begin{align}
    \hat{\bm{a}}_k^{(d)} = \mathbb{E}\big\llbracket{\bm{a}_k^{(d)}}\big\rrbracket,
\end{align}
\begin{align}
    \hat{\bm{\lambda}}_{k}^{(d)} = \bm{\lambda}_{k}^{(d)} - \frac{J_d S_{d+1}}{2} - \frac{J_{d-1} S_{d-1}}{2},
\end{align}
\begin{align}
    & \hat{\bm{b}}_k^{(d)} = \bm{b}_k^{(d)} + \frac{1}{2}\sum_{\ell=1}^{S_{d-1}} \mathbb{E}\big\llbracket\frac{1}{\bm{z}_\ell^{(d-1)}}\big\rrbracket \mathbb{E}\big\llbracket{\mathbm{G}_{\ell,k,:}^{(d-1)}}^T \bm{L}^{(d-1)} {\mathbm{G}_{\ell,k,:}^{(d-1)}}\big\rrbracket \nonumber \\
    & + \sum_{\ell=1}^{S_{d+1}} \mathbb{E}\big\llbracket\frac{1}{\bm{z}_\ell^{(d+1)}}\big\rrbracket\mathbb{E}\llbracket{\mathbm{G}_{\ell,k,:}^{(d)}}^T \bm{L}^{(d)} {\mathbm{G}_{\ell,k,:}^{(d)}}\rrbracket.
\end{align}

Similarly, by taking expectation on (\ref{eqn:TTjointapdx}) with respect to $\bm{a}^{(d)}$, the variational distribution of $\bm{a}^{(d)}$ is
\begin{align}
    & \ln q( \bm{a}^{(d)}) = \sum_{k=1}^{S_d} \bigg((\bm{c}_d + \frac{ \bm{\lambda}_k^{(d)}}{2}-1) \ln  \bm{a}_{k}^{(d)} \nonumber\\
    &- ( \bm{f}_k^{(d)} + \frac{\mathbb{E}\llbracket\bm{z}_k^{(d)}\rrbracket}{2}) \bm{a}_{k}^{(d)} \bigg) + \text{const},
\end{align}
in which there are only terms related with $\ln{\bm{a}_{k}^{(d)}}$ and $\bm{a}_{k}^{(d)}$, indicating that $q(\bm{a}_{k}^{(d)})$ is a Gamma distribution with parameters
\begin{align}
    \hat{\bm{c}}_k^{(d)} = \bm{c}_k^{(d)} + \frac{\hat{\bm{\lambda}}_{k}^{(d)}}{2},
\end{align}
\begin{align}
    \hat{\bm{f}}_k^{(d)} = \bm{f}_k^{(d)} + \frac{\mathbb{E}\llbracket\bm{z}_k^{(d)}\rrbracket}{2}.
\end{align}

{Next, we derive the updates for the outlier-related variables---$\mathbm E$ and $\mathbm U$. Taking expectations of \eqref{eqn:TTjointapdx} w.r.t. $\mathbm E$, the variational distribution for each element of $\mathbm E$ is Gaussian, given by
\begin{align} \label{eq:VI_Eposterior}
& \ln q(\mathbm E_{j_1\ldots j_D}) = 
    -\frac{1}{2} ( \mathbb E \llbracket \tau \rrbracket \mathbm O_{j_1\ldots j_D} + \mathbb E\llbracket \mathbm U_{j_1\ldots j_D} \rrbracket) \mathbm E_{j_1\ldots j_D}^2 \nonumber \\
    & + \mathbb E \llbracket \tau \rrbracket \mathbm O_{j_1\ldots j_D} (\mathbm Y_{j_1\ldots j_D} - \mathbb E \llbracket \mathbm G_{:,:,j_1}^{(1)}\rrbracket\ldots \mathbb E \llbracket \mathbm G_{:,:,j_D}^{(D)}\rrbracket) \mathbm E_{j_1\ldots j_D},
\end{align}
The variance and mean of $\mathbm E_{j_1\ldots j_D}$ are
\begin{align}
    \mathbm V_{j_1\ldots j_D} = ( \mathbb E \llbracket \tau \rrbracket \mathbm O_{j_1\ldots j_D} + \mathbb E\llbracket \mathbm U_{j_1\ldots j_D} \rrbracket)^{-1},
\end{align}
\begin{align}
    &\mathbm M_{j_1 \ldots j_D} = \nonumber \\
    & \mathbb E\llbracket \tau \rrbracket \mathbm O_{j_1\ldots j_D}\mathbm V_{j_1\ldots j_D}  (\mathbm Y_{j_1\ldots j_D} - \mathbb E \llbracket \mathbm G_{:,:,j_1}^{(1)}\rrbracket\ldots \mathbb E \llbracket \mathbm G_{:,:,j_D}^{(D)}\rrbracket),
\end{align}
respectively.
}

{Similarly, the variational distribution for $\mathbm U$ is derived by taking expectations of \eqref{eqn:TTjointapdx} with respect to $\mathbm U$. The log-density is
\begin{align}
    & \ln q(\mathbm U_{j_1\ldots j_D}) =  - (\frac{1}{2} \mathbb E \llbracket \mathbm E_{j_1\ldots j_D}^2 \rrbracket + \mathbm Q_{j_1\ldots j_D}) \mathbm U_{j_1\ldots j_D} \nonumber \\
    & + (\mathbm P_{j_1\ldots j_D} - \frac{1}{2}) \ln \mathbm U_{j_1\ldots j_D} ,
\end{align}
which corresponds to a Gamma distribution with the following updated parameters:
\begin{align}
    \hat{ \mathbm P }_{j_1\ldots j_D} &= \mathbm P_{j_1\ldots j_D} + \frac{1}{2}  ,\\
    \hat{ \mathbm Q }_{j_1\ldots j_D} &= \mathbm Q_{j_1\ldots j_D} + \frac{1}{2} \mathbb E \llbracket \mathbm E_{j_1\ldots j_D}^2 \rrbracket .
\end{align}
}

Finally, by taking expectation of (\ref{eqn:TTjointapdx}) with respect to $\tau$, its variational distribution is
\begin{align}
    &\ln{q\left( \tau \right)} \nonumber \\
    &= -\Bigg(\frac{1}{2} \bigg({\mathbb E \llbracket \left\| \bm{\mathcal{O}}\ast  ( \bm{\mathcal{Y}} - \mathbm E)\right\|_F^2 \rrbracket } - 2\sum_{j_1=1}^{J_1}\ldots \sum_{j_D=1}^{J_D}\bm{\mathcal{O}}_{j_1\ldots j_D} (\bm{\mathcal{Y}}_{j_1\ldots j_D} \nonumber \\
    & {- \mathbb E\llbracket \mathbm E_{j_1\ldots j_D} \rrbracket} ) \times \mathbb{E}\llbracket\bm{\mathcal{G}}_{:,:,j_1}^{(1)}\rrbracket\ldots \mathbb{E}\llbracket\bm{\mathcal{G}}_{:,:,j_D}^{(D)}\rrbracket + \sum_{j_1=1}^{J_1}\ldots \sum_{j_D=1}^{J_D} \bm{\mathcal{O}}_{j_1\ldots j_D}\nonumber\\
    &\quad \times \mathbb{E}\llbracket\bm{\mathcal{G}}_{:,:,j_1}^{(1)}\otimes \bm{\mathcal{G}}_{:,:,j_1}^{(1)}\rrbracket \ldots \mathbb{E}\llbracket\bm{\mathcal{G}}_{:,:,j_D}^{(D)}\otimes \bm{\mathcal{G}}_{:,:,j_D}^{(D)}\rrbracket \bigg) +b_\tau\Bigg)\tau \nonumber \\
    &  \quad +  \bigg(\frac{|\Omega|}{2}+{a}_{\tau} -1 \bigg) \ln \tau+\text{const}.
    \label{eqn:posteriortau}
\end{align}
which shows that $\tau$ follows a Gamma distribution, with parameters
\begin{align}
    \hat{a}_{\tau} = a_{\tau} + \frac{|\Omega|}{2},
\end{align}
and
\begin{align}
    &\hat{\beta}_{\tau} = \frac{1}{2} \bigg({\mathbb E \llbracket \left\| \bm{\mathcal{O}}\ast  (\bm{\mathcal{Y}} - \mathbm E) \right\|_F^2 \rrbracket }  - 2\sum_{j_1=1}^{J_1}\ldots \sum_{j_D=1}^{J_D}\bm{\mathcal{O}}_{j_1\ldots j_D} \nonumber\\
    & \times {(\bm{\mathcal{Y}}_{j_1\ldots j_D} - \mathbb E \llbracket \mathbm E_{j_1 \ldots j_D} \rrbracket)}\mathbb{E}\llbracket\bm{\mathcal{G}}_{:,:,j_1}^{(1)}\rrbracket\ldots \mathbb{E}\llbracket\bm{\mathcal{G}}_{:,:,j_D}^{(D)}\rrbracket  \nonumber \\
    &  + \sum_{j_1=1}^{J_1}\ldots \sum_{j_D=1}^{J_D} \bm{\mathcal{O}}_{j_1\ldots j_D}\mathbb{E}\llbracket\bm{\mathcal{G}}_{:,:,j_1}^{(1)}\otimes \bm{\mathcal{G}}_{:,:,j_1}^{(1)}\rrbracket \ldots \mathbb{E}\llbracket\bm{\mathcal{G}}_{:,:,j_D}^{(D)}\otimes \bm{\mathcal{G}}_{:,:,j_D}^{(D)}\rrbracket \bigg) +b_\tau.
\label{eqn:b_tau_update}
\end{align}
{
The only unknown term in \eqref{eqn:b_tau_update} is $\mathbb E \llbracket \left\| \bm{\mathcal{O}}\ast  (\bm{\mathcal{Y}} - \mathbm E) \right\|_F^2 \rrbracket$. 
Since the variational distribution of $\mathbm E$ is Gaussian, we can compute this expectation as follows:
\begin{align}
    &\mathbb E \llbracket \left\| \bm{\mathcal{O}}\ast  (\bm{\mathcal{Y}} - \mathbm E) \right\|_F^2 \rrbracket
    \nonumber \\
    = & \sum_{j_1=1}^{J_1} \ldots \sum_{j_D=1}^{J_D} \mathbm O_{j_1\ldots j_D} \Big((\mathbm Y_{j_1\ldots j_D} - \mathbb E\llbracket \mathbm E_{j_1\ldots j_D} \rrbracket)^2 + \mathbm V_{j_1\ldots j_D} \Big).
\end{align}
}

\end{appendices}

\end{document}